\pdfoutput=1

\documentclass[11pt,twoside,a4paper,cmspaper,final,collab]{cms-tdr}
\def\svnVersion{473933P}\def\cmsCernNoTag{CERN-EP-2018-069}\def\cmsCernDate{\today}\def\cmsMessage{Published in the Journal of High Energy Physics as \href{http://dx.doi.org/10.1007/JHEP08(2018)177}{\doi{10.1007/JHEP08(2018)177.}}}

\begin{document}\cmsNoteHeader{B2G-17-011}

\hyphenation{had-ron-i-za-tion}
\hyphenation{cal-or-i-me-ter}
\hyphenation{de-vices}
\RCS$HeadURL: svn+ssh://svn.cern.ch/reps/tdr2/papers/B2G-17-011/trunk/B2G-17-011.tex $
\RCS$Id: B2G-17-011.tex 470539 2018-08-01 15:22:29Z jmhogan $
\newlength\cmsFigWidth
\newlength\cmsTabSkip\setlength{\cmsTabSkip}{1ex}
\ifthenelse{\boolean{cms@external}}{\setlength\cmsFigWidth{0.85\columnwidth}}{\setlength\cmsFigWidth{0.4\textwidth}}
\ifthenelse{\boolean{cms@external}}{\providecommand{\cmsLeft}{top\xspace}}{\providecommand{\cmsLeft}{left\xspace}}
\ifthenelse{\boolean{cms@external}}{\providecommand{\cmsRight}{bottom\xspace}}{\providecommand{\cmsRight}{right\xspace}}

\providecommand{\PQT}{\ensuremath{\cmsSymbolFace{T}}\xspace}
\providecommand{\PQB}{\ensuremath{\cmsSymbolFace{B}}\xspace}
\providecommand{\PAQT}{\ensuremath{\overline{\cmsSymbolFace{T}}}\xspace}
\providecommand{\PAQB}{\ensuremath{\overline{\cmsSymbolFace{B}}}\xspace}
\newcommand{\ST}{\ensuremath{S_\mathrm{T}}\xspace}
\newcommand{\TTbar}{\ensuremath{\PQT\PAQT}\xspace}
\newcommand{\BBbar}{\ensuremath{\PQB\PAQB}\xspace}
\newcommand{\DRak}{\ensuremath{\Delta\mathrm{R}_{\text{min}}}\text{(leading AK8, other AK8)}}
\newcommand{\minMlb}{\ensuremath{\min[M(\ell, \cPqb)]}}
\newcommand{\minMlj}{\ensuremath{\min[M(\ell, \mathrm{j})]}}
\newcommand{\HTlep}{\ensuremath{\HT^{\text{lep}}}}
\providecommand{\cmsTable}[1]{\resizebox{\textwidth}{!}{#1}}
\providecommand{\NA}{\ensuremath{\text{---}}}
\providecommand{\CL}{CL\xspace}

\cmsNoteHeader{B2G-17-011}
\title{Search for vector-like $\PQT$ and $\PQB$ quark pairs in final states with leptons at $\sqrt{s} = 13$\TeV}

\date{\today}

\abstract{A search is presented for pair production of heavy vector-like $\PQT$ and $\PQB$ quarks in proton-proton collisions at $\sqrt{s} = 13\TeV$. The data sample corresponds to an integrated luminosity of 35.9\fbinv, collected with the CMS detector at the CERN LHC in 2016. Pair production of $\PQT$ quarks would result in a wide range of final states, since vector-like $\PQT$ quarks of charge 2$e$/3 are predicted to decay to $\cPqb\PW$, $\cPqt\PZ$, and $\cPqt\PH$. Likewise, vector-like $\PQB$ quarks are predicted to decay to $\cPqt\PW$, $\cPqb\PZ$, and $\cPqb\PH$. Three channels are considered, corresponding to final states with a single lepton, two leptons with the same sign of the electric charge, or at least three leptons. The results exclude $\PQT$ quarks with masses below 1140--1300\GeV and $\PQB$ quarks with masses below 910--1240\GeV for various branching fraction combinations, extending the reach of previous CMS searches by 200--600\GeV. 
}

\hypersetup{
pdfauthor={CMS Collaboration},
pdftitle={Search for vector-like T and B quark pairs in final states with leptons at sqrt(s) = 13 TeV},
pdfsubject={CMS},
pdfkeywords={CMS, physics, vector-like quarks}}

\maketitle

\section{Introduction}

The discovery of a Higgs boson~\cite{Aad20121,Chatrchyan201230,Chatrchyan:2013lba} (\PH) has further encouraged searches for new physics at the CERN LHC. Potentially divergent loop corrections to the Higgs boson mass require either significant fine tuning of the standard model (SM) parameters or new particles at the TeV scale. The existence of heavy top quark partners is particularly well motivated to cancel the largest corrections from SM top quark loops. In supersymmetric theories bosonic partners of the top quark serve this purpose, but in several other theories, such as little Higgs~\cite{PhysRevD.69.075002,Matsedonskyi2013} or composite Higgs~\cite{PhysRevD.75.055014, compHiggs, KAPLAN1991259, Dugan:1984hq} models, this role is filled by fermionic top quark partners. These heavy quark partners interact predominantly with the third generation of the SM quarks~\cite{vecQuarkMix,PhysRevLett.82.1628} and have vector-like transformation properties under the SM gauge group $\mathrm{ SU(2)_L\times U(1)_Y\times SU(3)_C }$ , inspiring the name ``vector-like quarks'' (VLQs). A heavy fourth generation of chiral quarks has been excluded by precision electroweak measurements from electron-positron collisions~\cite{LEP-2,PhysRevLett.109.241802} and by the measurement of Higgs-boson-mediated cross sections~\cite{Djouadi2012310,Chatrchyan:2013sfs}, but VLQs are not excluded by these experimental data.

\begin{figure}[hbtp]
\centering
\includegraphics[width=0.48\textwidth]{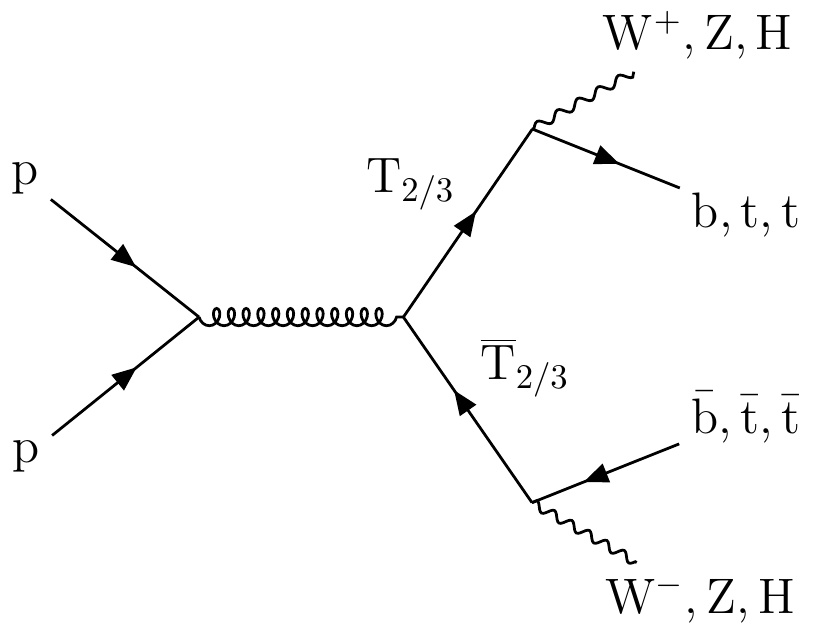}
\hspace{0.3cm}
\includegraphics[width=0.48\textwidth]{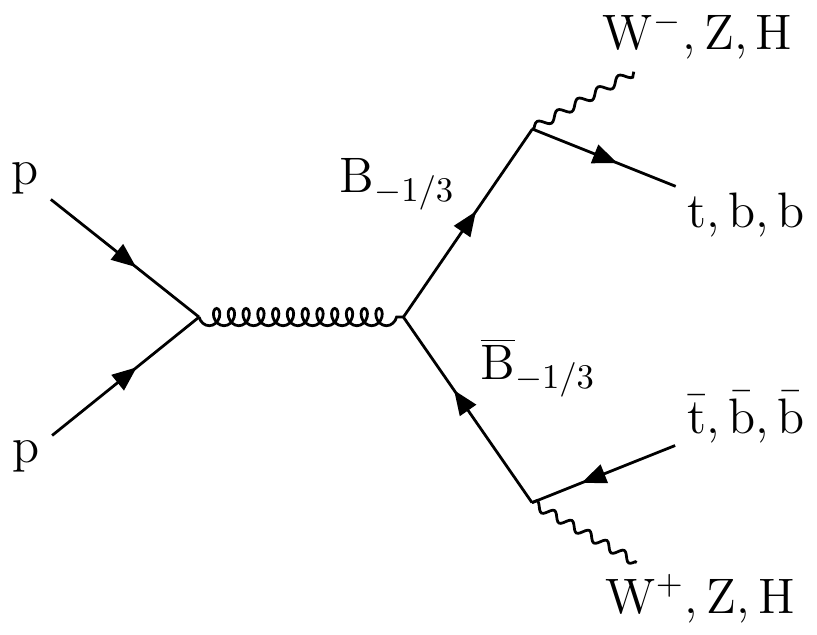}
\caption{Leading order Feynman diagrams showing pair production and decays of \TTbar (left) and \BBbar (right).}
\label{fig:diagrams}
\end{figure}

We search for a vector-like \PQT quark with charge 2$e$/3 that is produced in pairs with its antiquark, \PAQT, via the strong interaction in proton-proton collisions at $\sqrt{s} = 13$\TeV. Our search uses a data sample corresponding to an integrated luminosity of 35.9\fbinv, collected with the CMS detector in 2016. Many models in which VLQs appear assume that \PQT quarks may decay to three final states: \cPqb\PW, \cPqt\PZ, or \cPqt\PH~\cite{PhysRevD.88.094010}, as illustrated by the diagrams in Fig.~\ref{fig:diagrams}. The partial decay widths depend on the particular model \cite{DeSimone2013}, but for VLQ masses significantly larger than the \PW\ boson mass, as considered here, an electroweak singlet \PQT quark is expected to have branching fractions ($\mathcal{B}$) of 50\% for $\PQT\to\cPqb\PW$, and 25\% for both $\PQT\to\cPqt\PZ$ and \cPqt\PH~\cite{delAguila:1989rq, DeSimone2013}. A doublet \PQT quark decays only to \cPqt\PZ and \cPqt\PH, each with 50\% branching fraction. Although this search is optimized for \TTbar production, vector-like bottom (B) quark decays can produce similar final state signatures, as illustrated in Fig.~\ref{fig:diagrams} (right), and are also considered. A \PQB quark with charge $-e/3$ is expected to decay to \cPqt\PW, \cPqb\PH, or \cPqb\PZ with branching fractions equal to those of the corresponding \PQT quark decays to the same SM boson. In the interpretation of this search we assume that only one type of new particle is present, either the \PQT or the \PQB quark. The singlet branching fraction scenario is used as a benchmark for both \PQT and \PQB quarks.

Searches for pair-produced \PQT and \PQB quarks have been performed by both the ATLAS and CMS Collaborations at $\sqrt{s} = 7$\TeV~\cite{EXO-11-005,EXO-11-099,Aad:2012bdq}, 8\TeV~\cite{CMScombo2014,Run1anal,PhysRevD.92.112007,Aad2015} and at 13\TeV (with 2.6\fbinv and 36\fbinv of data)~\cite{Aaboud:2017qpr,Aaboud:2017zfn,B2G-16-024,B2G-17-003,Aaboud:2018xuw,Aaboud:2018uek,Aaboud:2018saj}. Previous searches by CMS in single lepton final states have excluded \PQT quark masses below 1295\GeV for $\mathcal{B}(\cPqb\PW)=100\%$~\cite{B2G-17-003}, and masses below 790 to 900\GeV for any possible choice of branching fractions to the three decay modes~\cite{B2G-16-024}. This search focuses on channels with exactly one lepton, a same-sign (SS) dilepton pair, and at least three leptons (trilepton). For background categorization, the latter two channels distinguish between leptons produced directly in decays of \PW, \PH, or \PZ bosons (prompt) and leptons produced from other sources (nonprompt), such as heavy flavor hadron decays.

This paper is organized as follows: Section~\ref{sec:cms} describes the CMS detector and how events are reconstructed, Section~\ref{sec:samples} describes the simulated background samples, and Section~\ref{sec:objects} describes the physics objects. In Sections~\ref{sec:1lep}--\ref{sec:TripleLeptonFinalState} we describe strategies for the three channels of the search, and in Section~\ref{sec:systs} we describe the systematic uncertainties. Lastly, in Sections~\ref{sec:Results}--\ref{sec:summary} we present our results and give a summary.

\section{The CMS detector and event reconstruction}
\label{sec:cms}

The central feature of the CMS detector is a superconducting solenoid of 6\unit{m} internal diameter, providing a magnetic field of 3.8\unit{T}. Within the solenoid volume are a silicon pixel and strip tracker, a lead tungstate crystal electromagnetic calorimeter (ECAL), and a brass and scintillator hadron calorimeter (HCAL), each composed of a barrel and two endcap sections. Forward calorimeters extend the pseudorapidity ($\eta$) coverage provided by the barrel and endcap detectors. Muons are measured in gas-ionization detectors embedded in the steel flux-return yoke outside the solenoid. A more detailed description of the CMS detector, together with a definition of the coordinate system used and the relevant kinematic variables, can be found in Ref.~\cite{Chatrchyan:2008zzk}.

A particle-flow (PF) algorithm aims to reconstruct and identify each individual particle in an event with an optimized combination of information from the various elements of the CMS detector~\cite{particleflow}. The energy of photons is directly obtained from the ECAL measurement. The energy of electrons is determined from a combination of the electron momentum at the primary interaction vertex as determined by the tracker, the energy of the corresponding ECAL cluster including the energy sum of all bremsstrahlung photons compatible with originating from the electron track.
The momentum of muons is obtained from the curvature of the corresponding track.
The energies of charged hadrons are determined from a combination of their momenta measured in the tracker and the matching ECAL and HCAL energy deposits, corrected for zero-suppression effects and for the response function of the calorimeters to hadronic showers. Finally, the energies of neutral hadrons are obtained from the corresponding corrected ECAL and HCAL energies.

Jets are reconstructed from the individual particles produced by the PF event algorithm (PF particles), clustered with the anti-\kt algorithm~\cite{Cacciari:2008gp, Cacciari:2011ma} with distance parameters of 0.4 (``AK4 jets'') and 0.8 (``AK8 jets''). Jet momentum is determined as the vector sum of all particle momenta in the jet, and is found from simulation to be within 5--15\% of the true momentum over the whole transverse momentum (\pt) spectrum and detector acceptance. Additional proton-proton interactions within the same or nearby bunch crossings (``pileup'') can contribute additional tracks and calorimetric energy depositions to the jet momentum. To mitigate this effect, tracks identified to be originating from pileup vertices are discarded, and an offset correction~\cite{Cacciari:2008gn} is applied to correct for remaining contributions. Jet energy corrections are derived from simulation, and are confirmed with in situ measurements of the energy balance in dijet, multijet, and photon/\PZ($\to \EE/\MM$) + jet events. A smearing of the jet energy is applied to simulated events to mimic detector resolution effects observed in data~\cite{Chatrchyan:2011ds}. Additional selection criteria are applied to each event to remove spurious jet-like features originating from isolated noise patterns in certain HCAL regions~\cite{JME-16-003}.

Events of interest are selected using a two-tiered trigger system~\cite{Khachatryan:2016bia}. The first level (L1), composed of custom hardware processors, uses information from the calorimeters and muon detectors to select events at a rate of around 100\unit{kHz} within a time interval of less than 4\mus. The second level, known as the high-level trigger (HLT), consists of a farm of processors running a version of the full event reconstruction software optimized for fast processing, and reduces the event rate to around 1\unit{kHz} before data storage.

\section{Simulated samples}\label{sec:samples}

To compare the SM expectation with 2016 collision data, samples of events of all relevant SM background processes and the \TTbar signal are simulated using the Monte Carlo (MC) method. Background processes are simulated using several matrix element generators and NNPDF3.0~\cite{NNPDF30} parton distribution functions (PDFs) at leading-order (LO) or next-to-leading-order (NLO). The \POWHEG~v2~\cite{Nason:2004rx,Frixione:2007vw,Alioli:2010xd,Frixione:2007nw} generator is used to simulate \ttbar events, single top quark events in the $t$-channel and \cPqt\PW\ channel, \cPqt\cPaqt\PH events, \PW\PZ events decaying to three leptons, and $\PZ\PZ$ events decaying to four leptons at NLO. The \MGvATNLO 2.2.2~\cite{MADGRAPH} generator with the FxFx matching scheme~\cite{FXFX} is used for NLO generation of \cPqt\cPaqt\PW\ events, as well as \cPqt\cPaqt\PZ events, $\ttbar\ttbar$ events, triboson events, and $s$-channel production of single top quark events. The \MGvATNLO 2.2.2 generator is used in LO mode with the MLM matching scheme~\cite{MLMmatching} to generate \PW+jets, Drell--Yan+jets, multijet events, and \PWp\PWp events.

Parton showering and the underlying event kinematics are simulated with \PYTHIA~8.212~\cite{Sjostrand:2006za,Sjostrand:2014zea}, using the underlying event tunes CUETP8M2T4~\cite{CUETP8M2T4} for \ttbar simulation and CUETP8M1~\cite{CUETP8M1} for all other processes. Diboson samples for use in the single-lepton channel are also generated at LO with \PYTHIA. Detector simulation for all MC samples is performed with \GEANTfour~\cite{GEANT4}. Additional inelastic \Pp\Pp\ collisions, both within the same bunch crossing as well as in the previous and following bunch crossings, are simulated in all samples. Weights are applied to simulated events so that the distribution of the number of pileup events agrees with data.

The simulated background samples are grouped into categories. In the single-lepton channel, the ``TOP'' group is dominated by \ttbar and includes single top samples; the ``EW'' group is dominated by the electroweak \PW+jets and includes Drell--Yan+jets, and diboson samples; and the ``QCD'' group includes quantum chromodynamics multijet samples. In the same-sign dilepton and trilepton channels the ``VV(V)'' group contains all \PW\PW, \PW\PZ, $\PZ\PZ$, and triboson samples, and the ``\cPqt\cPaqt+X'' group contains \cPqt\cPaqt\PW, \cPqt\cPaqt\PZ, \cPqt\cPaqt\PH, and $\ttbar\ttbar$ samples. Other backgrounds in these channels are estimated from data.

The \TTbar and \BBbar signals are simulated at LO using the \MGvATNLO generator interfaced with \PYTHIA~8.212 for parton showering and fragmentation. Signal with masses between 800 and 1800\GeV are simulated in steps of 100\GeV. A narrow width of 10\GeV is assumed for each generated \PQT and \PQB signal, independent of its mass. The corresponding theoretical cross sections, computed at next-to-NLO with the \textsc{Top++}2.0 program~\cite{TPRIMEXSEC,MITOV1,MITOV2,MITOV3,BARNREUTHER,NNLL}, are listed in Table~\ref{tab:signal}.

\renewcommand{\arraystretch}{1.2}
\begin{table}[htb]
  \centering
    \topcaption{Theoretical cross sections of \TTbar or \BBbar production, for various masses, assuming a width of 10\GeV at each mass point. The cross section uncertainties include contributions from uncertainties in the PDFs and uncertainties estimated by varying factorization and renormalization scales by a factor of two.}
    \begin{tabular}{c ccr@{\,$\pm$\,}>{$}l<{$}}
      \PQT/\PQB quark mass [{\GeVns}]  &   \multicolumn{4}{c}{Cross section [fb]} \\ \hline
      800 &&&  196 & 8 \\
      900 &&&  90 & 4 \\
      1000 &&&  44 & 2 \\
      1100 &&&  22 & 1 \\
      1200 &&&  11.8 & 0.6 \\
      1300 &&&  6.4 & ^{0.4}_{0.3} \\
      1400 &&&  3.5 & 0.2 \\
      1500 &&&  2.0 & 0.1 \\
      1600 &&&  1.15 & ^{0.09}_{0.07} \\
      1700 &&&  0.67 & ^{0.06}_{0.04} \\
      1800 &&&  0.39 & ^{0.04}_{0.03} \\[\cmsTabSkip]
    \end{tabular}
    \label{tab:signal}

\end{table}
\renewcommand{\arraystretch}{1.}

\section{Reconstruction methods}\label{sec:objects}

This search requires that selected events have at least one reconstructed \Pp\Pp\ interaction vertex within the luminous region (longitudinal position $\abs{z}< 24$ cm and radial position $\rho < 2$ cm)~\cite{Chatrchyan:2014fea}. The reconstructed vertex with the largest value of summed physics-object $\pt^2$ is taken to be the primary \Pp\Pp\ interaction vertex. The physics objects are the jets, clustered using the jet finding algorithm~\cite{Cacciari:2008gp,Cacciari:2011ma} with the tracks assigned to the vertex as inputs, and the associated missing transverse momentum, taken as the negative vector sum of the \pt of those jets. Each event must have at least one charged lepton (electron or muon) candidate that is reconstructed within the detector acceptance region of $\abs{\eta} < 2.5$ (2.4) for electrons (muons), excluding the barrel-endcap transition region ($1.44<\abs{\eta}<1.57$) for electrons.

Events containing leptons are initially selected using the HLT. For the single-lepton channel events must pass a set of triggers requiring one electron or muon with $\pt > 15$\GeV and jets with \pt that sums to at least 450\GeV. A secondary set of triggers selects events with one isolated electron ($\pt > 35$\GeV) or one muon ($\pt > 50$\GeV). For the SS dilepton channel events must pass triggers based on double lepton combinations, with momentum thresholds that varied over time. The dielectron trigger requires two electrons with $\pt > 37$ and 27\GeV. Triggers for electron-muon events have a variety of thresholds: both leptons with $\pt > 30$\GeV, or one lepton with $\pt > 37$\GeV and the other flavor lepton with $\pt > 27$\GeV. The dimuon trigger requires one muon with $\pt > 30$\GeV, and another muon with $\pt > 11$\GeV. For the trilepton channel, dilepton triggers with lower momentum thresholds were used to select events with isolated leptons. The dielectron channel requires an electron with $\pt > 23$\GeV and another electron with $\pt > 12$\GeV. Events with both lepton flavors are selected with triggers that require one lepton with $\pt > 23$\GeV and a different flavor lepton with $\pt > 8$\GeV. The dimuon trigger selects events featuring one muon with $\pt > 17$\GeV and another muon with $\pt > 8$\GeV.

Dedicated event filters remove events that are affected by: known noise patterns in the HCAL, accelerator-induced particles traveling along the beam direction at large radius (up to 5m), anomalously high energy deposits in certain ECAL ``superclusters"~\cite{Khachatryan:2015hwa}, ECAL cell triggers that are not performing optimally, and muon candidates with large track uncertainties matched to misreconstructed tracks or charged hadrons.

Electrons are reconstructed~\cite{Khachatryan:2015hwa} taking into account track quality, association between the track and electromagnetic shower, shower shape, and the likelihood of the electron being produced in a photon conversion in the detector. A multivariate discriminant is used to identify well-reconstructed electrons at two quality levels: a tight level with ${\approx}88$\% efficiency (${\approx}4$\% misidentification efficiency) and a loose level with ${\approx}95$\% efficiency (${\approx}5$\% misidentification efficiency).

Muons are reconstructed using information from both the CMS silicon tracker and the muon spectrometer in a global fit, matching deposits in the silicon tracker with deposits in the muon detector~\cite{Chatrchyan:2012xi}. Identification algorithms consider the global fit $\chi^2$ value, the number or fraction of deposits in the trackers and muon detectors, track kinks, and the distance between the track from the silicon tracker and the primary interaction vertex. We consider two quality levels: a tight level with ${\approx}97$\% efficiency, and a loose level with ${\approx}100$\% efficiency, in the barrel region of the detector. Both levels have a hadronic misidentification efficiency of ${<}1$\%.

The large Lorentz boost of the decay products of the \PQT quarks can produce final-state leptons that are in close proximity to hadronic activity, and are similar to background events with jets that contain a lepton from semileptonic hadron decays. The isolation of a lepton from surrounding particles is evaluated using a variable $I_{\text{mini}}$, defined as the \pt sum of PF particles within a \pt-dependent cone around the lepton, corrected for the effects of pileup using the effective area of the cone~\cite{Cacciari:2008gn} and divided by the lepton \pt. The radius of the isolation cone in $\eta-\phi$ space, $\mathcal{R}$, is determined by:
\begin{equation}
\mathcal{R} = \frac{10\:\GeVns}{\min(\max(\pt,50\GeV),200\GeV)}.
\end{equation}
Using a \pt-dependent cone size allows for greater efficiency at high energies when jets and leptons are more likely to overlap. The reconstructed electrons and muons must have $I_{\text{mini}} < 0.1 $ to be labeled tight, and $I_{\text{mini}} < 0.4$ to be labeled loose. Scale factors to describe efficiency differences between data and MC simulation for the lepton reconstruction, identification, and isolation algorithms are calculated using the ``tag-and-probe" method~\cite{Chatrchyan:2012xi}, and are applied to simulated events.

All AK4 jets with $\pt > 30$\GeV that lie within the tracker acceptance of $\abs{\eta} < 2.4$ are considered in this search (unless otherwise noted, ``jets'' refers to AK4 jets). Additional selection criteria are applied to reject events containing noise and mismeasured jets. Leptons that pass tight identification and isolation requirements in the single-lepton channel, or loose requirements in the SS dilepton and the trilepton channels are removed from jets that have an angular separation of $\Delta\mathrm{R} = \sqrt{\smash[b]{(\Delta\eta)^2+(\Delta\phi)^2}}
 < 0.4$ with the leptons (where $\phi$ is azimuthal angle in radians), before jet energy corrections are applied. This is done by matching PF particles in the lepton and jet collections and subtracting the four-momentum of a matched lepton candidate from the jet four-momentum. In the SS dilepton and the trilepton channels, loose leptons, as well as tight leptons, are removed from jets because these leptons are used to estimate nonprompt lepton backgrounds.

The missing transverse momentum vector \ptvecmiss is defined as the projection onto the plane perpendicular to the beam axis of the negative vector sum of the momenta of all reconstructed PF objects in an event. Its magnitude is referred to as \ptmiss. The energy scale corrections applied to jets are propagated to \ptmiss. We define \HT as the scalar \pt sum of all reconstructed jets in the event that have $\pt>30$\GeV and $\abs{\eta}< 2.4$. In addition, we define the \ST as the scalar sum of \ptmiss, the \pt of leptons, and the \HT in the event.

This search relies on techniques to analyze the internal structure of jets and to identify the parton that created the jet. Jets are tagged as \cPqb\ quark jets using a multivariate discriminant, specifically the combined secondary vertex (CSVv2) algorithm~\cite{BTV-16-002}, which uses  information about secondary vertices within the jet. For simulated \ttbar events, our requirement on this discriminant has an efficiency for tagging true \cPqb\ quark jets of ${\approx}65\%$, averaged over jets with $\pt > 30$\GeV. The efficiency for falsely tagging light-quark or gluon jets, measured in multijet event data, is ${\approx}1\%$. Efficiency differences in data and simulation are corrected by applying scale factors, which are functions of jet \pt and flavor~\cite{BTV-16-002}.

Heavy VLQ decays can produce top quarks and \PW, \PZ, or Higgs bosons with high momenta, causing their decay products to merge into a single AK8 jet. The ``$N$-subjettiness'' algorithm~\cite{Thaler:2010tr} creates jet shape variables, $\tau_N$, that quantify the consistency of the jet's internal structure with an $N$-prong hypothesis. Ratios of $\tau_N/\tau_{N-1}$ are powerful discriminants between jets predicted to have $N$ internal energy clusters and jets predicted to have fewer clusters. Techniques called ``pruning'' or ``softdrop''~\cite{PRUNING, SOFTDROP, Dasgupta:2013ihk} remove soft and wide-angle radiation from the jet so that the mass of its primary constituents can be measured more accurately. The softdrop algorithm identifies two smaller subjets within the AK8 jet, and these can be identified as \cPqb\ quark subjets using the same algorithm as that applied to AK4 jets. The AK8 jets are reconstructed independently of AK4 jets, so they will frequently overlap. Unless otherwise stated, jet multiplicity criteria assume that AK4 and AK8 jets are clustered independently and may share constituents.

An AK8 jet is labeled as \PW\ tagged if it has $\pt > 200$\GeV, $\abs{\eta}<2.4$, pruned jet mass between 65 and 105\GeV, and the ratio of $N$-subjettiness variables $\tau_2/\tau_1 < 0.6$. These requirements yield a \PW\ tag efficiency of 60--70\%, depending on AK8 jet momentum. The pruned mass distribution in simulation is smeared such that the resolution of the \PW\ mass peak matches the resolution observed in data~\cite{JME-16-003}. Scale factors describing efficiency differences between data and simulation for the $\tau_2/\tau_1$ selection are applied to the AK8 jets matched to true boosted hadronic \PW\ boson decays~\cite{JME-16-003}. An AK8 jet is labeled as \PH tagged if it has $\pt > 300\GeV$, $\abs{\eta}< 2.4$, pruned mass between 60 and 160\GeV, and at least one \cPqb-tagged subjet. Having a larger mass than the \PW\ boson mass, the Higgs boson requires more momentum for the \cPqb\ quarks to merge into one AK8 jet. This algorithm exploits the large branching fraction of the Higgs boson to \bbbar pairs and has an efficiency of ${\approx}65\%$. If an AK8 jet is both \PH and \PW\ tagged, the \PH tag is given precedence.

\section{Single-lepton channel}\label{sec:1lep}
The single-lepton channel includes events with exactly one charged lepton. Boosted hadronic decay products of \PW\ and Higgs bosons are identified in AK8 jets and used to categorize events. This final state is highly sensitive to \TTbar production with at least one $\PQT\to\cPqb\PW$ or $\PQT\to\cPqt\PH$ decay, as well as \BBbar production with at least one $\PQB\to\cPqt\PW$ or $\PQB\to\cPqb\PH$ decay.

\subsection{Event selection and categorization}\label{sec:SLselection}

Each event must have one electron or muon that passes the tight selection requirements described previously. The tight lepton must have $\pt>60\GeV$, and events with extra leptons passing the loose quality requirements with $\pt>10\GeV$ and $\abs{\eta}<2.5$ (2.4) for electrons (muons) are rejected. We also require \ptmiss of at least 75\GeV to account for the presence of a neutrino from a \PW\ boson decay and to reduce multijet background events.

Each selected event must have at least three jets with $\pt>300$, 150, and 100\GeV. Events must also have at least two AK8 jets with $\pt > 200$\GeV and $\abs{\eta}<2.4$, which are permitted to overlap with the AK4 jets. The requirement of at least two AK8 jets is highly efficient for signal in all decay modes (${>}98\%$) and reduces the background contribution. 

Events are divided into 16 categories based on lepton flavor and the presence of \PH-, \PW-, and \cPqb-tagged jets:
\begin{itemize}
\item \textbf{H2b}: events with one or more \PH-tagged jets with two \cPqb-tagged subjets each;
\item \textbf{H1b}: events failing the H2b criterion, but having one or more \PH-tagged jets with only one \cPqb-tagged subjet;
\item \textbf{W1}: events with zero \PH-tagged jets but at least one \PW-tagged jet;
\item \textbf{W0}: events with zero \PH-tagged jets and zero \PW-tagged jets.
\end{itemize}
In both the H1b and H2b categories, we require an extra \cPqb-tagged jet that does not overlap with the \PH-tagged jet, since signal events with a Higgs boson always contain at least one top quark decay as well. In the W0 category we require a fourth jet with $\pt>30$\GeV and $\abs{\eta}<2.4$. Events in the W0 and W1 categories are subcategorized by the number of \cPqb-tagged jets (1, 2, ${\geq}$3).

Discrepancies in the modeling of top quark momentum are corrected by applying a weight that depends on the generated top quark \pt~\cite{TOP-16-008} to simulated \ttbar events. Discrepancies observed in \HT-binned \MADGRAPH samples are corrected by applying a scaling function that describes the observed difference in the \HT spectrum between binned and inclusive simulations~\cite{B2G16005,B2G16006,B2G-16-024}.

To maximize signal efficiency in the search regions, and to create signal-depleted control regions, we calculate the minimum angular separation between the highest \pt AK8 jet and any other AK8 jet in the event. In background processes there are often only two AK8 jets, usually emitted back to back from each other. In signal processes there are typically more than two AK8 jets and the minimum separation will be significantly smaller. The search region is therefore defined by requiring $0.8<\DRak<3.0$, and the control region by requiring $\DRak>3.0$. Signal efficiencies in the search region for the singlet decay mode are 9--15\%, increasing with VLQ mass.

\begin{figure}[hbp]
\centering
\includegraphics[width=0.49\textwidth]{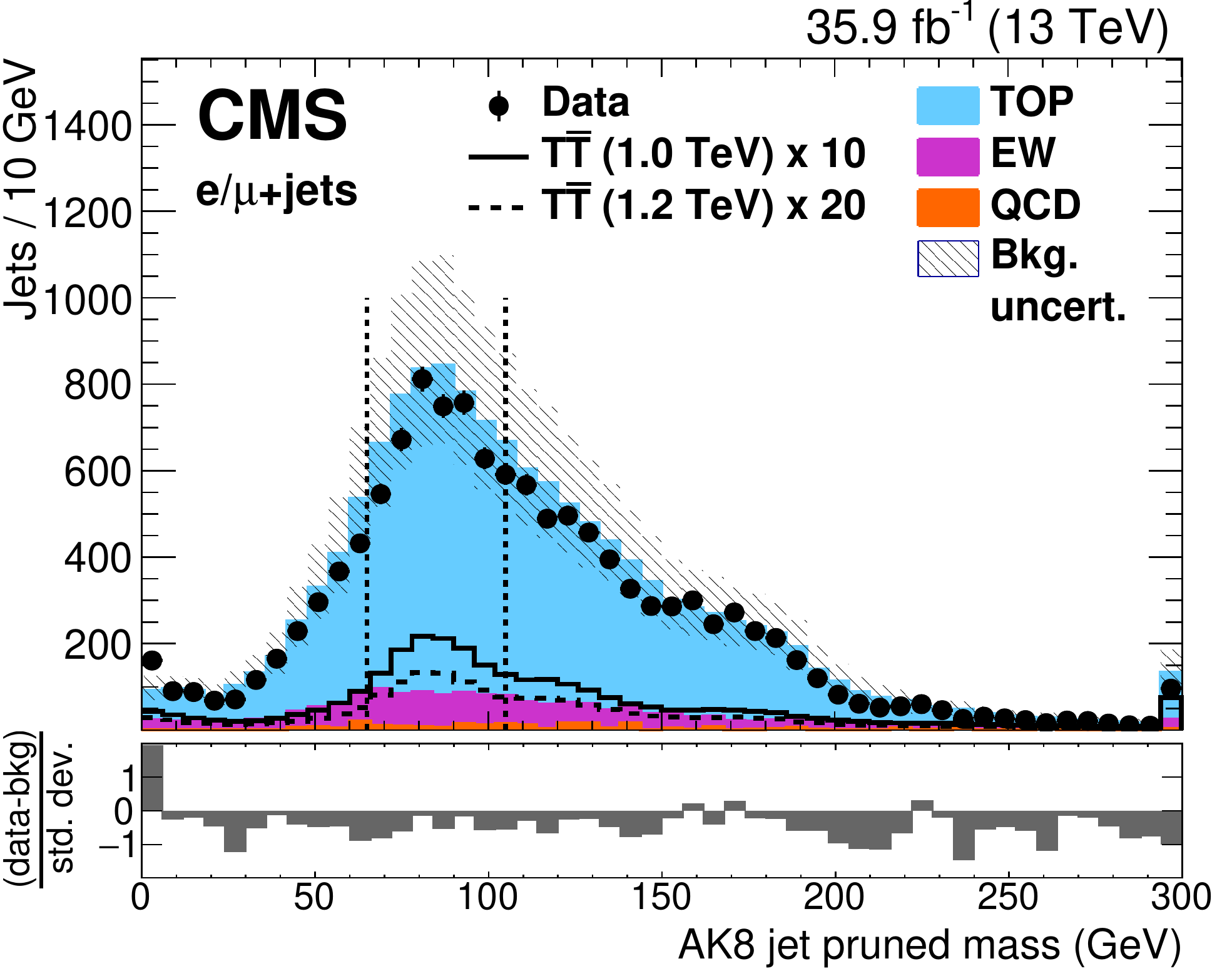}
\includegraphics[width=0.49\textwidth]{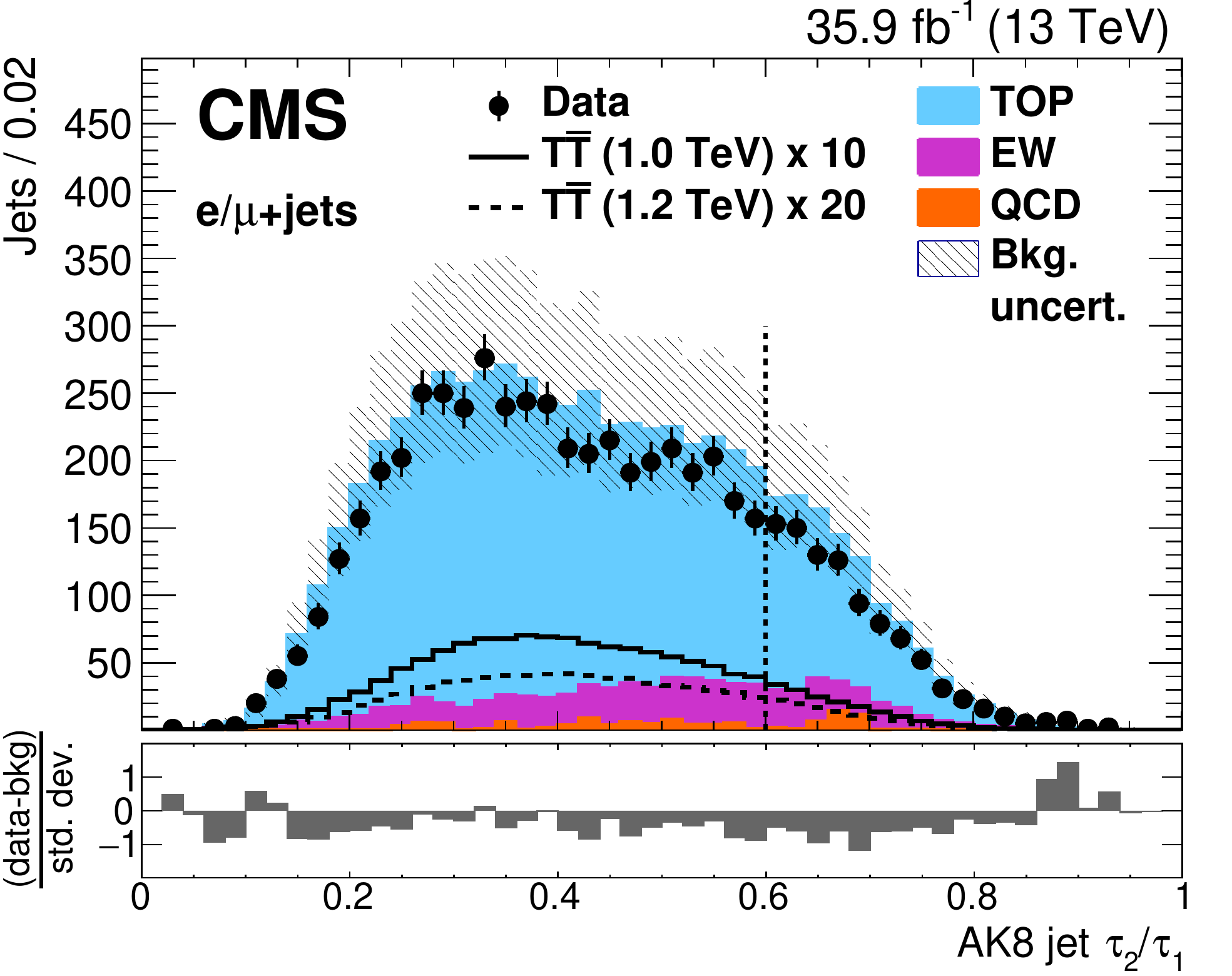}
\includegraphics[width=0.49\textwidth]{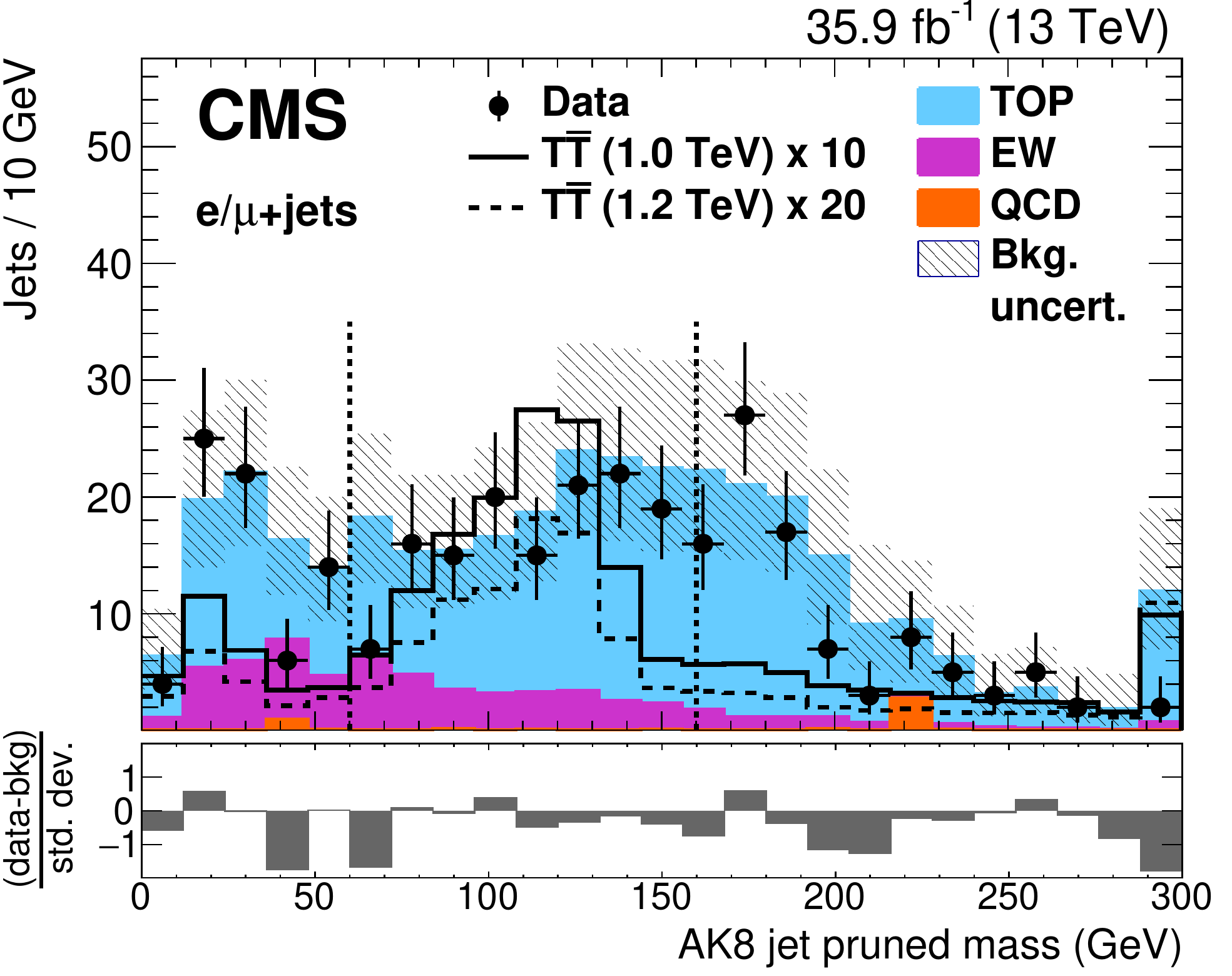}
\includegraphics[width=0.49\textwidth]{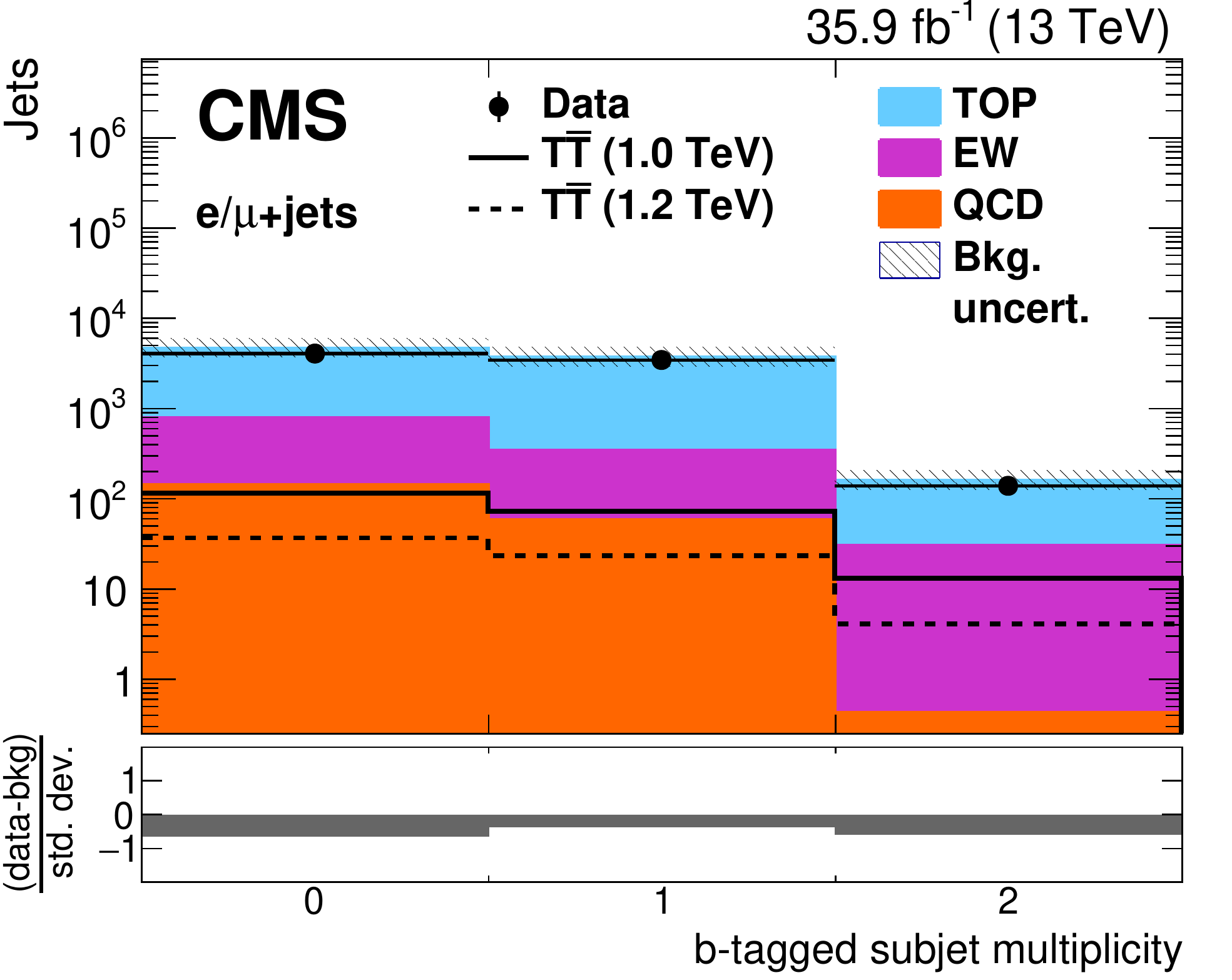}
\caption{Distributions of \PW\ and \PH tagging input variables after all selection requirements, before the fit to data: pruned mass in AK8 jets with $\tau_2/\tau_1 < 0.6$ (upper left), $N$-subjettiness $\tau_2/\tau_1$ ratio in AK8 jets with pruned mass between 65--105\GeV (upper right), pruned mass in AK8 jets with two \cPqb-tagged subjets (lower left), and number of \cPqb-tagged subjets in AK8 jets with pruned mass in the range 60--160\GeV (lower right). Vertical dashed lines mark the selection windows for each distribution. The black points are the data and the filled histograms show the simulated background distributions, grouped into categories as described in Section~\ref{sec:samples}. The expected signal is shown by solid and dotted lines for \PQT quark masses of 1.0 and 1.2\TeV. The final bin includes overflow events. Uncertainties, indicated by the hatched area, include both statistical and systematic components. The lower panel shows the difference between data and background divided by the total uncertainty.}
\label{fig:wtags}
\end{figure}

Figure~\ref{fig:wtags} shows distributions of \PW\ tagging input variables after all selection requirements: pruned mass in AK8 jets with $\tau_2/\tau_1 < 0.6$, showing a clear \PW\ boson contribution in signal events, and $\tau_2/\tau_1$ in AK8 jets with pruned mass inside the mass window of 65--105\GeV. The distribution of $\tau_2/\tau_1$ shows that background processes with primarily one-prong jets, such as \PW+jets or multijet events, are concentrated at higher values, while signal events and top quark decays tend toward lower values. Figure~\ref{fig:wtags} also shows the pruned mass in AK8 jets with two \cPqb-tagged subjets, and the number of \cPqb-tagged subjets in AK8 jets with a pruned mass within the range 60--160\GeV. The \PH tag algorithm is efficient for both $\PH\to\bbbar$ and $\PZ\to\bbbar$ decays. The systematic difference between data and background (bkg) is due to known issues with jet momentum distributions in the \ttbar simulation that are only partially corrected by applying the top quark momentum weight~\cite{TTDIFFXSEC}. The residual difference is described by the uncertainty in the renormalization and factorization energy scales, discussed further in Section~\ref{sec:systs}.

To search for VLQ events in the W0 and W1 categories, we analyze the minimum mass constructed from the lepton ($\ell$) and a \cPqb-tagged jet, labeled \minMlb. This distribution provides strong discrimination between \ttbar events and signal events with a $\PQT\to\cPqb\PW$ decay. Reconstructing the mass of two out of three leptonic SM top quark decay products, namely the lepton and \cPqb\ quark jet, produces a sharp edge below the top quark mass, while $\PQT\to\cPqb\PW$ decays will produce a similar edge near the \PQT mass. Since the \PH tagged categories have relatively few $\PQT\to\cPqb\PW$ decays, the \ST distribution is used as the search variable in these categories. Compared to other possibilities, such as using \ST as the search variable in all categories, this combination of discriminating variables provides the best sensitivity to \PQT quark production in the 1\TeV mass range in the singlet branching fraction scenario. Distributions of \minMlb\ and \ST in the search regions are shown in Section~\ref{sec:Results}.

\subsection{Background modeling}\label{sec:SLbackground}

Backgrounds are modeled from simulation in this channel and we perform a closure test in a control region, categorizing events as done in the search regions. The control region is defined by requiring $\DRak>3$. Further selection criteria are applied to form regions with significant amounts of \PH-tagged jets, \PW+jets events, or \ttbar events. To form the \ttbar control region, events from the W1 and W0 categories are split according to lepton flavor and \cPqb\ tag content: 1, 2 or ${\geq}3$ \cPqb-tagged jets. In the \PW+jets control region, events from the W1 and W0 categories without \cPqb-tagged jets are categorized based on \PW\ tag content: zero or at least one \PW-tagged jets. The \PH-tagged jet control region includes events from the H1b or H2b categories, split according to lepton flavor and number of \cPqb-tagged jets (0 or ${\geq}1$) that do not overlap any \PH-tagged jet. Signal efficiencies in the control regions are negligibly small (${<}1\%$)
 for both \TTbar and \BBbar production.

The comparison between data and simulation in these regions is used to evaluate the level of remaining differences after the event selection, efficiency corrections, and generator-level corrections, such as the differences in the rate of misidentified \PW- or \PH-tagged jets. In all control regions the data agree with simulation, within the systematic uncertainties described in Section~\ref{sec:systs}.

To provide background-dominated regions in the statistical interpretation of the results, the control regions are aggregated into fewer categories. These aggregate regions target the \ttbar + jets background in events with zero \PH-tagged jets and at least one \cPqb-tagged jet, the \PW+jets background in events without any \PH- or \cPqb-tagged jets, and misidentified \PH-tagged jets in events with at least one \PH-tagged jet and any number of \cPqb-tagged jets (including zero). To constrain the uncertainty in the renormalization and factorization energy scale for the background events, the \HT distribution is used in these categories. Predicted and observed event yields in the control regions are listed in Table~\ref{tab:CRht}.

\begin{table}[htbp]
\centering
\topcaption{Predicted and observed event yields in the aggregated control region categories of the single-lepton channel. Uncertainties include both statistical and systematic components.}
\begin{tabular}{l ccr@{\,$\pm$\,}lcc ccr@{\,$\pm$\,}lcc  ccr@{\,$\pm$\,}lcc}
Sample                & \multicolumn{6}{c}{0 \PH, ${\geq}0$ \PW, 0 \cPqb} & \multicolumn{6}{c}{0 \PH, ${\geq}0$ \PW, ${\geq}1$ \cPqb} & \multicolumn{6}{c}{${\geq}1$ \PH, ${\geq}0$ \PW, ${\geq}1$ \cPqb}      \\
\hline

\TTbar (1.0\TeV)      &&& 1.99 & 0.15      &&&&& 6.94 & 0.37      &&&&& 3.63 & 0.22 &&  \rule{0pt}{2.5ex}\\
\TTbar (1.2\TeV)      &&& 0.65 & 0.05      &&&&& 2.08 & 0.11      &&&&& 0.94 & 0.06 && \\ [\cmsTabSkip]
\BBbar (1.0\TeV)      &&& 1.73 & 0.14      &&&&& 6.55 & 0.36      &&&&& 2.94 & 0.20 && \\
\BBbar (1.2\TeV)      &&& 0.64 & 0.04      &&&&& 1.94 & 0.10      &&&&& 0.82 & 0.05 && \\ [\cmsTabSkip]
TOP                   &&& 1120 & 220       &&&&& 2830 & 580       &&&&& 2360 & 370  &&   \\
EW                    &&& 3050 & 510       &&&&& 580 & 100        &&&&& 195 & 34    &&   \\
QCD                   &&& 322 & 73         &&&&& 116 & 30         &&&&& 47 & 18     &&   \\[\cmsTabSkip]
Total bkg             &&& 4490 & 580       &&&&& 3520 & 600       &&&&& 2600 & 370  &&   \\
Data                  & \multicolumn{6}{c}{4420} & \multicolumn{6}{c}{3409} & \multicolumn{6}{c}{2476}    \\
Data/bkg              &&& 0.99 & 0.13      &&&&& 0.97 & 0.16      &&&&& 0.95 & 0.14 &&  \\[\cmsTabSkip]
\end{tabular}
\label{tab:CRht}
\end{table}

\section{Same-sign dilepton channel}\label{sec:SSDLState}

The SS dilepton channel attempts to make use of a unique feature of VLQ signals, namely the presence of prompt SS dilepton pairs. In \TTbar production SS lepton pairs are most common in events having at least one $\PQT\to\cPqt\PH$ decay, where the Higgs boson decays to a pair of \PW\ bosons. Since at least one \PW\ boson is produced in the decay of the other \PQT quark, at least four \PW\ bosons are present in the final state, two of each charge. In \BBbar production SS lepton pairs are more frequent, arising from events with at least one $\PQB\to\cPqt\PW$ decay, since at least one other \PW\ boson is produced in the decay of the other \PQB quark.

\subsection{Event selection and categorization}

We require events to have exactly two leptons with the same electric charge that are within the detector acceptance ($\abs{\eta} < 2.4$).
Different triggers were used during early and late 2016 data taking, with different \pt requirements for the leptons.
We require the leading (subleading) lepton to have \pt greater than 40 (35)\GeV for the early data set and greater than 40 (30)\GeV for the later data set. The two leptons must pass the tight identification and isolation requirements described in Section~\ref{sec:objects} and the events are divided into three categories based on the flavors: $\Pe\Pe$, $\Pe\mu$, and $\mu\mu$.

After requiring two tight SS leptons, we apply additional selection criteria to reduce the background rate. To remove quarkonia decays we require $M(\ell,\ell) > 20$\GeV. To remove \PZ boson decays we reject dielectron events with invariant lepton pair mass $76.1 < M(\ell,\ell) < 106.1$\GeV. This cut is not applied to dimuon events because muons have a negligibly small rate of charge misidentification. We require the number of jets to be ${\geq}4$ and the scalar sum of selected jet and lepton transverse momenta, \HTlep\, to exceed 1200\GeV.

To search for VLQ events in data in the SS dilepton channel, we perform a counting experiment using the yield of events passing the selections. Signal efficiencies for this channel after applying all selection criteria are 0.42 (0.5)\% for a singlet \PQT quark of mass 1.0 (1.2)\TeV.

\subsection{Background modeling}\label{sec:ssdlbkg}

We consider three categories of backgrounds associated with this channel: SM processes with SS dilepton signatures; opposite-sign (OS) prompt leptons misreconstructed as SS leptons; and nonprompt leptons from heavy flavor hadron decays, jets misidentified as leptons, or photons converting to electrons. Leptons from tau decays are likely to be interpreted as prompt electrons or muons, whereas hadronic tau decays are likely to be considered to be nonprompt leptons.
The background contribution from prompt SS dilepton processes is obtained from simulated samples in the VV(V) and $\ttbar+\mathrm{X}$ groups.

Prompt OS dileptons can contribute background events when one lepton is assigned the wrong charge, leading to an SS dilepton final state. Muon reconstruction in CMS provides very reliable charge identification, leading to a very small rate of charge misidentification that is considered negligible for this search. The rate of charge misidentification for electrons is derived from a data sample dominated by $\PZ\to\Pe\Pe$ decays, by computing the ratio of SS dilepton events to all events. Misidentification efficiencies are derived as a function of $\abs{eta}$ for electrons with $\pt < 100\GeV$, $100 < \pt < 200\GeV$, and $\pt > 200\GeV$. The values are about 1\% in the barrel region and about 5\% in the endcap region. The number of SS dilepton events arising from charge misidentification is estimated by weighting the number of observed OS dilepton events that pass all other selection criteria, by the misidentification efficiency per electron.

Same-sign dilepton events arising from the presence of one or more nonprompt leptons is the primary reducible background. Two components of this background are jets misidentified as leptons and nonprompt leptons that pass tight isolation criteria. This contribution is estimated using the  ``tight-to-loose'' method~\cite{susy2011}, in which events with one or more loose leptons are weighted by the tight-loose ratios expected for prompt and nonprompt leptons. The efficiency for prompt leptons to pass the tight selection criteria, or ``prompt lepton efficiency,'' is determined using events with a lepton pair invariant mass within 10\GeV of the \PZ boson mass. For muons the average prompt efficiency, found to be generally constant over \pt and $\eta$, is $0.943 \pm 0.001$. For electrons the prompt efficiency depends on \pt and ranges from 0.80 to 0.95.

The ``misidentified lepton efficiency,'' or efficiency for nonprompt leptons to pass the tight selection criteria, is determined using a data sample enriched in nonprompt leptons. The fraction of prompt leptons from \PW\ and \PZ boson decays is reduced by requiring exactly one loose lepton per event, low \ptmiss, and that the lepton and \ptmiss be inconsistent with a \PW\ boson decay (transverse mass ${<}25\GeV$). At least one jet is required with large angular separation from the lepton ($\Delta\mathrm{R} > 1.0$). Events are rejected if the invariant mass of any jet-lepton combination is compatible with the \PZ boson mass, within 10\GeV. Misidentified lepton efficiencies are then measured as a function of lepton \pt and $\eta$, with values ranging from 0.17--0.25 for electrons and 0.16--0.33 for muons.

\begin{figure}[hbtp]
\centering
\includegraphics[width=0.49\textwidth]{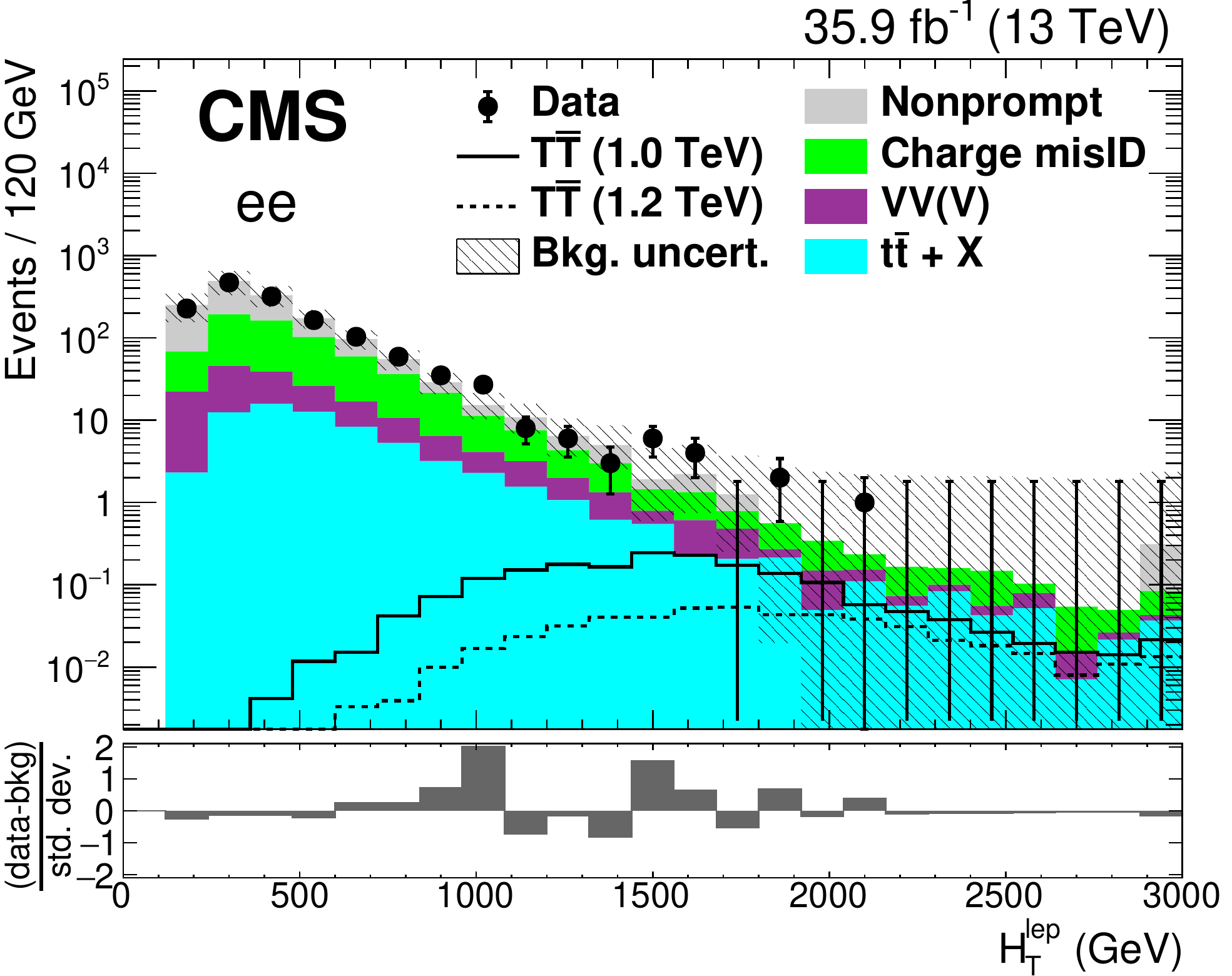}
\includegraphics[width=0.49\textwidth]{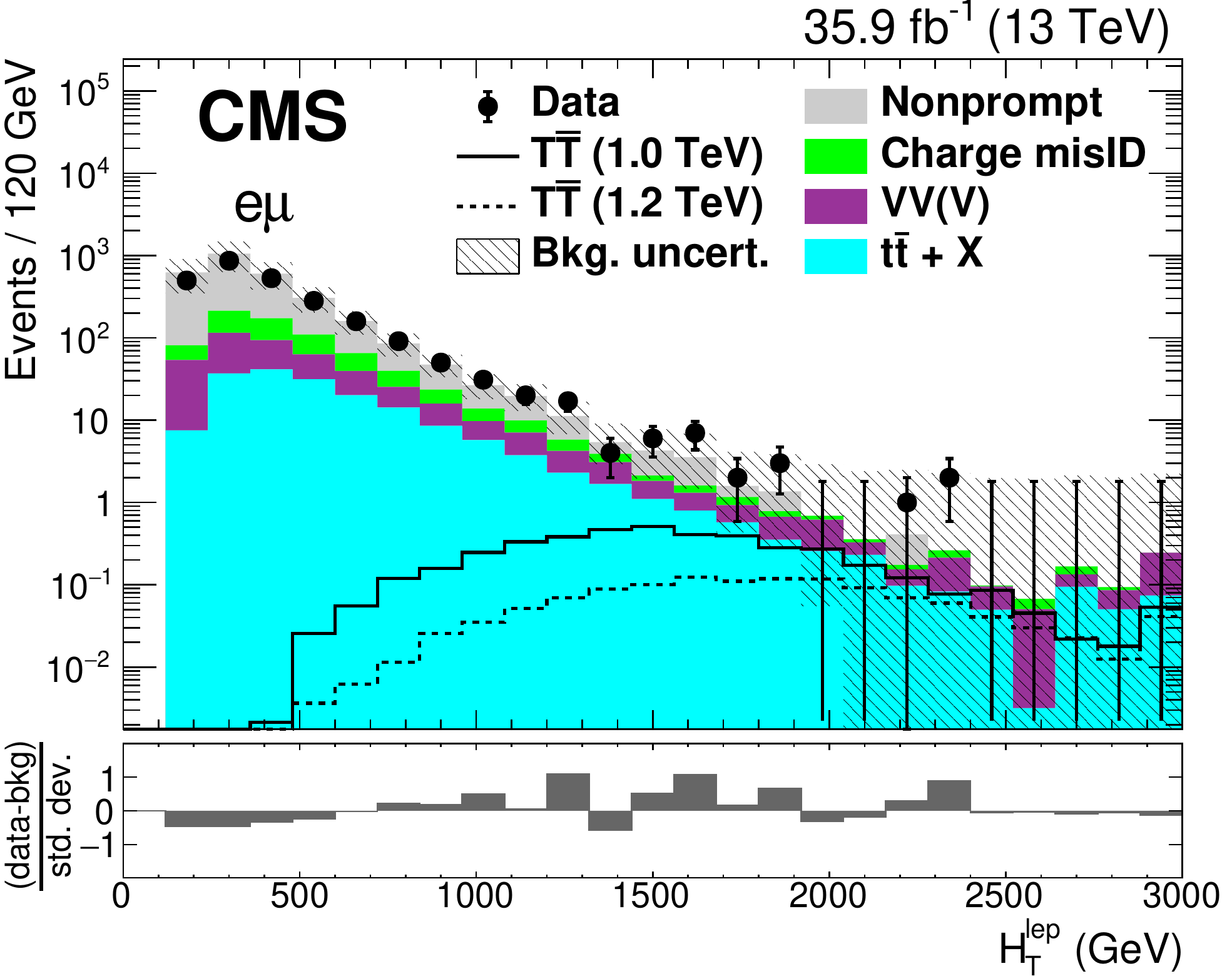}
\includegraphics[width=0.49\textwidth]{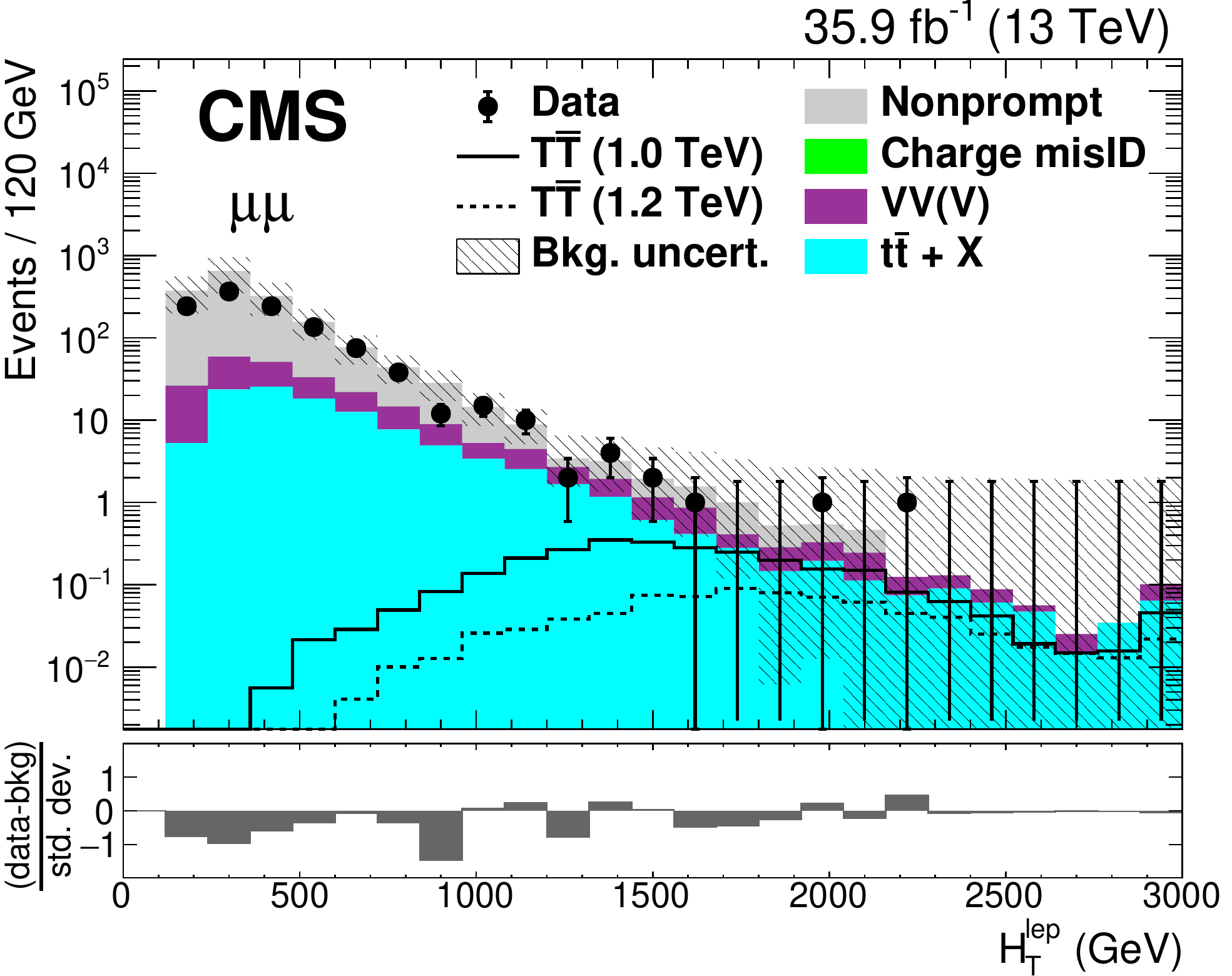}
\includegraphics[width=0.49\textwidth]{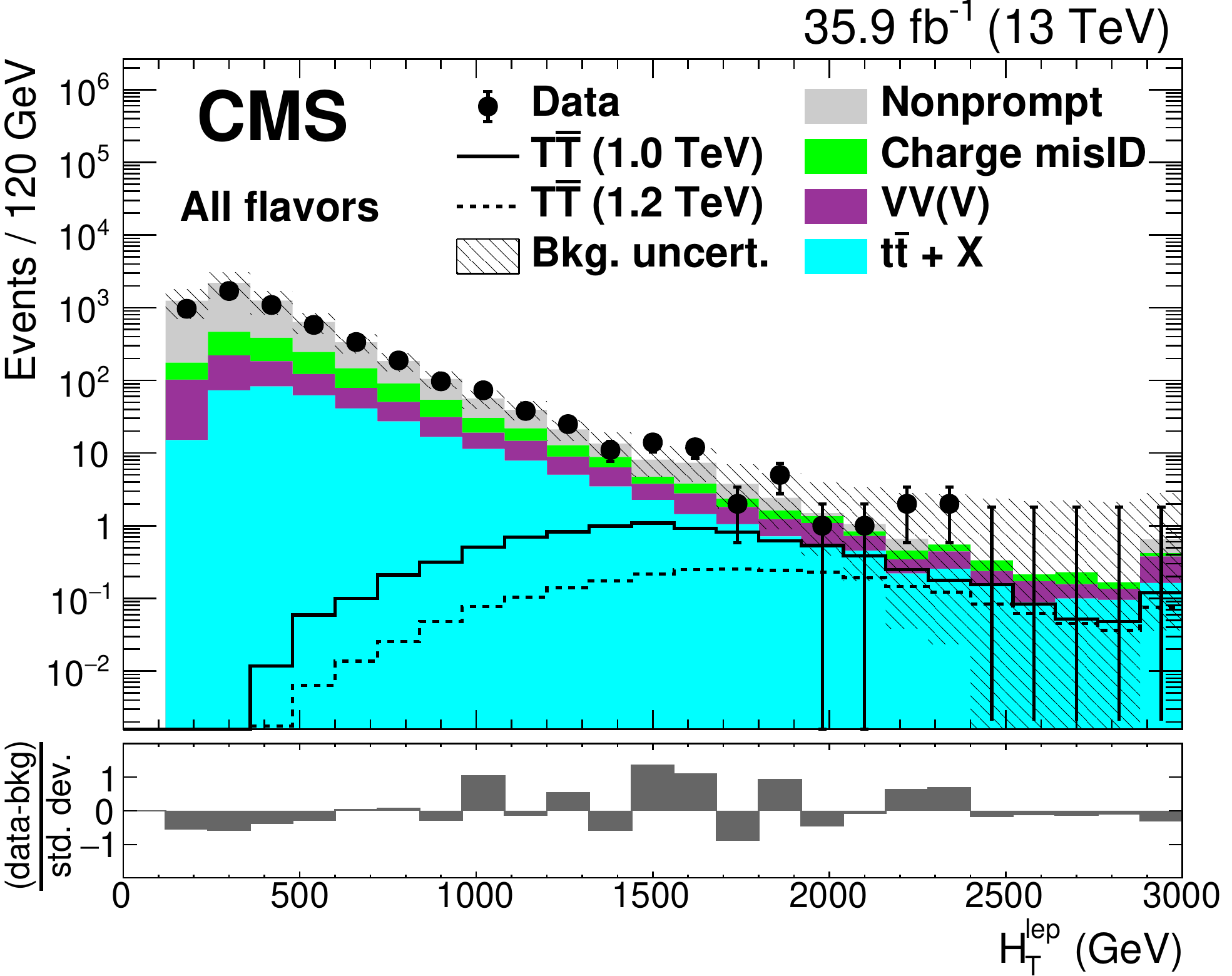}
\caption{The \HTlep\ distributions before the fit to data in events with at least two jets after the SS dilepton selection and \PZ-boson/quarkonia vetoes, for the $\Pe\Pe$, $\Pe\mu$, and $\mu\mu$ categories, and for the combination of all categories. The black points are the data and the filled histograms show the background distributions, with simulated backgrounds grouped into categories as described in Section~\ref{sec:samples}. The expected signal is shown by solid and dotted lines for \PQT quark masses of 1.0 and 1.2\TeV. The final bin includes overflow events. Uncertainties, indicated by the hatched area, include both statistical and systematic components. The lower panel shows the difference between data and background divided by the total uncertainty. Accepted events are required to have $\HTlep>1200\GeV$.}
\label{fig:SSdilep_HT}
\end{figure}

Figure~\ref{fig:SSdilep_HT} shows the the full spectrum of \HTlep\ distributions for SS dilepton events in the different lepton flavor categories, where two or more jets are required in each event.

\section{Trilepton channel}\label{sec:TripleLeptonFinalState}

The trilepton final state is highly sensitive to VLQ pair production with at least one $\PQT\to\cPqt\PZ$, $\PQB\to\cPqb\PZ$, or $\PQB\to\cPqt\PW$ decay, all of which can produce two or more prompt leptons. When combined with the decay of the other \PQT or \PQB quark, three or more prompt leptons can exist in the final state, a signature that is rare in SM processes.

\subsection{Event selection and categorization}

We select events with at least three leptons, each with $\pt>30\GeV$, that pass the tight identification and isolation requirements described in Section~\ref{sec:objects}. The background from nonprompt leptons is estimated in a control sample with less restrictive selection criteria, including events with three leptons that pass the loose identification and isolation requirements.
Leptons are sorted first based on tight or loose quality, and then based on \pt in descending order. The events are divided into the four categories of the flavors ($\Pe$ or $\mu$) of the first three leptons: $\Pe\Pe\Pe$, $\Pe\Pe\mu$, $\Pe\mu\mu$, and $\mu\mu\mu$.

Additionally, to reject background events with leptons originating from low-mass resonances, no OS same flavor lepton pair with invariant mass $M(\ell,\ell)_{\mathrm{OS}} < 20$\GeV is allowed. We also require the events to have at least three jets with $\pt>30\GeV$ and $\abs{\eta}<2.4$, at least one of which is \cPqb-tagged, since top quark decays and/or \cPqb\ quarks are expected in signal events. Lastly, we require $\ptmiss>20$\GeV. These requirements create a sample with many leptons from \PZ and \PW\ boson decays, together with several jets to account for hadronic decays products of the \PQT or \PAQT. To search for VLQ events in data in the trilepton channel, we use the \ST distribution to discriminate the signal from the background. With respect to the expected number of events before any selections, the signal efficiencies for this channel after all selections are 0.65\,(0.66)\% for a singlet \PQT quark of mass 1.0\,(1.2)\TeV.

We define a signal-depleted control region for the purpose of calculating misidentified lepton efficiencies. This control region is defined using the initial selection requirements above, except that we require exactly two jets instead of at least three. Processes containing nonprompt leptons contribute almost equally to this region and to the signal region.

\subsection{Background modeling}\label{sec:Trilep_BackgroundEstimation}

Backgrounds are divided into two categories, prompt and nonprompt. The prompt category contains events originating from SM processes capable of producing three or more prompt leptons in the final state. These include the \PW\PZ, $\PZ\PZ$, and triboson processes in the ``VV(V)'' group, and the \cPqt\cPaqt\PZ and \cPqt\cPaqt\PW\ processes in the ``\cPqt\cPaqt+V'' group. We use simulation to predict the yields of these background processes. The nonprompt category contains events with nonprompt leptons that pass the tight lepton identification and isolation criteria, and jets misidentified as leptons, such as trilepton events coming from \ttbar or \PZ+jets processes. We use a three-lepton extended version of the tight-to-loose technique to estimate the rate of nonprompt background events.

\subsection{Prompt and misidentified lepton efficiencies}\label{sec:FR}

Prompt lepton efficiencies are the same in the same-sign dilepton and trilepton channels. Misidentified lepton efficiencies are obtained from measurements in the control region using events with exactly three leptons. The misidentified lepton efficiencies are obtained by calculating the minimum of a $\chi^2$ statistic from fits of the predicted background to data. The predicted background is the sum of the nonprompt background estimate and the prompt MC background. Specifically, we use the bins ($i$) of the lepton \pt distribution to calculate $\chi^2$:
\begin{equation} \label{eq:chiSq}
\chi^{2} (r) = \sum\limits_{i} \frac{\bigl[ N^{i}_{\text{data}} - (N^{i}_{\mathrm{NP}}(r)+ N^{i}_{\mathrm{MC}}) \bigr]^2}{N^{i}_{\mathrm{NP}}(r)+ N^{i}_{\mathrm{MC}}},
\end{equation}
where $r$ represents the prompt and misidentified lepton efficiencies, $N_{\text{data}}$ is the number of events observed in data, $N_{\mathrm{NP}}(r)$ is the number of nonprompt background events (as a function of $r$) estimated from data, $N_{\mathrm{MC}}$ is the number of prompt background events estimated from MC simulation.

Using Eq.~(\ref{eq:chiSq}), we calculate $\chi^2$ for each of the four flavor categories, while varying both the misidentified electron and muon efficiencies from 0.01 to 0.5, and sum the individual terms. The minimum of the $\chi^2$ per degree of freedom is found to be $1.58$ which corresponds to misidentified electron and muon efficiencies of $0.20 \pm 0.02$ and $0.14 \pm 0.01$, respectively. The uncertainties are the standard deviations of a Gaussian probability distribution constructed from the $\chi^2$ values.

\begin{figure}[hbp]
\centering
\includegraphics[width=0.49\textwidth]{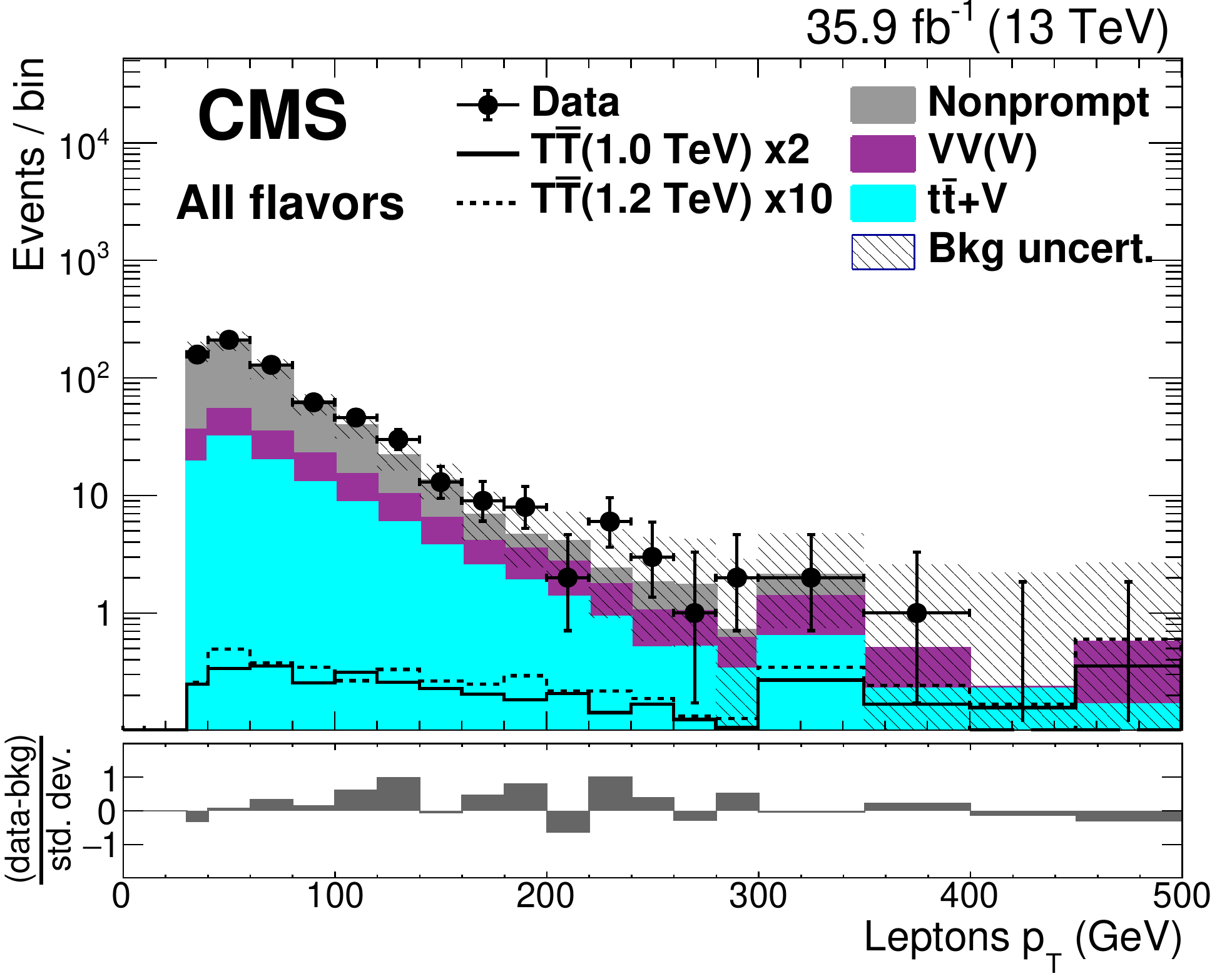}
\includegraphics[width=0.49\textwidth]{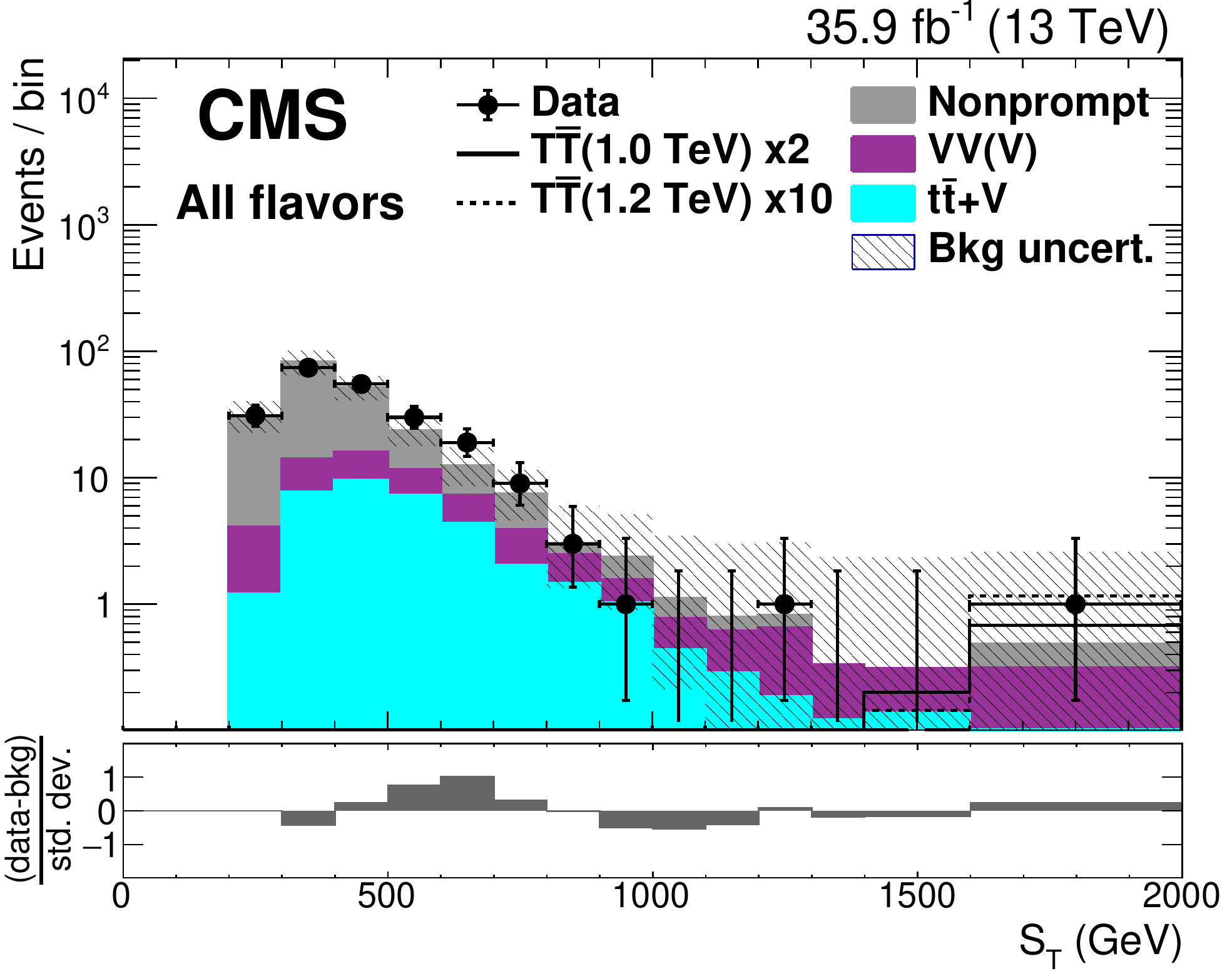}
\caption{Distributions of lepton \pt (left) and \ST (right) before the fit to data in the control region of the trilepton channel. The black points are the data (horizontal bars indicate the bin width) and the filled histograms show the background distributions, with simulated backgrounds grouped into categories as described in Section~\ref{sec:samples}. The expected signal is shown by solid and dotted lines for \PQT quark masses of 1.0 and 1.2\TeV. The final bin includes overflow events. Uncertainties, indicated by the hatched area, include both statistical and systematic components. The lower panel shows the difference between data and background divided by the total uncertainty.}
\label{fig:Trilepton_KinDistribAfterPresel_CR}
\end{figure}

Figure~\ref{fig:Trilepton_KinDistribAfterPresel_CR} shows distributions of lepton \pt and \ST in the control regions, where the nonprompt background is estimated using the misidentified lepton efficiencies that correspond to the minimum of the $\chi^2$.

We perform a closure test for the nonprompt background estimation by measuring the misidentified lepton efficiencies in a \ttbar MC sample. This measurement is used to predict the number of events with three tight leptons, two of which are prompt and one nonprompt. The following discrepancies are observed between the number of observed and predicted events: 28\% in the $\Pe\Pe\Pe$ channel, 31\% in the $\Pe\Pe\mu$ channel, 17\% in the $\Pe\mu\mu$ channel, and 20\% in the $\mu\mu\mu$ channel. In addition, we perform misidentification efficiency measurements using the $\chi^2$ minimization method described above in the \ttbar MC sample in both the control region and signal region selections. We observe that there is a change of 0.04 in the misidentified electron efficiency between regions and negligible change in the misidentified muon efficiency. The change in the misidentified electron efficiency is assigned as a systematic uncertainty. The misidentified lepton efficiencies are taken to be \pt-independent and any dependency on $\eta$ is included as a systematic uncertainty.

\section{Systematic uncertainties}\label{sec:systs}

We consider sources of systematic uncertainties that can affect the normalization and/or the shape of expected background distributions. A summary of the systematic uncertainties and how they are applied to signal and background samples can be found in Table~\ref{tab:systs}.

\begin{table}[htbp]
  \centering
    \topcaption{Summary of values for normalization uncertainties and dependencies for shape uncertainties. The symbol $\sigma$ denotes one standard deviation of the uncertainty and ``env'' denotes an envelope of values. Background from opposite-sign dilepton events is denoted ``OS'', background from nonprompt leptons is denoted ``NP'', while other backgrounds modeled from simulation are denoted ``MC''. For signals, theoretical uncertainties are labeled as ``Shape'' for shape-based searches, and ``Accept.'' for counting experiments. Additionally, ``CR'' denotes control region  and ``RMS'' denotes root mean square.}\label{tab:systs}
\cmsTable{
    \begin{tabular}{lccccccc}
      \multirow{2}{*}{Source} & \multirow{2}{*}{Uncertainty}  & \multicolumn{2}{c}{$1\ell$} & \multicolumn{2}{c}{SS $2\ell$} & \multicolumn{2}{c}{${\geq}3\ell$} \\
                        &   & Sig & Bkg & Sig & Bkg & Sig & Bkg \\
      \hline
      Integrated luminosity                & 2.5\%                  & Yes & MC & Yes & MC & Yes & MC \\
      Reconstruction            & 1\%                    & Yes & MC & Yes & MC & Yes & MC \\
      Identification            & 2\%(\Pe), 3\%($\mu$)   & Yes & MC & Yes & MC & Yes & MC\\
      Isolation (\Pe$,\mu$)       & 1\%                    & Yes & MC & Yes & MC & Yes & MC\\
      Trigger (\Pe\ or $\mu$)    & $\pm \sigma(\pt,\eta)$ & Yes & MC & \NA  & \NA & \NA  & \NA \\
      Trigger ($\ell\ell$)    	& 3\% 	                 & \NA  & \NA  & Yes & MC & \NA  & \NA \\
      Trigger ($\ell\ell\ell$)  & 3\%                    & \NA  & \NA  & \NA  & \NA & Yes & MC \\ [\cmsTabSkip]
      Charge misid. rate        & 30\%                   & \NA  & \NA  & No  & OS & \NA  & \NA  \\
      $\ell$ misid. efficiency        & 50\%                   & \NA  & \NA  & No  & NP & \NA  & \NA  \\ [\cmsTabSkip]
      $\ell$ misid. efficiency                 & 4--30\%              & \NA  & \NA  & \NA  & \NA & No & NP \\
      $\mu$ misid. efficiency $\eta$ dep.  & 12--33\%             & \NA  & \NA  & \NA  & \NA & No & NP \\
      NP method closure         & 17--31\%             & \NA  & \NA  & \NA  & \NA & No & NP \\
      NP method in CR           & 2--35\%              & \NA  & \NA  & \NA  & \NA & No & NP \\
      Prompt $\ell$ efficiency        & 2--9\% (\Pe), 1--7\% ($\mu$)& \NA & \NA  & \NA & \NA & No & NP \\ [\cmsTabSkip]
      Pileup           & $\sigma_{\text{inel.}}\pm4.6$\% & Yes & MC & Yes & MC & Yes & MC\\
      Jet energy scale & $\pm \sigma(\pt,\eta)$            & Yes & MC & Yes & MC & Yes & MC\\
      Jet energy res.  & $\pm \sigma(\eta)$                & Yes & MC & Yes & MC & Yes & MC\\
      \HT scaling      & env(upper, lower fits)           & No  & \PW+jets & \NA & \NA & \NA & \NA\\ [\cmsTabSkip]
      \cPqb\ tag: \cPqb                   & $\pm \sigma(\pt)$      & Yes & MC & \NA & \NA & Yes & MC \\
      \cPqb\ tag: light               & $\pm \sigma$           & Yes & MC & \NA & \NA & Yes & MC \\
      \PW\ tag: $\tau_2/\tau_1$     & $\pm \sigma$           & Yes & MC & \NA & \NA & \NA  & \NA \\
      \PW\ tag: $\tau_2/\tau_1$ \pt & $\pm \sigma(\pt)$      & Yes & MC & \NA & \NA & \NA  & \NA \\
      \PW/\PH tag: mass scale        & $\pm \sigma(\pt,\eta)$ & Yes & MC & \NA & \NA & \NA  & \NA \\
      \PW/\PH tag: mass res.         & $\pm \sigma(\eta)$     & Yes & MC & \NA & \NA & \NA  & \NA \\
      \PH tag: propagation         & 5\%                    & Yes & MC & \NA & \NA & \NA  & \NA \\ [\cmsTabSkip]
      Renorm./fact. scale & env($\times$2,\,$\times$0.5) & Shape & MC & Accept. & MC & Shape & MC\\
      PDF                 & RMS(replicas)                & Shape & MC & Accept. & MC & Shape & MC\\
      VV rate             & 15\%                         & No    & VV  & \NA      & \NA & \NA    & \NA\\
      Single \cPqt\PW\ rate      & 16\%                         & No    & \cPqt\PW  & \NA      & \NA & \NA    & \NA\\[\cmsTabSkip]
    \end{tabular}
    }
\end{table}

The uncertainty in the integrated luminosity is 2.5\%~\cite{LUM-17-001} and is applied to all samples. Lepton reconstruction, identification, and isolation efficiency scale factor uncertainties are applied based on the number of leptons in each channel. Trigger efficiency uncertainties in each channel are independent, and are applied as a function of lepton flavor, \pt, and $\abs{\eta}$ in the single lepton channel, and as flat percentages in the SS dilepton and trilepton channels.
In the single-lepton channel a 15\% uncertainty is applied to the cross section of diboson samples~\cite{SMP-14-016,SMP-16-002,SMP-16-001}, and a 16\% uncertainty is applied to single $\cPqt\PW$ production.

In the SS dilepton channel, closure tests are performed in the \ttbar MC simulation by comparing the predicted nonprompt background using the tight-to-loose method and the observed nonprompt background from truth information, based on which an uncertainty of 50\% is applied for the nonprompt background yield. An uncertainty of 30\% is applied to the OS prompt background to account for possible \pt variations in the rate of charge misidentification within the \pt bins, and for differences in rates of charge misidentification calculated in Drell--Yan versus \ttbar MC.

In the trilepton channel, an uncertainty in the nonprompt background yield is calculated by varying the misidentified lepton efficiencies by their uncertainties of 0.04 for electrons and 0.01 for muons. These are obtained by summing in quadrature the statistical uncertainties and the systematic uncertainties due to the possible discrepancies between misidentification efficiencies measured in the control region and in the signal region. It results in an uncertainty of 12--30\% (4--12\%) in the nonprompt background yield. From the closure test described in Section~\ref{sec:TripleLeptonFinalState}, we also apply an uncertainty of 17--31\% in the nonprompt background yield based on discrepancies between the tight-to-loose method prediction and the observed yields in simulation.
As an additional source of systematic uncertainty, we evaluate the remaining difference in yield between the background estimate and data in the control region, using the misidentification efficiencies measured in that region. These differences range from 2\% in the $\mu\mu\mu$ channel to 35\% in the $\Pe\Pe\Pe$ channel.
In the SS dilepton channel the muon fake rate can be modeled by a quadratic dependence on $\eta$, while the trilepton channel uses an $\eta$-independent value. The change in trilepton nonprompt background yield if an $\eta$-dependent muon fake rate is adopted is 12--33\%, and an additional uncertainty is applied to take account of this. Finally, the prompt lepton efficiencies were calculated in a control sample selected using a trigger with less stringent lepton isolation requirements than those in the triggers used to select the trilepton channel events. Because the true prompt efficiency in the trilepton channel is expected to be slightly higher than the values used for the SS dilepton channel, an uncertainty is assigned by comparing the trilepton nonprompt background yields with yields obtained when using prompt lepton efficiencies of unity. These uncertainties ranges from 2--9\%\,(1--7\%) in the nonprompt background yield for electrons (muons), with the smallest values in the categories with only one lepton of a given flavor and the largest uncertainties in the same-flavor channels.

Uncertainties affecting both the shape and normalization of the distributions in multiple channels include uncertainties related to the jet energy scale, jet energy resolution, and \cPqb\ tagging and light-parton mistag rates~\cite{BTV-16-002}. The uncertainty due to the pileup simulation is evaluated by adjusting the total inelastic cross section ($\sigma_{\text{inel.}}$) used to calculate the correction by $\pm$4.6\%~\cite{FSQ-15-005}.

The uncertainties in the PDFs used in MC simulations are evaluated from the set of NNPDF3.0 MC replicas~\cite{NNPDF30}. Renormalization and factorization energy scale uncertainties are calculated by varying the corresponding scales up and down (both independently and simultaneously) by a factor of two and taking the envelope, or largest spread, of all observed variations as the uncertainty. These theoretical uncertainties are applied to the signal simulations primarily as shape uncertainties. The normalization uncertainty is small and associated with changes in acceptance. For backgrounds, the full theoretical uncertainties are applied. All common uncertainties are treated as correlated across the three analysis channels.

In the single-lepton channel we also associate shape uncertainties with the \PW\ tagging scale factors for the pruned mass scale and smearing, the $\tau_2$/$\tau_1$ selection efficiency, and its \pt dependence~\cite{JME-16-003}. An uncertainty of 5\% is applied to account for the effects of propagating corrections derived from the \PW\ mass peak to the Higgs mass peak. These corrections are anticorrelated between categories with and without \PH tags. The uncertainty in the generator-level top quark \pt reweighting is estimated as the difference between weighted and unweighted distributions. This uncertainty is excluded from fits because of strong correlations with the renormalization and factorization energy scale uncertainties. The uncertainty in the \HT scaling procedure is the difference between scaling functions obtained by fitting the inclusive-to-binned \HT ratio after shifting values up or down by their statistical uncertainties.

\section{Results}\label{sec:Results}

The strongest overall sensitivity to \TTbar and \BBbar production is achieved by combining the three leptonic channels, since each channel is sensitive to different VLQ decay modes. Table~\ref{tab:sigeffs} shows the selection efficiency for all three channels in each \TTbar or \BBbar decay mode, with respect to the total number of expected events for a given decay mode (\eg, $\cPqt\PH\cPqt\PH$). The most sensitive decay modes for each channel are noted in bold. Comparing efficiencies across \TTbar decay modes, the single-lepton channel has the highest efficiency for decay modes with at least one $\PQT\to\cPqb\PW$ decay, the SS dilepton channel is sensitive to $\PQB\to\cPqt\PW$ decays, and the trilepton channel has high efficiency for decay modes with at least one $\PQT\to\cPqt\PZ$ decay.

\begin{table}[htbp]
\centering
\topcaption{Signal efficiencies in the single-lepton, same-sign dilepton, and trilepton channels, split into the six possible final states of both \TTbar and \BBbar production, for three mass points. Efficiencies, stated in percent, are calculated with respect to the expected number of events in the corresponding decay mode, before any selection. The most sensitive decay modes for each channel are noted in bold. The efficiency for $\cPqb\PW\cPqb\PW$ events in the same-sign dilepton and trilepton channels is negligible, as is the efficiency for $\cPqb\PZ\cPqb\PZ$ events in the same-sign dilepton channel.}\label{tab:sigeffs}
\begin{tabular}{l r@{.}l cr@{.}lc  r@{.}l l r@{.}l cr@{.}lc  r@{.}l}
\multicolumn{9}{ c }{\TTbar (1.0\TeV)}                                         & \multicolumn{9}{ c }{\BBbar (1.0\TeV)}\\
Decay mode & \multicolumn{2}{c}{$1\ell$} & \multicolumn{4}{c}{SS$2\ell$} & \multicolumn{2}{c}{${\geq} 3\ell$} & Decay mode & \multicolumn{2}{c}{$1\ell$} & \multicolumn{4}{c}{SS$2\ell$} & \multicolumn{2}{c}{${\geq} 3\ell$} \\
\hline
$\cPqt\PH\cPqt\PH$ & 9&1 && \textbf{1}&\textbf{1} && 0&74                      & $\cPqb\PH\cPqb\PH$ & 2&9 && 0&16 && 0&08 \\
$\cPqt\PH\cPqt\PZ$ & 8&4 && \textbf{0}&\textbf{78} && \textbf{1}&\textbf{50}   & $\cPqb\PH\cPqb\PZ$ & 1&8 && 0&05 && 0&22 \\
$\cPqt\PH\cPqb\PW$ & \textbf{11}&\textbf{0} && 0&61 && 0&29                    & $\cPqb\PH\cPqt\PW$ & \textbf{11}&\textbf{2} && \textbf{0}&\textbf{61} && 0&31 \\
$\cPqt\PZ\cPqt\PZ$ & 7&4 && 0&45 && \textbf{1}&\textbf{92}                     & $\cPqb\PZ\cPqb\PZ$ & 1&0 && 0&02 && 0&25 \\
$\cPqt\PZ\cPqb\PW$ & 9&2 && 0&34 && 0&88                                       & $\cPqb\PZ\cPqt\PW$ & 9&2 && 0&23 && \textbf{ 0}&\textbf{89} \\
$\cPqb\PW\cPqb\PW$ & \textbf{10}&\textbf{8} && 0&02 && \multicolumn{2}{c}{\NA} & $\cPqt\PW\cPqt\PW$ & \textbf{12}&\textbf{3} && \textbf{2}&\textbf{5} && \textbf{1}&\textbf{28} \\[\cmsTabSkip]
\multicolumn{9}{ c }{\TTbar (1.2\TeV)}                                         & \multicolumn{9}{ c }{\BBbar (1.2\TeV)}\\
Decay mode & \multicolumn{2}{c}{$1\ell$} & \multicolumn{4}{c}{SS$2\ell$} & \multicolumn{2}{c}{${\geq} 3\ell$} & Decay mode & \multicolumn{2}{c}{$1\ell$} & \multicolumn{4}{c}{SS$2\ell$} & \multicolumn{2}{c}{${\geq} 3\ell$} \\
\hline
$\cPqt\PH\cPqt\PH$ & 10&9 && \textbf{1}&\textbf{4} && 0&81                     & $\cPqb\PH\cPqb\PH$ & 3&2 && 0&19 && 0&08 \\
$\cPqt\PH\cPqt\PZ$ & 10&1 && \textbf{0}&\textbf{93} && \textbf{1}&\textbf{48}  & $\cPqb\PH\cPqb\PZ$ & 2&0 && 0&08 && 0&19 \\
$\cPqt\PH\cPqb\PW$ & \textbf{12}&\textbf{4} && 0&71 && 0&31                    & $\cPqb\PH\cPqt\PW$ & \textbf{12}&\textbf{6} && \textbf{0}&\textbf{73} && 0&29 \\
$\cPqt\PZ\cPqt\PZ$ & 8&8 && 0&53 && \textbf{1}&\textbf{98}                     & $\cPqb\PZ\cPqb\PZ$ & 1&0 && 0&03 && 0&20 \\
$\cPqt\PZ\cPqb\PW$ & 10&4 && 0&27 && 0&87                                      & $\cPqb\PZ\cPqt\PW$ & 10&4 && 0&28 && \textbf{0}&\textbf{87} \\
$\cPqb\PW\cPqb\PW$ & \textbf{11}&\textbf{4} && 0&04 && \multicolumn{2}{c}{\NA} & $\cPqt\PW\cPqt\PW$ & \textbf{14}&\textbf{1} && \textbf{2}&\textbf{8} && \textbf{1}&\textbf{33} \\[\cmsTabSkip]
\multicolumn{9}{ c }{\TTbar (1.4\TeV)}                                         & \multicolumn{9}{ c }{\BBbar (1.4\TeV)}\\
Decay mode & \multicolumn{2}{c}{$1\ell$} & \multicolumn{4}{c}{SS$2\ell$} & \multicolumn{2}{c}{${\geq} 3\ell$} & Decay mode & \multicolumn{2}{c}{$1\ell$} & \multicolumn{4}{c}{SS$2\ell$} & \multicolumn{2}{c}{${\geq} 3\ell$} \\
\hline
$\cPqt\PH\cPqt\PH$ & 11&7 && \textbf{1}&\textbf{5} && 0&81                     & $\cPqb\PH\cPqb\PH$ & 3&2 && 0&19 && 0&07 \\
$\cPqt\PH\cPqt\PZ$ & 10&8 && \textbf{0}&\textbf{95} && \textbf{1}&\textbf{47}  & $\cPqb\PH\cPqb\PZ$ & 2&0 && 0&07 && 0&18 \\
$\cPqt\PH\cPqb\PW$ & \textbf{13}&\textbf{3} && 0&49 && 0&30                    & $\cPqb\PH\cPqt\PW$ & \textbf{13}&\textbf{4} && \textbf{0}&\textbf{75} && 0&29 \\
$\cPqt\PZ\cPqt\PZ$ & 9&3 && 0&29 && \textbf{1}&\textbf{87}                     & $\cPqb\PZ\cPqb\PZ$ & 1&0 && 0&02 && 0&20 \\
$\cPqt\PZ\cPqb\PW$ & 10&9 && 0&75 && 0&85                                      & $\cPqb\PZ\cPqt\PW$ & 11&0 && 0&29 && \textbf{ 0}&\textbf{81} \\
$\cPqb\PW\cPqb\PW$ & \textbf{11}&\textbf{8} && 0&03 && \multicolumn{2}{c}{\NA} & $\cPqt\PW\cPqt\PW$ & \textbf{15}&\textbf{4} && \textbf{3}&\textbf{05} && \textbf{1}&\textbf{36} \\[\cmsTabSkip]
\end{tabular}
\end{table}

\begin{figure}[hbtp]
\centering
\includegraphics[width=0.49\textwidth]{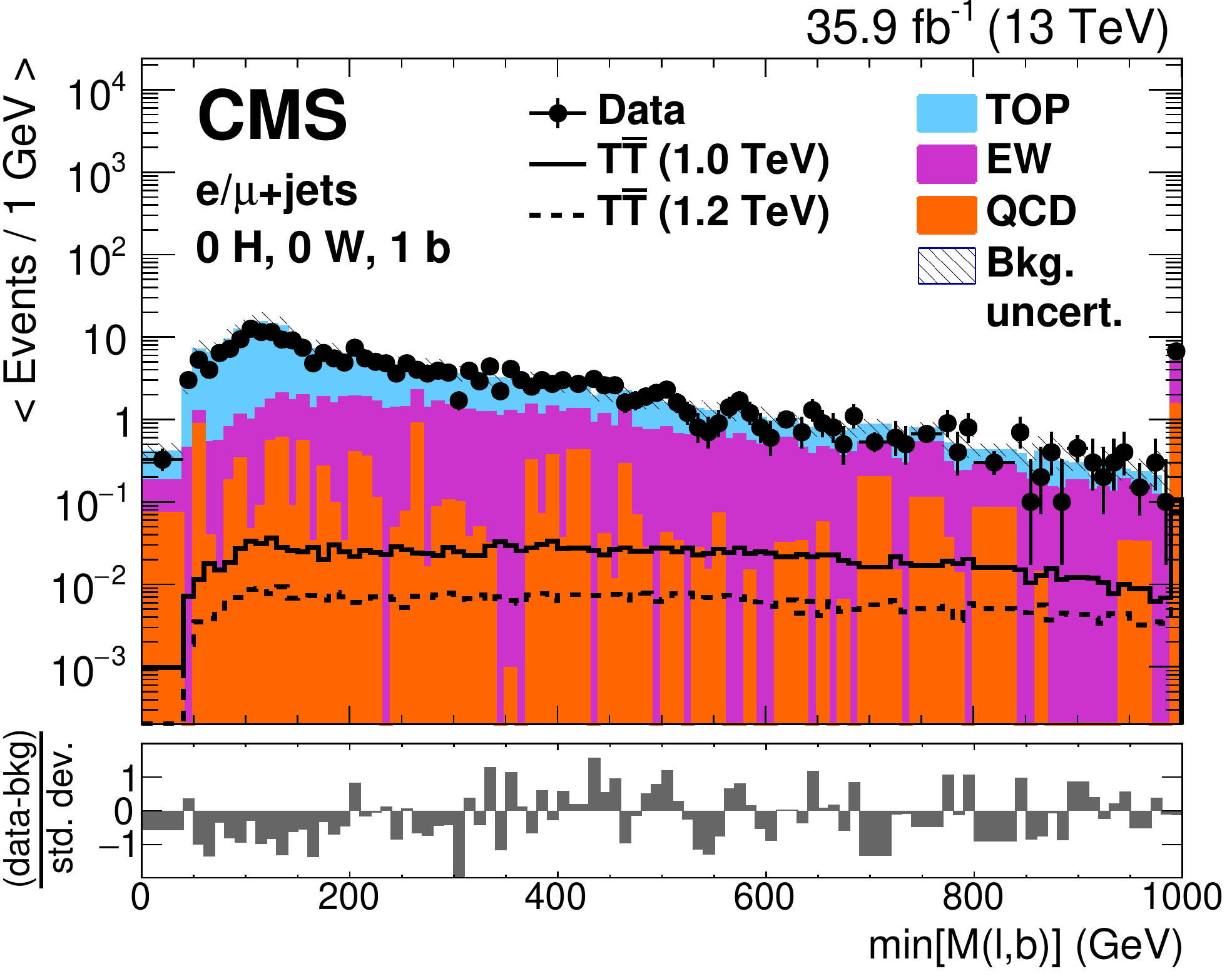}
\includegraphics[width=0.49\textwidth]{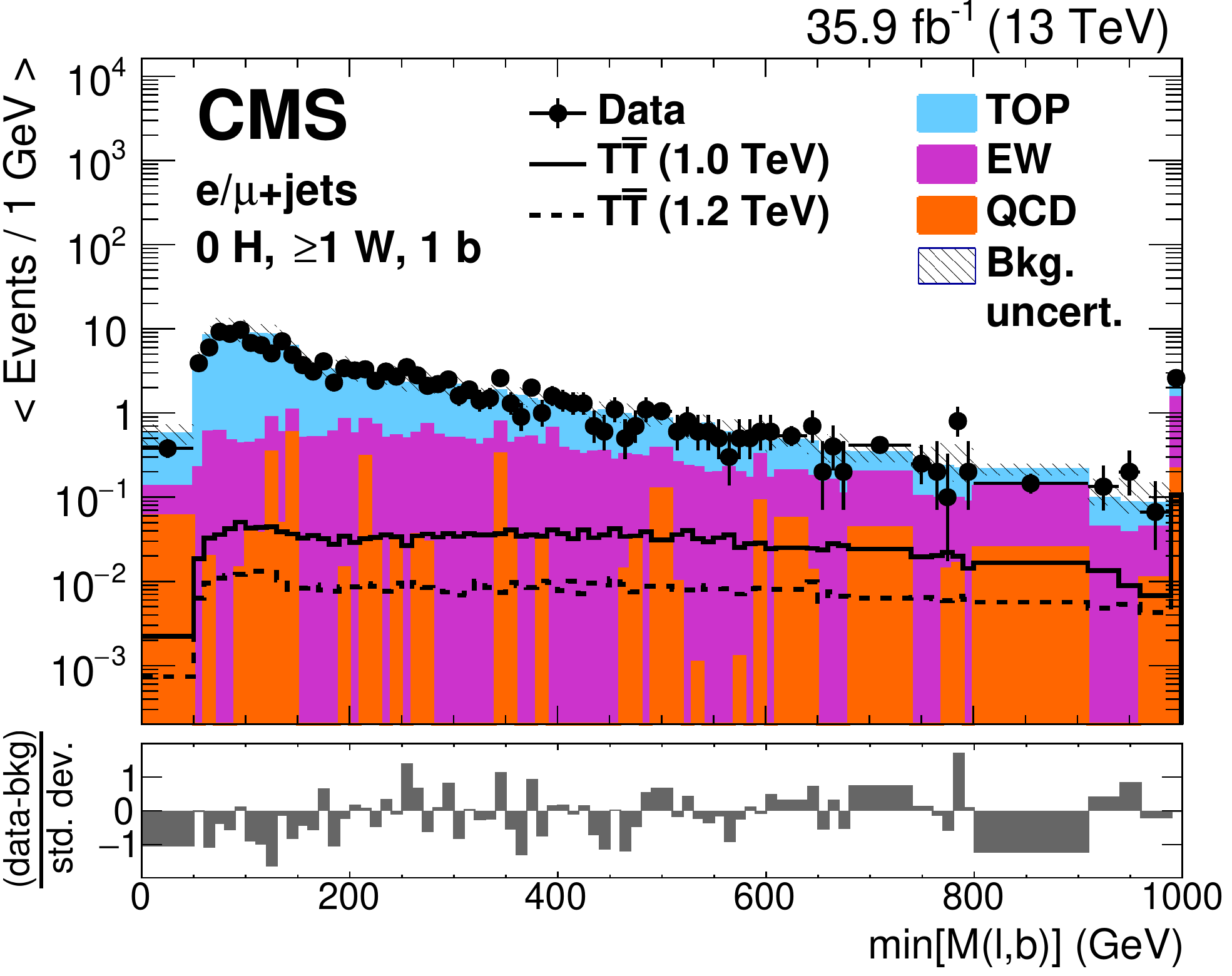}
\includegraphics[width=0.49\textwidth]{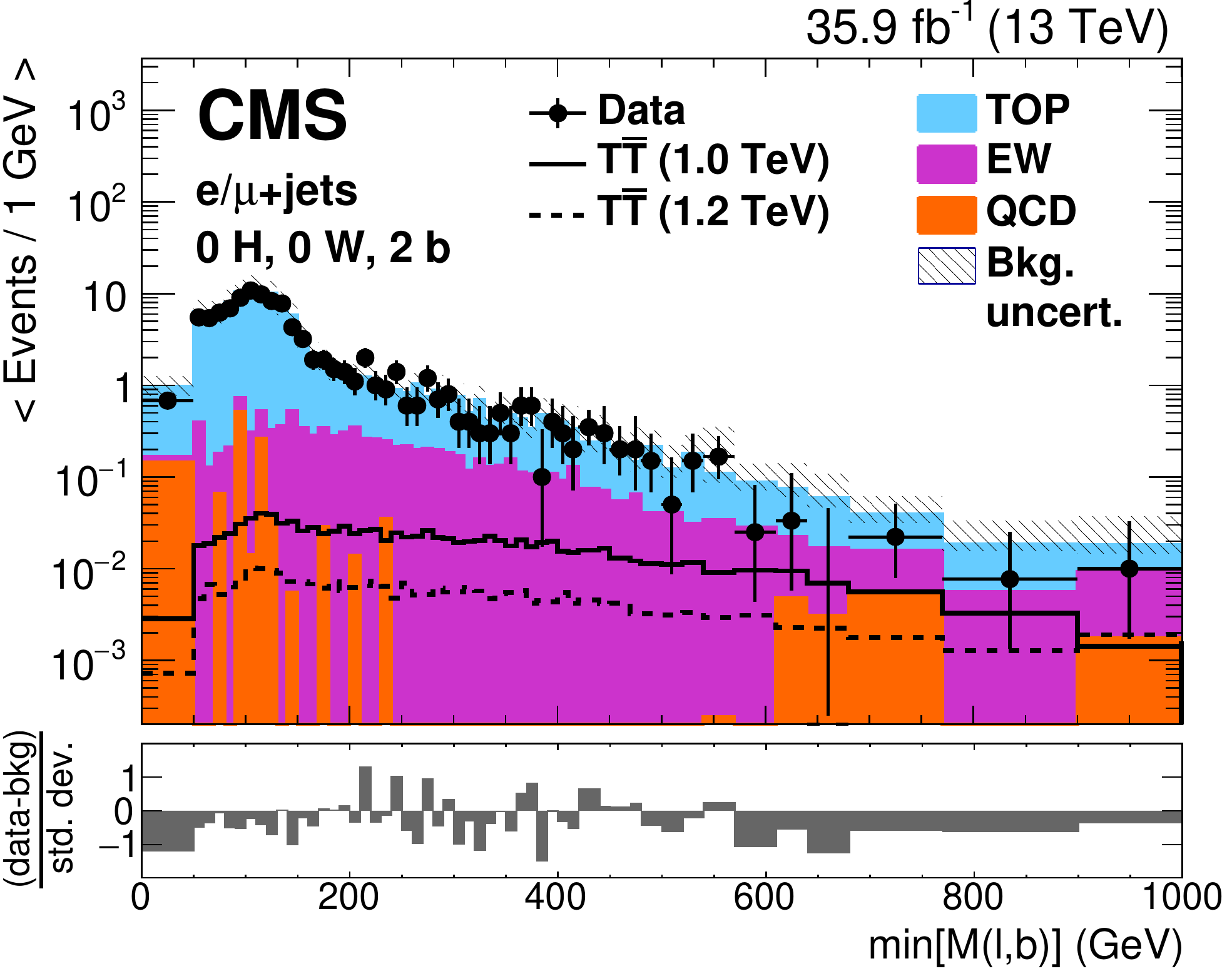}
\includegraphics[width=0.49\textwidth]{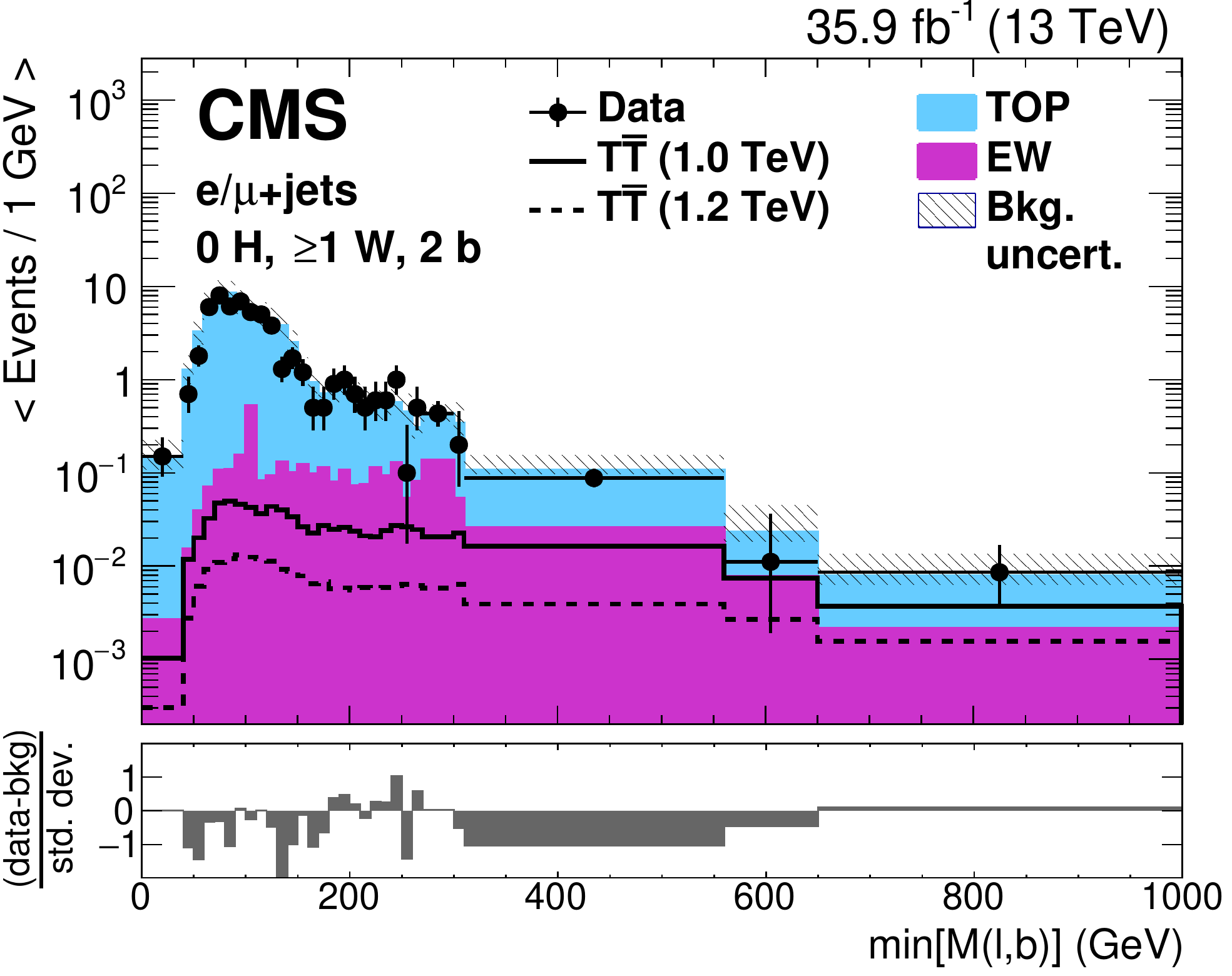}
\includegraphics[width=0.49\textwidth]{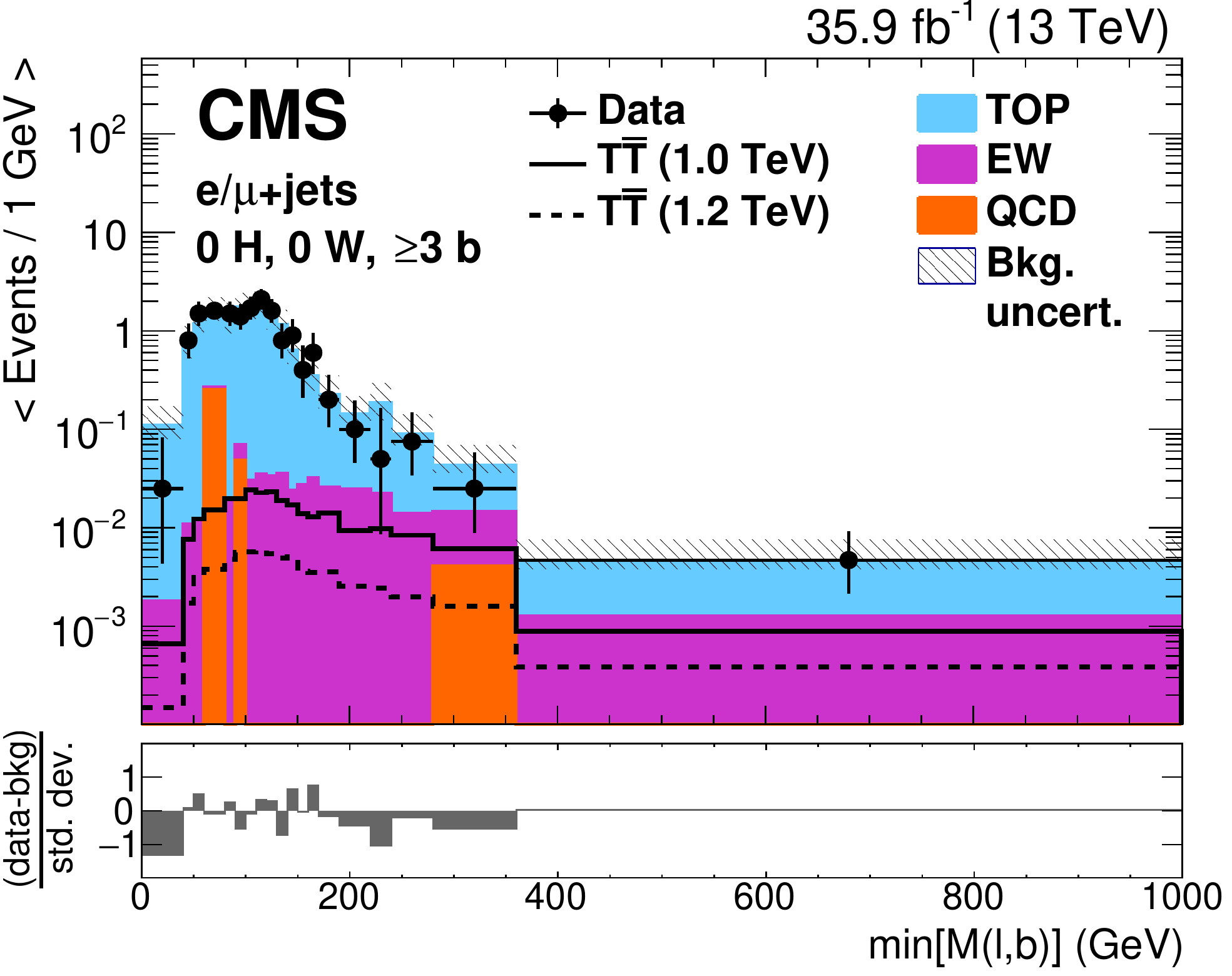}
\includegraphics[width=0.49\textwidth]{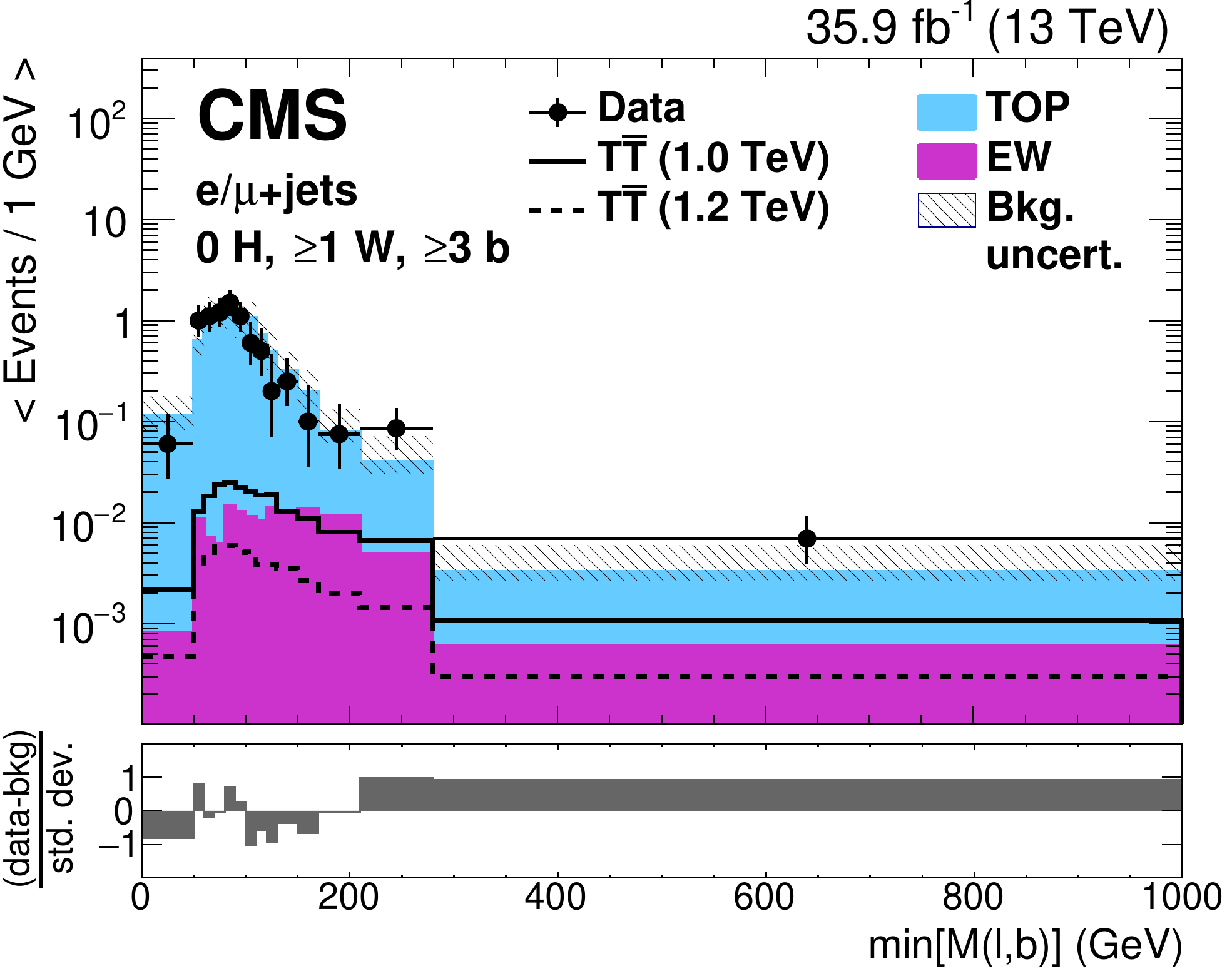}
\caption{Distributions of \minMlb\ before the fit to data in the single-lepton W0 (left) or W1 (right) categories with 1, 2, or ${\geq}3$ (upper to lower) \cPqb-tagged jets. The black points are the data (horizontal bars indicate the bin width) and the filled histograms show the simulated background distributions, grouped into categories as described in Section~\ref{sec:samples}. The expected signal is shown by solid and dotted lines for \PQT quark masses of 1.0 and 1.2\TeV. The final bin includes overflow events. Uncertainties, indicated by the hatched area, include both statistical and systematic components The lower panel shows the difference between data and background divided by the total uncertainty.}
\label{fig:cats}

\end{figure}

\begin{figure}[hbtp]
\centering
\includegraphics[width=0.49\textwidth]{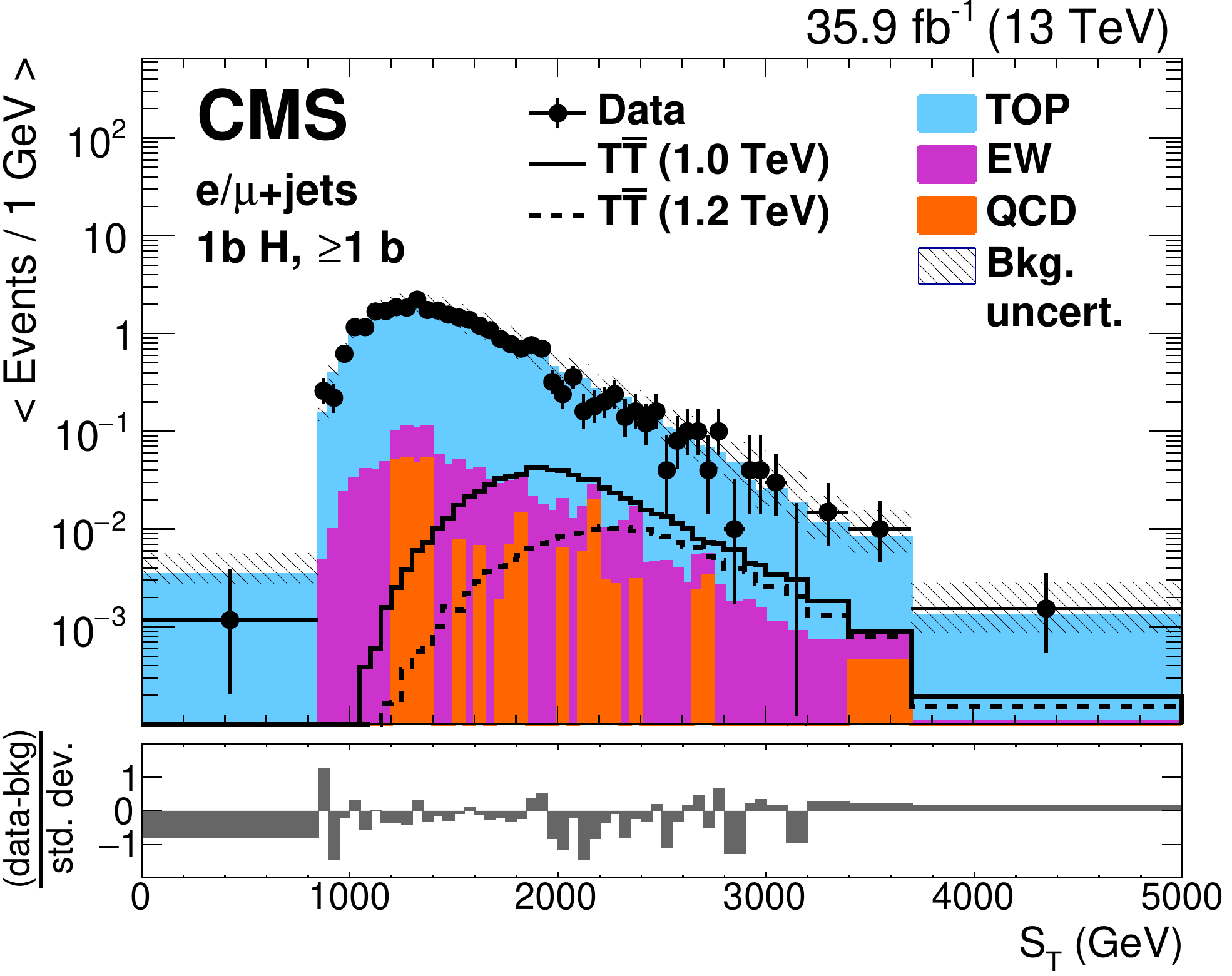}
\includegraphics[width=0.49\textwidth]{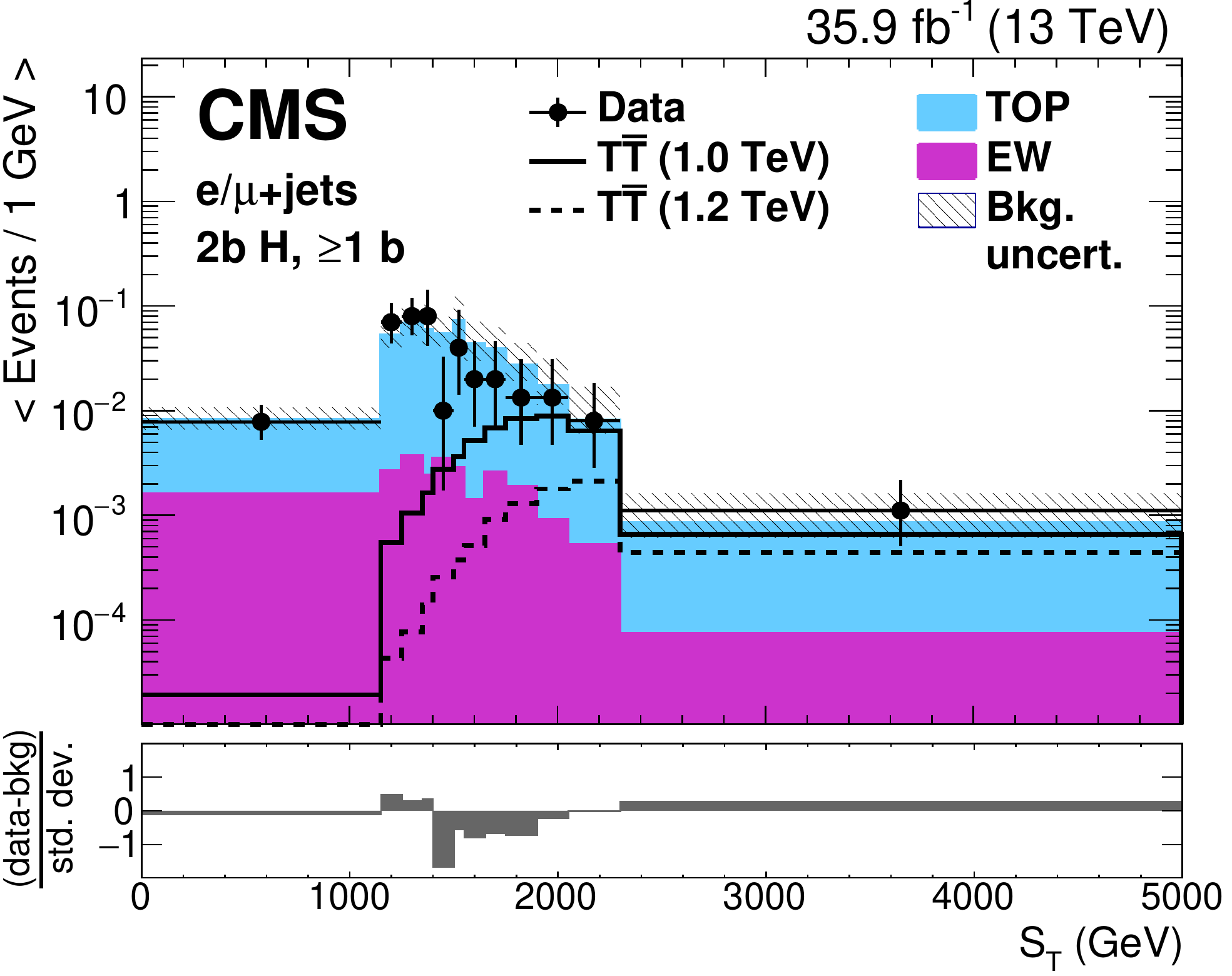}
\caption{Distributions of \ST before the fit to data in the single-lepton H1b (left) or H2b (right) categories. Uncertainties, indicated by the hatched area, include both statistical and systematic components. The black points are the data (horizontal bars indicate the bin width) and the filled histograms show the simulated background distributions, grouped into categories as described in Section~\ref{sec:samples}. The final bin includes overflow events. The expected signal is shown by solid and dotted lines for \PQT quark masses of 1.0 and 1.2\TeV. The lower panel shows the difference between data and background divided by the total uncertainty.}
\label{fig:cats2}

\end{figure}

\begin{figure}[hbtp]
\centering
\includegraphics[width=0.49\textwidth]{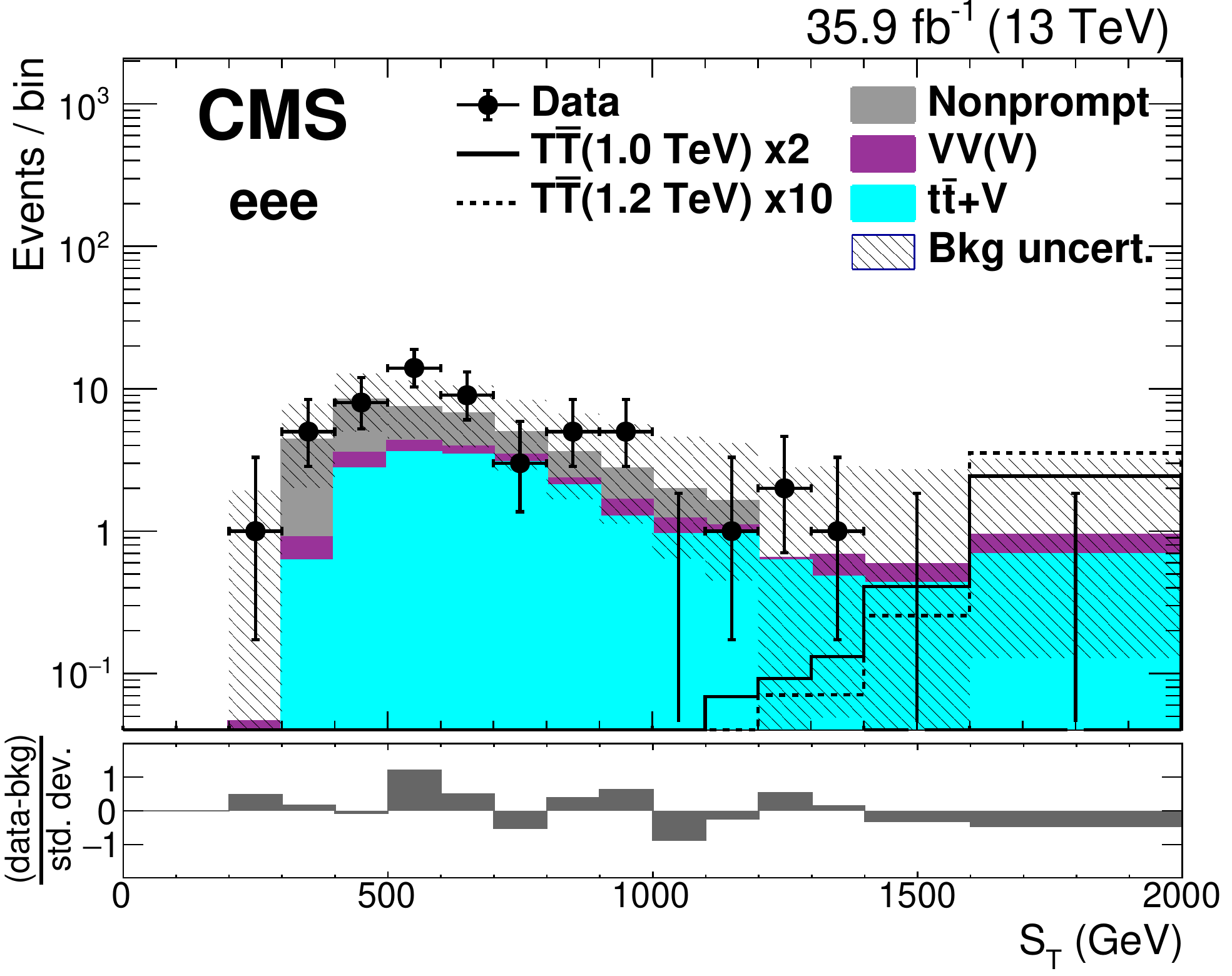}
\includegraphics[width=0.49\textwidth]{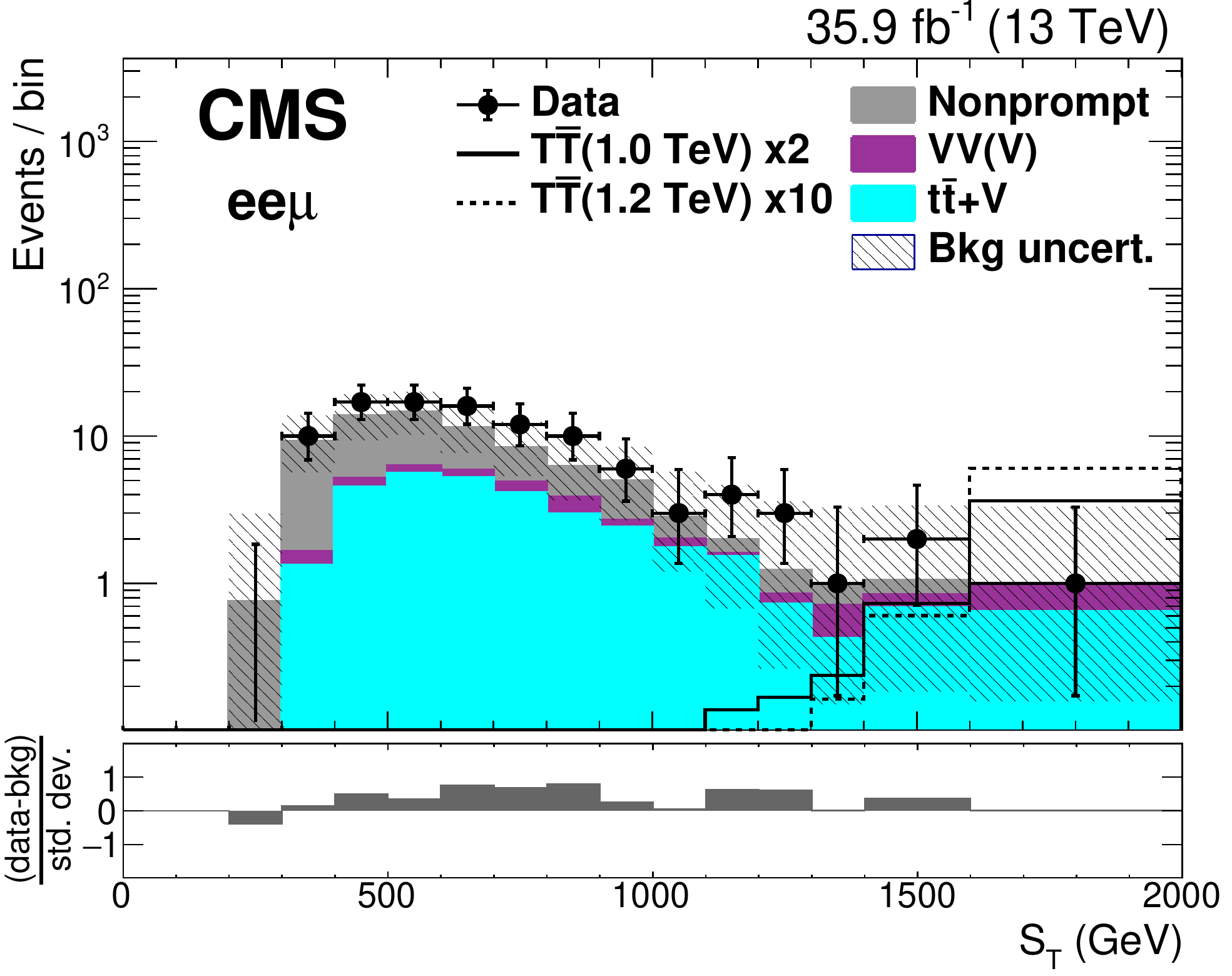}
\includegraphics[width=0.49\textwidth]{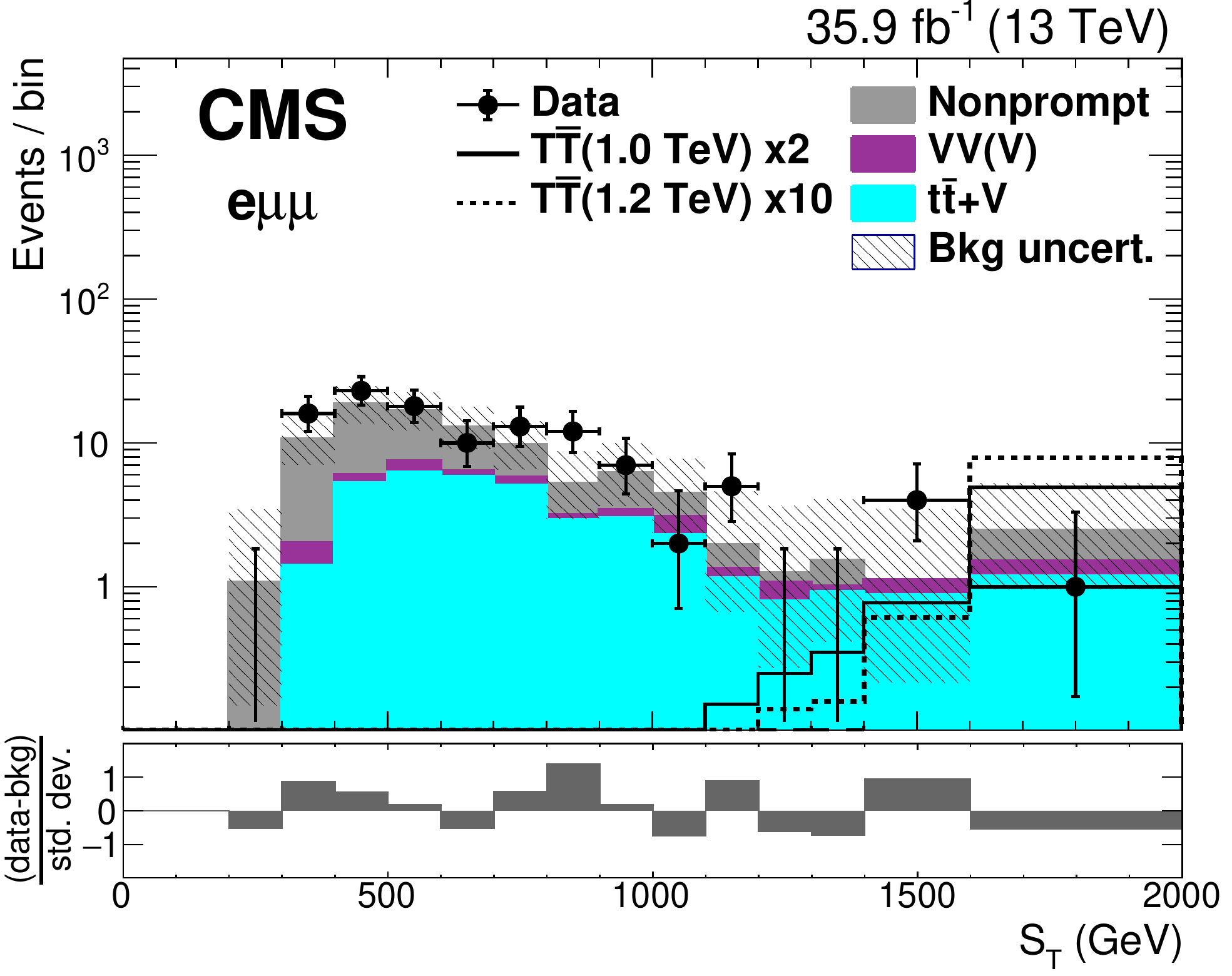}
\includegraphics[width=0.49\textwidth]{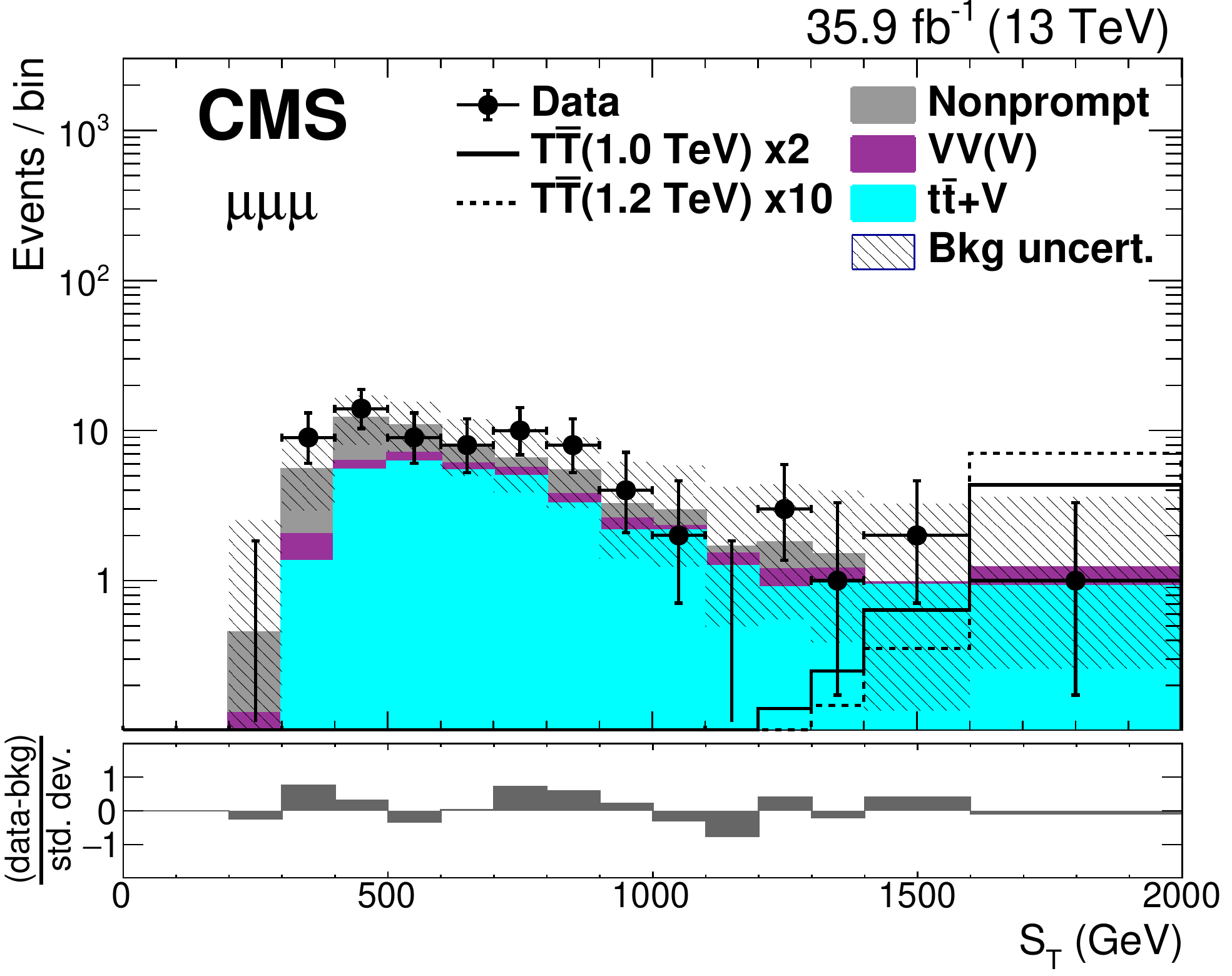}
\caption{Distributions of \ST in the trilepton final state before the fit to data, in the four flavor categories. The black points are the data (horizontal bars indicate the bin width) and the filled histograms show the background distributions, with simulated backgrounds grouped into categories as described in Section~\ref{sec:samples}. The expected signal is shown by solid and dotted lines for \PQT quark masses of 1.0 and 1.2\TeV. The final bin includes overflow events. Uncertainties, indicated by the hatched area, include both statistical and systematic components. The lower panel shows the difference between data and background divided by the total uncertainty.}\label{fig:yieldmain}
\end{figure}

Distributions of \minMlb\ and \ST in the search regions are shown in Figs.~\ref{fig:cats} and~\ref{fig:cats2} for the single-lepton channel categories. The distributions are binned such that the simulated background has a statistical uncertainty of ${<}30\%$ in each bin. Figure~\ref{fig:yieldmain} shows the \ST distribution in each category of the trilepton channel. The slight excess of data in the low \ST region is within the systematic uncertainty in the misidentified lepton efficiencies that describes the rate difference between the control and signal regions. Predicted and observed event yields for the single-lepton, SS dilepton, and trilepton channels are listed in Tables~\ref{tab:nevents1L}--\ref{tab:nevents3L}. The \PQT quark distributions and event yields are for the singlet branching fraction benchmark. No significant excess of data above the background prediction is observed.

\begin{table}[htbp]
\centering
\topcaption{Numbers of predicted and observed events for signal region categories of the single-lepton channel before the fit to data. Uncertainties include both statistical and systematic components.}
\begin{tabular}{l ccr@{\,$\pm$\,}lcc ccr@{\,$\pm$\,}lcc ccr@{\,$\pm$\,}lcc}
Sample  & \multicolumn{6}{c}{0 \PH, 0 \PW, 1 \cPqb} & \multicolumn{6}{c}{0 \PH, 0 \PW, 2 \cPqb} & \multicolumn{6}{c}{0 \PH, 0 \PW, ${\geq}3$ \cPqb} \\
\hline
\TTbar (1.0\TeV)      &&& 21.5 & 1.2       &&&&& 12.87 & 0.74       &&&&& 4.41 & 0.29  &&    \rule{0pt}{2.5ex}   \\
\TTbar (1.2\TeV)      &&& 6.48 & 0.36      &&&&& 3.68 & 0.21        &&&&& 1.22 & 0.08  &&    \\ [\cmsTabSkip]
TOP                   &&& 2030 & 420       &&&&& 1070 & 230         &&&&& 172 & 38     &&    \\
EW                    &&& 720 & 120        &&&&& 94 & 16            &&&&& 7.2 & 1.4    &&    \\
QCD                   &&& 117 & 31         &&&&& 18.1 & 9.7         &&&&& 5.9 & 5.2    &&    \\[\cmsTabSkip]
Total bkg             &&& 2870 & 450       &&&&& 1180 & 230         &&&&& 185 & 38     &&    \\
Data                  & \multicolumn{6}{c}{2598} & \multicolumn{6}{c}{1054} & \multicolumn{6}{c}{182} \\
Data/bkg              &&& 0.90 & 0.14      &&&&& 0.89 & 0.17        &&&&& 0.98 & 0.22  &&    \\[\cmsTabSkip]
Sample                & \multicolumn{6}{c}{0 \PH, ${\geq}1$ \PW, 1 \cPqb} & \multicolumn{6}{c}{0 \PH, ${\geq}1$ \PW, 2 \cPqb} & \multicolumn{6}{c}{0 \PH, ${\geq}1$ \PW, ${\geq}3$ \cPqb} \\
\hline

\TTbar (1.0\TeV)      &&& 27.7 & 1.4       &&&&& 13.91 & 0.73       &&&&& 3.75 & 0.22  &&  \rule{0pt}{2.5ex}    \\
\TTbar (1.2\TeV)      &&& 8.22 & 0.43      &&&&& 3.84 & 0.20        &&&&& 0.92 & 0.06  &&  \\[\cmsTabSkip]
TOP                   &&& 1410 & 290       &&&&& 660 & 130          &&&&& 95 & 21      &&      \\
EW                    &&& 291 & 47         &&&&& 38.1 & 7.6         &&&&& 2.68 & 0.58  &&      \\
QCD                   &&& 36 & 13          &&&&& 6.6 & 6.5          &&&&& \multicolumn{2}{c}{$<$1} &&\\[\cmsTabSkip]
Total bkg             &&& 1730 & 290       &&&&& 700 & 140          &&&&& 98 & 21      &&      \\
Data                  & \multicolumn{6}{c}{1589} & \multicolumn{6}{c}{594} & \multicolumn{6}{c}{96} \\
Data/bkg              &&& 0.92 & 0.16      &&&&& 0.84 & 0.17        &&&&& 0.98 & 0.23  &&      \\[\cmsTabSkip]
Sample                & \multicolumn{6}{c}{H1b, ${\geq}0$ \PW, ${\geq}1$ \cPqb} & \multicolumn{6}{c}{H2b, ${\geq}0$ \PW, ${\geq}1$ \cPqb} \\
\hline

\TTbar (1.0\TeV)      &&& 36.7 & 2.0        &&&&& 7.92 & 0.59 &&&&  \rule{0pt}{2.5ex}     \\
\TTbar (1.2\TeV)      &&& 11.18 & 0.60      &&&&& 2.39 & 0.19 &&&&      \\[\cmsTabSkip]
TOP                   &&& 1510 & 300        &&&&& 49 & 11     &&&&   \\
EW                    &&& 46.9 & 8.1        &&&&& 4.2 & 1.5   &&&&    \\
QCD                   &&& 14.4 & 6.3        &&&&& \multicolumn{2}{c}{${<}1$} && \\[\cmsTabSkip]
Total bkg             &&& 1570 & 300        &&&&& 53 & 11     &&&& \\
Data                  & \multicolumn{6}{c}{1488} & \multicolumn{6}{c}{44}  \\
Data/bkg              &&& 0.95 & 0.18       &&&&& 0.83 & 0.21 &&&& \\[\cmsTabSkip]
\end{tabular}
\label{tab:nevents1L}
\end{table}

\begin{table}[htbp]
\centering
\topcaption{Numbers of predicted and observed events for lepton flavor categories in the same-sign dilepton channel before the fit to data. Uncertainties include both statistical and systematic components.}
\begin{tabular}{l cr@{\,$\pm$\,}lc cr@{\,$\pm$\,}lc  cr@{\,$\pm$\,}lc}
Sample  & \multicolumn{4}{c}{$\Pe\Pe$} & \multicolumn{4}{c}{$\Pe\mu$} & \multicolumn{4}{c}{$\mu\mu$} \\
\hline

\TTbar (1.0\TeV) && 1.34 & 0.08  &&& 3.11 & 0.18 &&& 2.12 & 0.12 &\rule{0pt}{2.5ex}\\
\TTbar (1.2\TeV) && 0.42 & 0.02  &&& 1.00 & 0.06 &&& 0.66 & 0.04 &\\  [\cmsTabSkip]
Prompt SS        && 4.03 & 0.57  &&& 10.2 & 1.4  &&& 5.79  & 0.82& \\
Nonprompt        && 4.6 & 2.6    &&& 10.6 & 5.6  &&& 5.4  & 3.0  &\\
Charge misid.    && 4.1 & 1.3    &&& 2.61 & 0.81 &&& \multicolumn{2}{c}{\NA} &\\[\cmsTabSkip]
Total bkg        && 12.8 & 3.0   &&& 23.4 & 5.8  &&& 11.2 & 3.1  & \\
Data             & \multicolumn{4}{c}{12} & \multicolumn{4}{c}{31} & \multicolumn{4}{c}{9} \\
Data/bkg         && 0.94 & 0.35  &&& 1.33 & 0.41 &&& 0.80 & 0.35 & \\[\cmsTabSkip]
\end{tabular}
\label{tab:nevents2L}
\end{table}

\begin{table}[htbp]
\centering
\topcaption{Numbers of predicted and observed events for lepton flavor categories in the trilepton channel before the fit to data. Uncertainties include both statistical and systematic components.}
\begin{tabular}{l cr@{\,$\pm$\,}lc cr@{\,$\pm$\,}lc  cr@{\,$\pm$\,}lc cr@{\,$\pm$\,}lc}
Sample  & \multicolumn{4}{c}{$\Pe\Pe\Pe$} & \multicolumn{4}{c}{$\Pe\Pe\mu$} & \multicolumn{4}{c}{$\Pe\mu\mu$} & \multicolumn{4}{c}{$\mu\mu\mu$} \\
\hline
\TTbar (1.0\TeV)  	&& 1.60 & 0.14  	&&& 2.54 & 0.18  	&&& 3.32 & 0.23  	&&& 2.79 & 0.23  & \rule{0pt}{2.5ex}\\
\TTbar (1.2\TeV)  	&& 0.40 & 0.03  	&&& 0.71 & 0.05  	&&& 0.90 & 0.06  	&&& 0.78 & 0.06  & \\[\cmsTabSkip]
VV(V)            	&& 4.32 & 0.77  	&&& 5.44 & 0.78  	&&& 6.52 & 0.93  	&&& 5.89 & 0.89  & \\
\ttbar+V          	&& 20.9 & 2.9           &&& 31.9 & 4.1  	&&& 37.0 & 4.7  	&&& 35.8 & 5.0   & \\
Nonprompt        	&& 19 & 11  	        &&& 41 & 18  	        &&& 51 & 15  	        &&& 20.0 & 8.4   & \\[\cmsTabSkip]
Total bkg         	&& 44 & 11              &&& 78 & 19  	        &&& 94 & 15             &&& 61.7 & 9.8   & \\
Data  			& \multicolumn{4}{c}{54} & \multicolumn{4}{c}{102} & \multicolumn{4}{c}{111} & \multicolumn{4}{c}{71}\\
Data/bkg		&& 1.22 & 0.35  	&&& 1.31 & 0.34  	&&& 1.18 & 0.22  	&&& 1.15 & 0.23  & \\[\cmsTabSkip]
\end{tabular}
\label{tab:nevents3L}
\end{table}

Using the \textsc{Theta} program~\cite{THETA}, we calculate Bayesian credible intervals~\cite{PDGSTATS} to set 95\% \CL upper limits on the production cross section of \TTbar at each simulated mass point, for various branching fraction scenarios. Limits are calculated in a simultaneous fit to binned marginal likelihoods from the \minMlb\ and \ST distributions for the 16 single-lepton signal-region categories, \HT distributions for the 6 single-lepton aggregate control regions, event yields for the SS dilepton channel, and \ST distributions for the 4 trilepton categories. Statistical uncertainties in the background estimates are treated using the Barlow--Beeston light method~\cite{BBLITE1,BBLITE2}. Other systematic uncertainties are treated as nuisance parameters, as listed in Table~\ref{tab:systs}. Normalization uncertainties are given log-normal priors, and shape uncertainties with shifted templates are given Gaussian priors with a mean of zero and width of one. The signal cross section is assigned a flat prior distribution.

\begin{figure}[hbtp]
\centering
\includegraphics[width=0.49\textwidth]{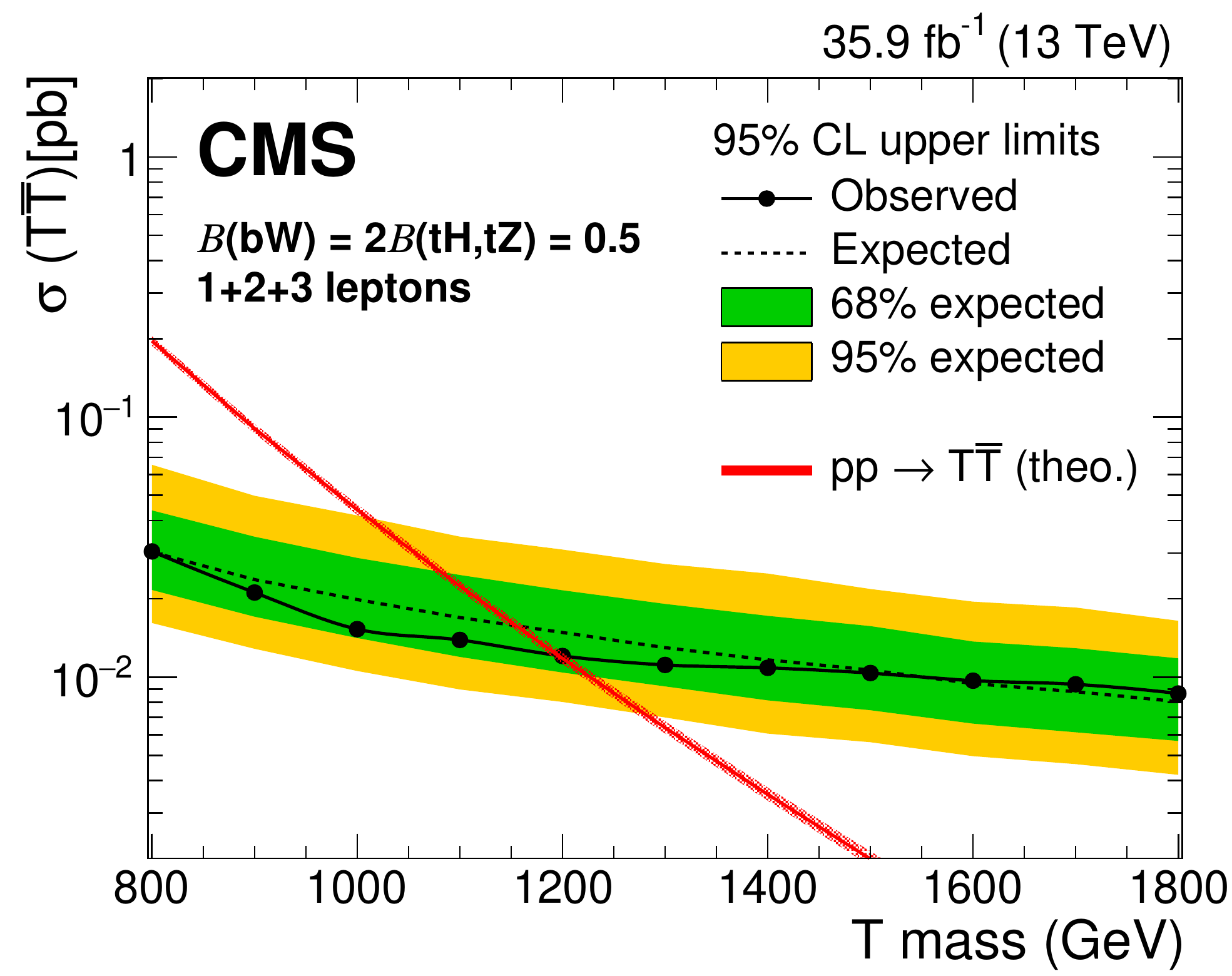}
\includegraphics[width=0.49\textwidth]{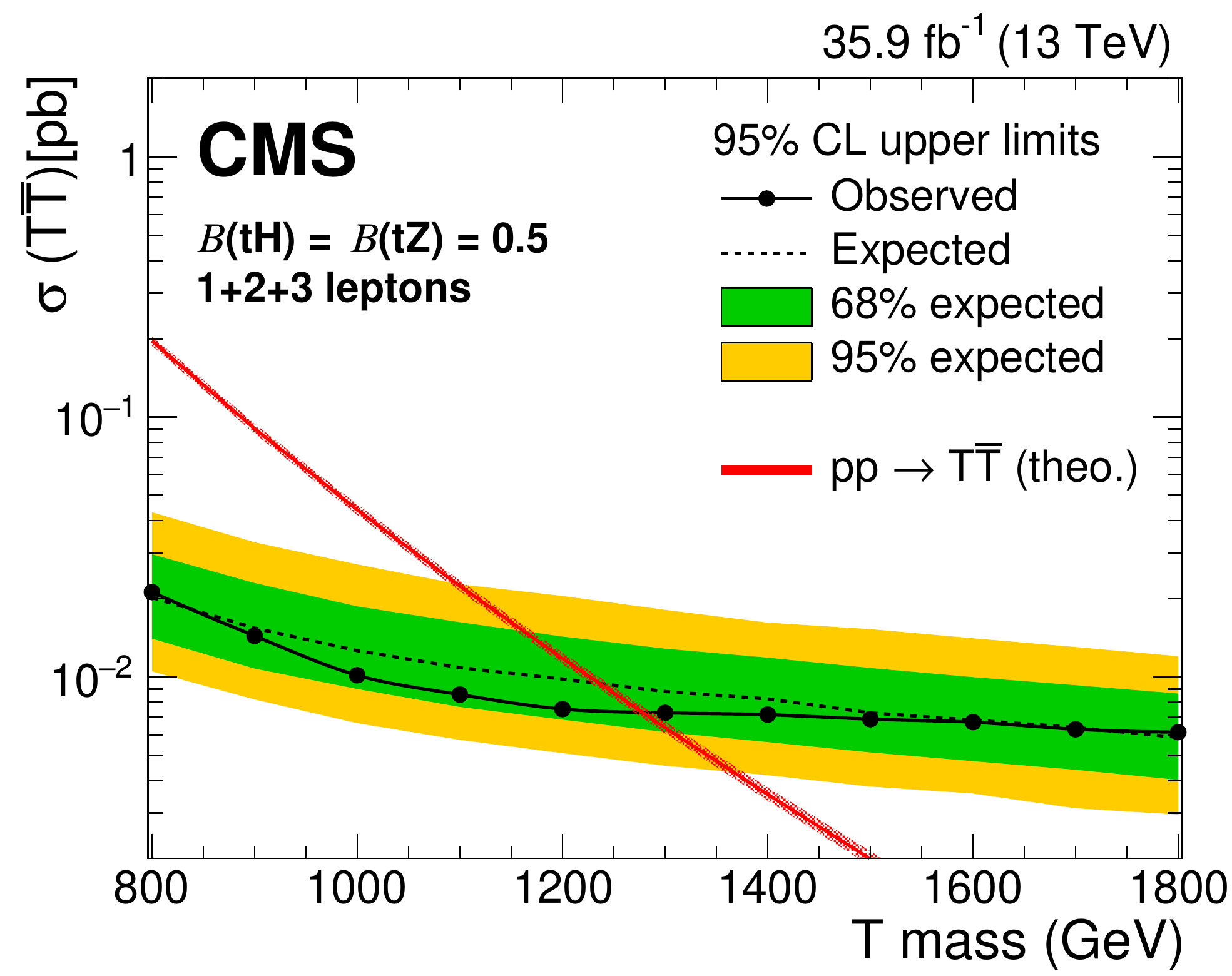}
\includegraphics[width=0.49\textwidth]{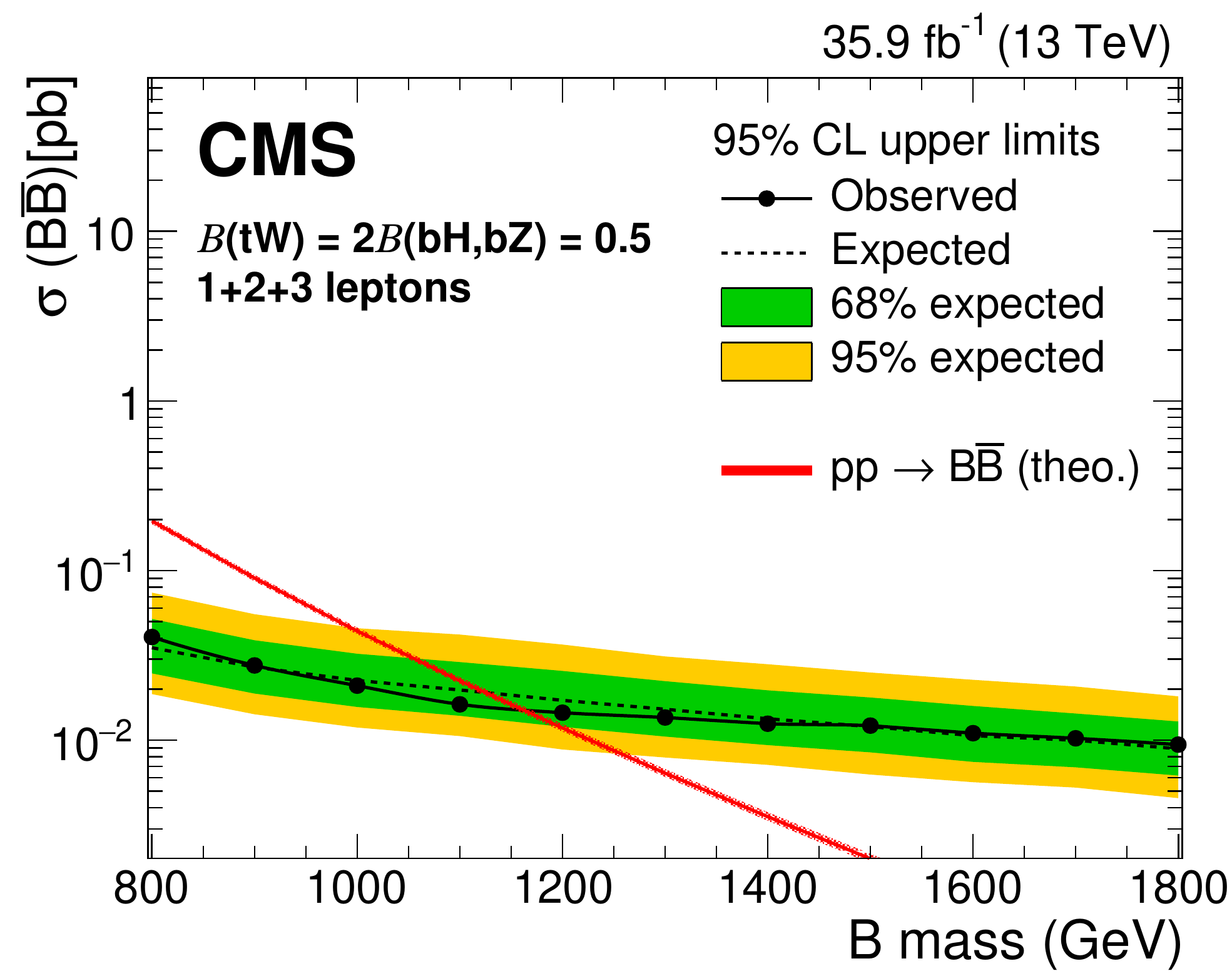}
\includegraphics[width=0.49\textwidth]{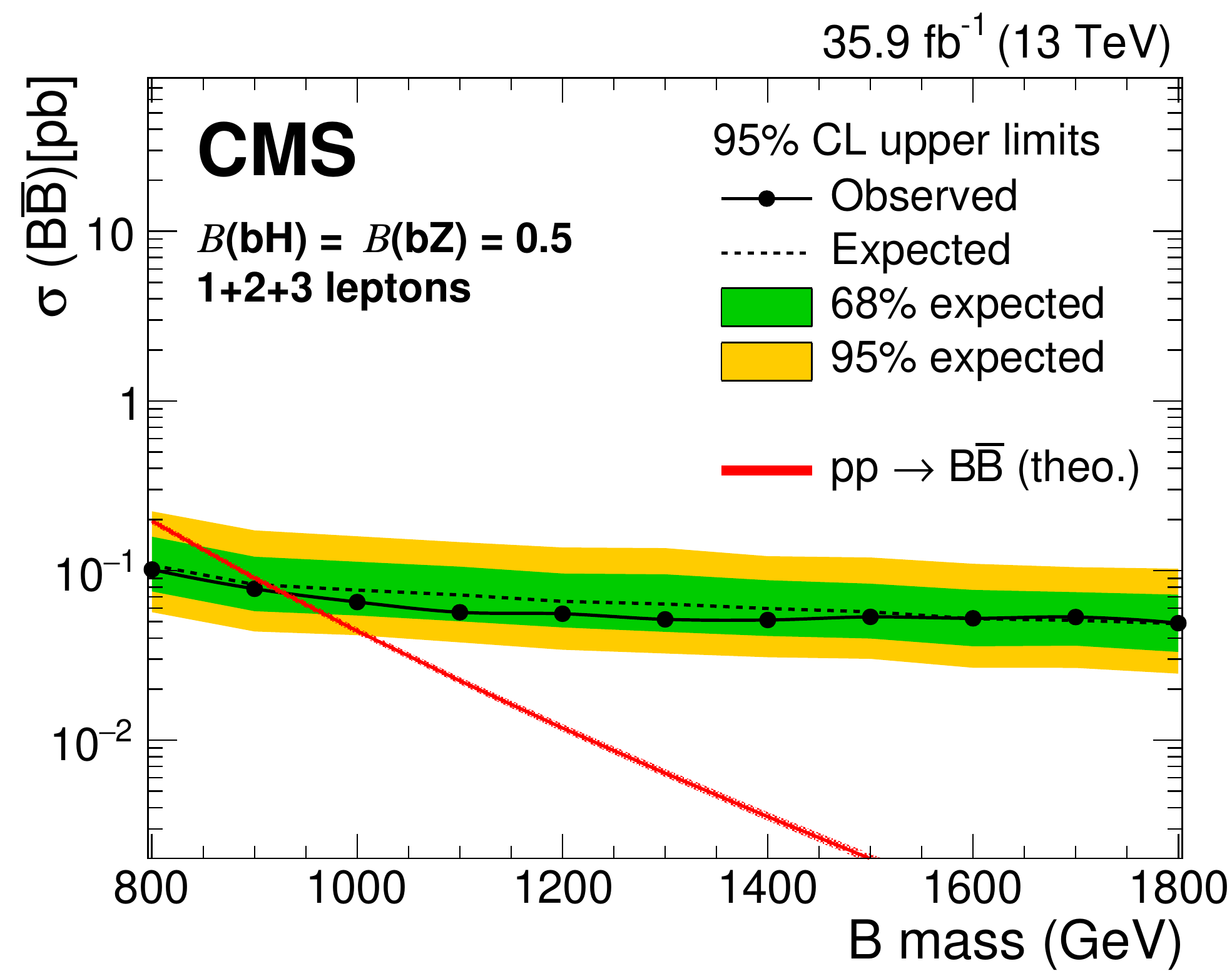}
\caption{The 95\% \CL expected and observed upper limits on the cross section of \TTbar (upper row) and \BBbar (lower row) production after combining all channels for the singlet (left) and doublet (right) branching fraction scenarios. The predicted cross sections are shown by the red curve, with the uncertainty indicated by the width of the line.}
\label{fig:benchmarkBRs}
\end{figure}

\begin{figure}[hbtp]
\centering
\includegraphics[width=0.49\textwidth]{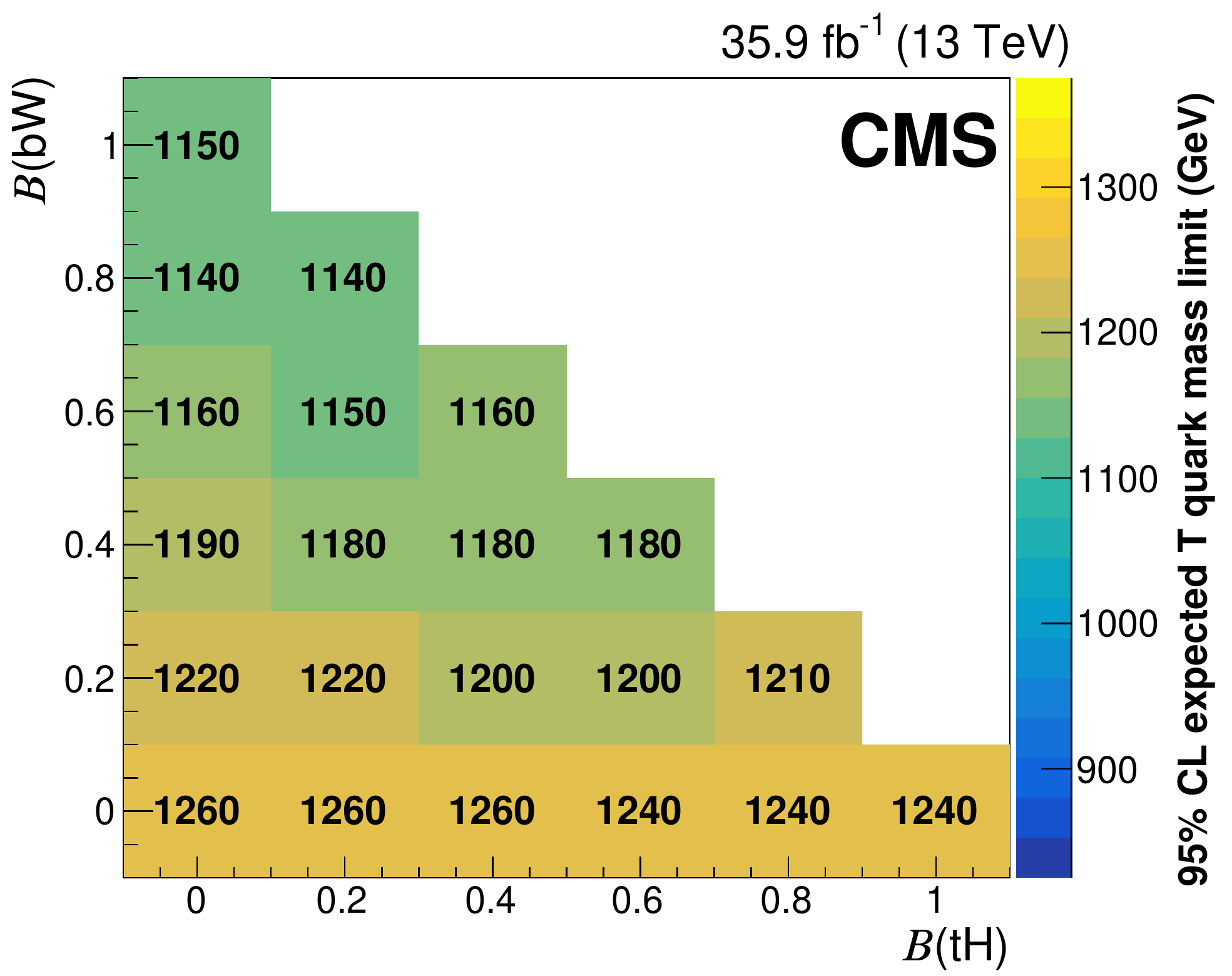}
\includegraphics[width=0.49\textwidth]{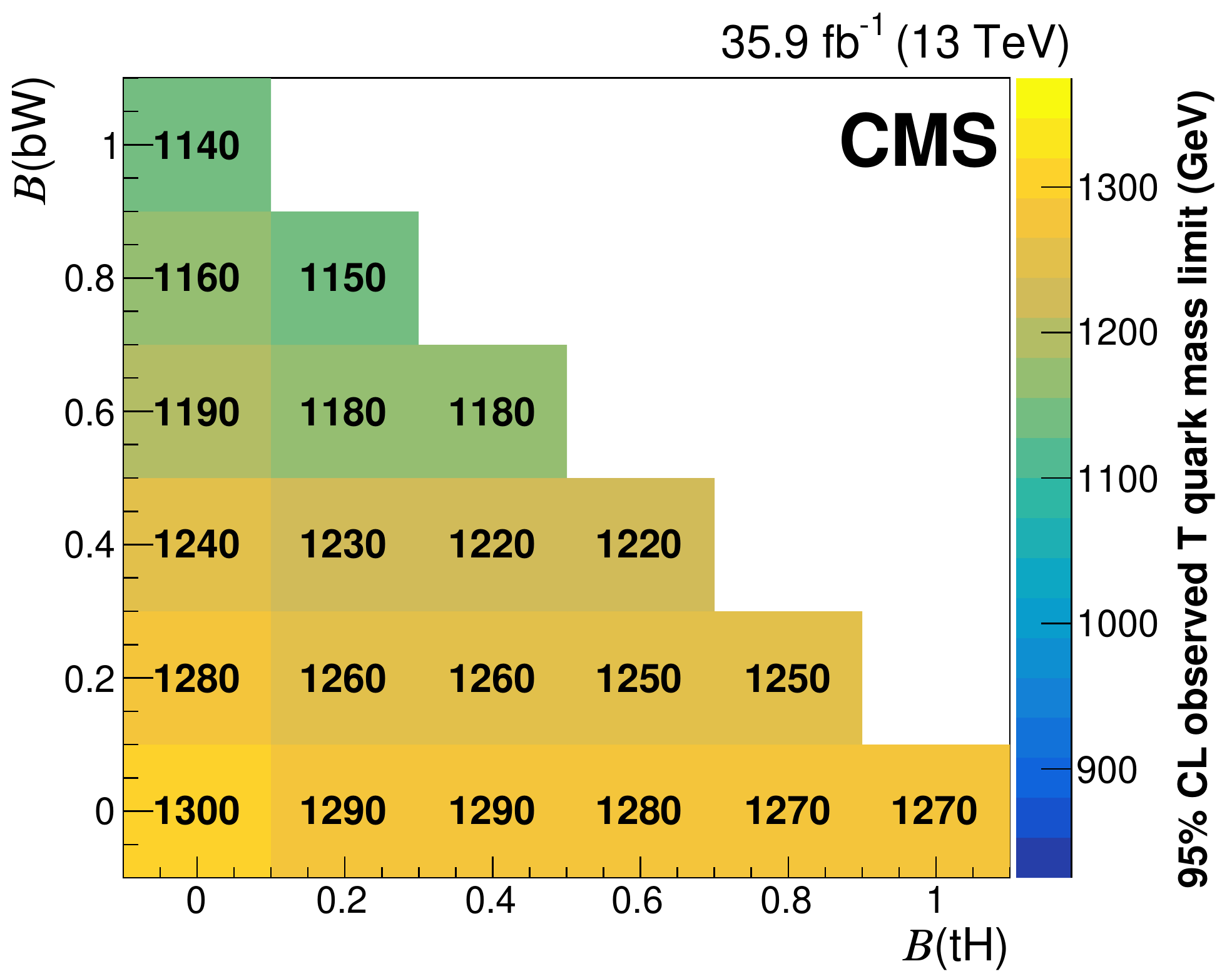}
\includegraphics[width=0.49\textwidth]{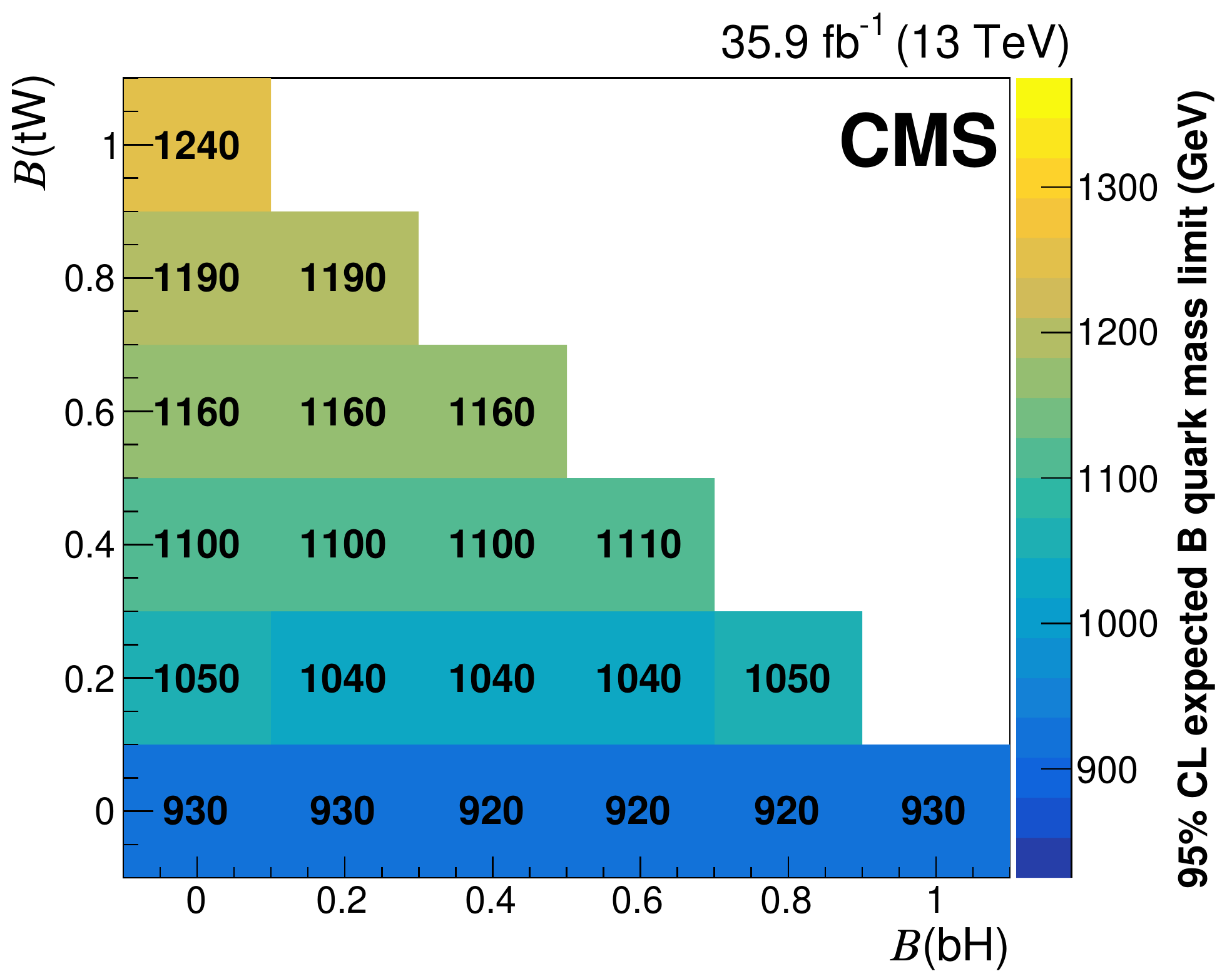}
\includegraphics[width=0.49\textwidth]{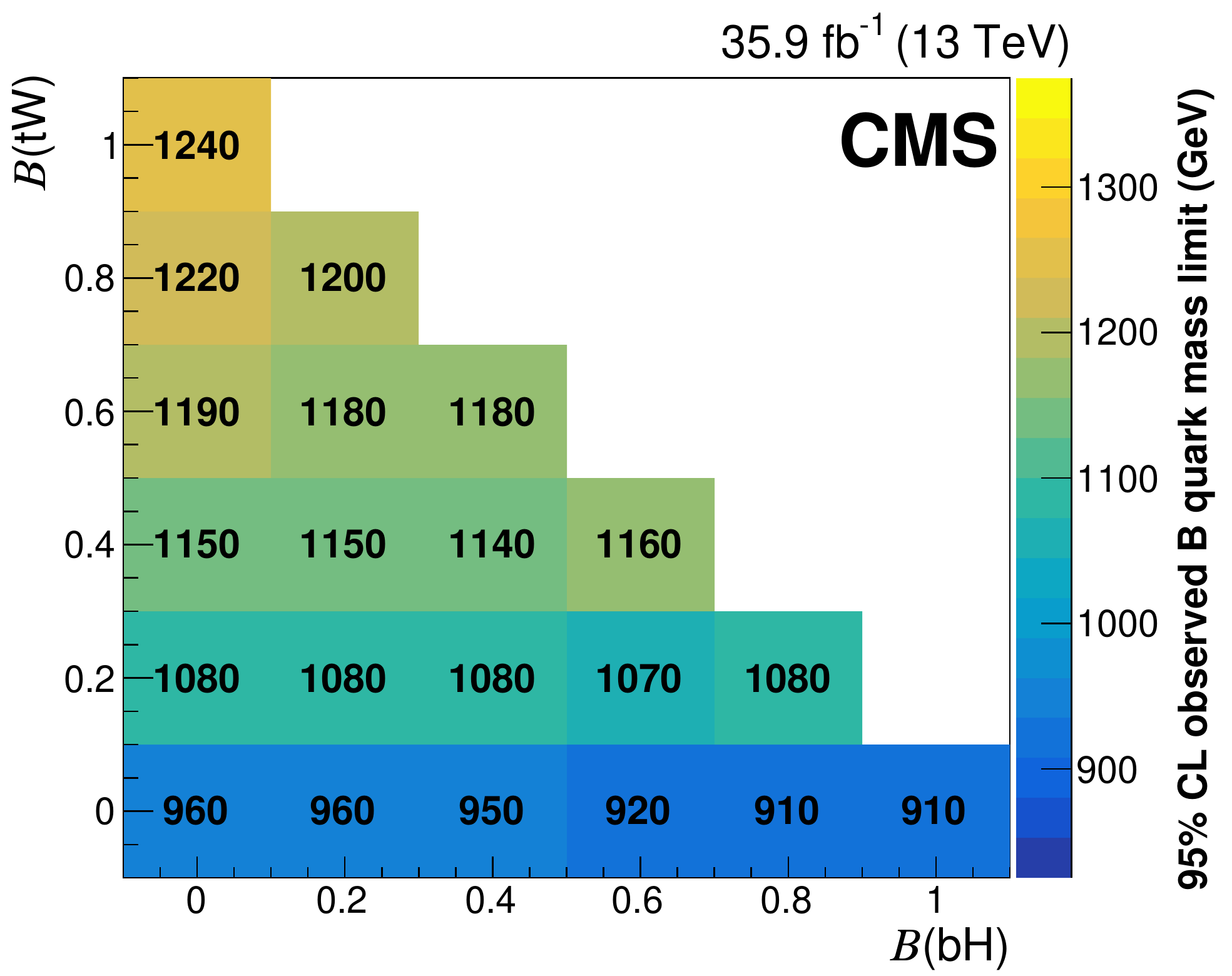}
\caption{The 95\% \CL expected (left) and observed (right) lower limits on the \PQT quark (upper row) and \PQB quark (lower row) mass, expressed in GeV, after combining all channels for various branching fraction scenarios.}
\label{fig:scanBRs}
\end{figure}

Figure~\ref{fig:benchmarkBRs} shows 95\% \CL upper limits on the production of \PQT and \PQB quarks in the benchmark branching fraction scenarios. We exclude singlet \PQT quark masses below 1200\GeV\,(1160\GeV expected), doublet \PQT quark masses below 1280\GeV\,(1240\GeV expected), singlet \PQB quark masses below below 1170\GeV\,(1130\GeV expected), and doublet \PQB quark masses below 940\GeV (920\GeV expected). Masses below 800\GeV were excluded in previous searches. For \PQT and \PQB quark masses in the range 800--1800\GeV, cross sections smaller than 30.4--9.4\unit{fb}\,(21.2--6.1\unit{fb}) and 40.6--9.4\unit{fb}\,(101--49.0\unit{fb}) are excluded for the singlet (doublet) scenario. Figure~\ref{fig:scanBRs} shows the expected and observed limits for scans over many possible \PQT and \PQB quark branching fraction scenarios. Based on the branching factions, lower limits on \PQT and \PQB quark masses range from 1140 to 1300\GeV, and from 910 to 1240\GeV.

\clearpage

\section{Summary}\label{sec:summary}

A search has been presented for pair-produced vector-like \PQT and \PQB quarks in a data sample of proton-proton collisions recorded during 2016 by the CMS experiment, and corresponding to an integrated luminosity of 35.9\fbinv. The search is performed in channels with one lepton, two same-sign leptons, or at least three leptons in the final state and makes use of techniques to identify Lorentz-boosted hadronically decaying \PW\ and Higgs bosons. Combining these channels, we exclude \PQT (\PQB) quarks at 95\% confidence level with masses below 1200 (1170)\GeV in the singlet branching fraction scenario and 1280 (940)\GeV in the doublet branching fraction scenario. For other branching fraction scenarios this search excludes \PQT (\PQB) quark masses below 1140--1300\GeV (910--1240\GeV). This represents an improvement in sensitivity of typically 200--600\GeV, compared to previous CMS results.

\begin{acknowledgments}
We congratulate our colleagues in the CERN accelerator departments for the excellent performance of the LHC and thank the technical and administrative staffs at CERN and at other CMS institutes for their contributions to the success of the CMS effort. In addition, we gratefully acknowledge the computing centres and personnel of the Worldwide LHC Computing Grid for delivering so effectively the computing infrastructure essential to our analyses. Finally, we acknowledge the enduring support for the construction and operation of the LHC and the CMS detector provided by the following funding agencies: BMWFW and FWF (Austria); FNRS and FWO (Belgium); CNPq, CAPES, FAPERJ, and FAPESP (Brazil); MES (Bulgaria); CERN; CAS, MoST, and NSFC (China); COLCIENCIAS (Colombia); MSES and CSF (Croatia); RPF (Cyprus); SENESCYT (Ecuador); MoER, ERC IUT, and ERDF (Estonia); Academy of Finland, MEC, and HIP (Finland); CEA and CNRS/IN2P3 (France); BMBF, DFG, and HGF (Germany); GSRT (Greece); NKFIA (Hungary); DAE and DST (India); IPM (Iran); SFI (Ireland); INFN (Italy); MSIP and NRF (Republic of Korea); LAS (Lithuania); MOE and UM (Malaysia); BUAP, CINVESTAV, CONACYT, LNS, SEP, and UASLP-FAI (Mexico); MBIE (New Zealand); PAEC (Pakistan); MSHE and NSC (Poland); FCT (Portugal); JINR (Dubna); MON, RosAtom, RAS and RFBR (Russia); MESTD (Serbia); SEIDI, CPAN, PCTI and FEDER (Spain); Swiss Funding Agencies (Switzerland); MST (Taipei); ThEPCenter, IPST, STAR, and NSTDA (Thailand); TUBITAK and TAEK (Turkey); NASU and SFFR (Ukraine); STFC (United Kingdom); DOE and NSF (USA).

\hyphenation{Rachada-pisek} Individuals have received support from the Marie-Curie programme and the European Research Council and Horizon 2020 Grant, contract No. 675440 (European Union); the Leventis Foundation; the A. P. Sloan Foundation; the Alexander von Humboldt Foundation; the Belgian Federal Science Policy Office; the Fonds pour la Formation \`a la Recherche dans l'Industrie et dans l'Agriculture (FRIA-Belgium); the Agentschap voor Innovatie door Wetenschap en Technologie (IWT-Belgium); the F.R.S.-FNRS and FWO (Belgium) under the ``Excellence of Science - EOS" - be.h project n. 30820817; the Ministry of Education, Youth and Sports (MEYS) of the Czech Republic; the Lend\"ulet (``Momentum") Programme and the J\'anos Bolyai Research Scholarship of the Hungarian Academy of Sciences, the New National Excellence Program \'UNKP, the NKFIA research grants 123842, 123959, 124845, 124850 and 125105 (Hungary); the Council of Science and Industrial Research, India; the HOMING PLUS programme of the Foundation for Polish Science, cofinanced from European Union, Regional Development Fund, the Mobility Plus programme of the Ministry of Science and Higher Education, the National Science Center (Poland), contracts Harmonia 2014/14/M/ST2/00428, Opus 2014/13/B/ST2/02543, 2014/15/B/ST2/03998, and 2015/19/B/ST2/02861, Sonata-bis 2012/07/E/ST2/01406; the National Priorities Research Program by Qatar National Research Fund; the Programa Estatal de Fomento de la Investigaci{\'o}n Cient{\'i}fica y T{\'e}cnica de Excelencia Mar\'{\i}a de Maeztu, grant MDM-2015-0509 and the Programa Severo Ochoa del Principado de Asturias; the Thalis and Aristeia programmes cofinanced by EU-ESF and the Greek NSRF; the Rachadapisek Sompot Fund for Postdoctoral Fellowship, Chulalongkorn University and the Chulalongkorn Academic into Its 2nd Century Project Advancement Project (Thailand); the Welch Foundation, contract C-1845; and the Weston Havens Foundation (USA).
\end{acknowledgments}
\bibliography{auto_generated}

\providecommand{\href}[2]{#2}\begingroup\raggedright\begin{thebibliography}{10}%
\makeatletter
\providecommand{\hrefCMSnoop }[0]{\@secondoftwo}%
\makeatother
\providecommand{\doi}{\texttt{doi:}\begingroup \urlstyle{tt}\Url}

\bibitem{Aad20121}
\hrefCMSnoop {}{{ATLAS Collaboration}, ``{Observation of a new particle in the
  search for the standard model Higgs boson with the ATLAS detector at the
  LHC}'',} \textit{ Phys. Lett. B} \textbf{ 716} (2012) 1,
  \href{http://dx.doi.org/10.1016/j.physletb.2012.08.020}{\doi{10.1016/j.physletb.2012.08.020}},
\href{http://www.arXiv.org/abs/1207.7214}{\texttt{arXiv:1207.7214}}.

\bibitem{Chatrchyan201230}
\hrefCMSnoop {}{{CMS Collaboration}, ``{Observation of a new boson at a mass of
  125 GeV with the CMS experiment at the LHC}'',} \textit{ Phys. Lett. B}
  \textbf{ 716} (2012) 30,
  \href{http://dx.doi.org/10.1016/j.physletb.2012.08.021}{\doi{10.1016/j.physletb.2012.08.021}},
\href{http://www.arXiv.org/abs/1207.7235}{\texttt{arXiv:1207.7235}}.

\bibitem{Chatrchyan:2013lba}
\hrefCMSnoop {}{{CMS Collaboration}, ``{Observation of a new boson with mass
  near 125 GeV in pp collisions at $\sqrt{s}$ = 7 and 8 TeV}'',} \textit{ JHEP}
  \textbf{ 06} (2013) 081,
  \href{http://dx.doi.org/10.1007/JHEP06(2013)081}{\doi{10.1007/JHEP06(2013)081}},
\href{http://www.arXiv.org/abs/1303.4571}{\texttt{arXiv:1303.4571}}.

\bibitem{PhysRevD.69.075002}
\hrefCMSnoop {}{M.~Perelstein, M.~E. Peskin, and A.~Pierce, ``Top quarks and
  electroweak symmetry breaking in little {Higgs} models'',} \textit{ Phys.
  Rev. D} \textbf{ 69} (2004) 075002,
  \href{http://dx.doi.org/10.1103/PhysRevD.69.075002}{\doi{10.1103/PhysRevD.69.075002}},
  \href{http://www.arXiv.org/abs/hep-ph/0310039}{\texttt{arXiv:hep-ph/0310039}}.

\bibitem{Matsedonskyi2013}
\hrefCMSnoop {}{O.~Matsedonskyi, G.~Panico, and A.~Wulzer, ``{Light top
  partners for a light composite Higgs}'',} \textit{ JHEP} \textbf{ 01} (2013)
  164,
  \href{http://dx.doi.org/10.1007/JHEP01(2013)164}{\doi{10.1007/JHEP01(2013)164}},
\href{http://www.arXiv.org/abs/1204.6333}{\texttt{arXiv:1204.6333}}.

\bibitem{PhysRevD.75.055014}
\hrefCMSnoop {}{R.~Contino, L.~Da~Rold, and A.~Pomarol, ``{Light custodians in
  natural composite Higgs models}'',} \textit{ Phys. Rev. D} \textbf{ 75}
  (2007) 055014,
  \href{http://dx.doi.org/10.1103/PhysRevD.75.055014}{\doi{10.1103/PhysRevD.75.055014}},
\href{http://www.arXiv.org/abs/hep-ph/0612048}{\texttt{arXiv:hep-ph/0612048}}.

\bibitem{compHiggs}
\hrefCMSnoop {}{R.~Contino, T.~Kramer, M.~Son, and R.~Sundrum,
  ``{Warped/composite phenomenology simplified}'',} \textit{ JHEP} \textbf{ 05}
  (2007) 074,
  \href{http://dx.doi.org/10.1088/1126-6708/2007/05/074}{\doi{10.1088/1126-6708/2007/05/074}},
\href{http://www.arXiv.org/abs/hep-ph/0612180}{\texttt{arXiv:hep-ph/0612180}}.

\bibitem{KAPLAN1991259}
\hrefCMSnoop {}{D.~B. Kaplan, ``Flavor at {SSC} energies: A new mechanism for
  dynamically generated fermion masses'',} \textit{ Nucl. Phys. B} \textbf{
  365} (1991) 259,
  \href{http://dx.doi.org/10.1016/S0550-3213(05)80021-5}{\doi{10.1016/S0550-3213(05)80021-5}}.

\bibitem{Dugan:1984hq}
\hrefCMSnoop {}{M.~J. Dugan, H.~Georgi, and D.~B. Kaplan, ``Anatomy of a
  composite higgs model'',} \textit{ Nucl. Phys. B} \textbf{ 254} (1985) 299,
\href{http://dx.doi.org/10.1016/0550-3213(85)90221-4}{\doi{10.1016/0550-3213(85)90221-4}}.

\bibitem{vecQuarkMix}
\hrefCMSnoop {}{J.~A. Aguilar-Saavedra, ``{Mixing with vector-like quarks:
  constraints and expectations}'',} \textit{ EPJ Web Conf.} \textbf{ 60} (2013)
  16012,
  \href{http://dx.doi.org/10.1051/epjconf/20136016012}{\doi{10.1051/epjconf/20136016012}},
\href{http://www.arXiv.org/abs/1306.4432}{\texttt{arXiv:1306.4432}}.

\bibitem{PhysRevLett.82.1628}
\hrefCMSnoop {}{F.~del Aguila, J.~A. Aguilar-Saavedra, and R.~Miquel,
  ``{Constraints on top couplings in models with exotic quarks}'',} \textit{
  Phys. Rev. Lett.} \textbf{ 82} (1999) 1628,
  \href{http://dx.doi.org/10.1103/PhysRevLett.82.1628}{\doi{10.1103/PhysRevLett.82.1628}},
\href{http://www.arXiv.org/abs/hep-ph/9808400}{\texttt{arXiv:hep-ph/9808400}}.

\bibitem{LEP-2}
\hrefCMSnoop {}{{ALEPH, DELPHI, L3, and OPAL Collaborations}, ``{Electroweak
  Measurements in Electron-Positron Collisions at W-Boson-Pair Energies at
  LEP}'',} \textit{ Phys. Rept.} \textbf{ 532} (2013) 119,
  \href{http://dx.doi.org/10.1016/j.physrep.2013.07.004}{\doi{10.1016/j.physrep.2013.07.004}},
\href{http://www.arXiv.org/abs/1302.3415}{\texttt{arXiv:1302.3415}}.

\bibitem{PhysRevLett.109.241802}
O.~Eberhardt\hrefCMSnoop {}{ {et~al.}, ``Impact of a {Higgs} boson at a mass of
  126 {GeV} on the standard model with three and four fermion generations'',}
  \textit{ Phys. Rev. Lett.} \textbf{ 109} (2012) 241802,
  \href{http://dx.doi.org/10.1103/PhysRevLett.109.241802}{\doi{10.1103/PhysRevLett.109.241802}},
  \href{http://www.arXiv.org/abs/1209.1101}{\texttt{arXiv:1209.1101}}.

\bibitem{Djouadi2012310}
\hrefCMSnoop {}{A.~Djouadi and A.~Lenz, ``{Sealing the fate of a fourth
  generation of fermions}'',} \textit{ Phys. Lett. B} \textbf{ 715} (2012) 310,
  \href{http://dx.doi.org/10.1016/j.physletb.2012.07.060}{\doi{10.1016/j.physletb.2012.07.060}},
\href{http://www.arXiv.org/abs/1204.1252}{\texttt{arXiv:1204.1252}}.

\bibitem{Chatrchyan:2013sfs}
\hrefCMSnoop {}{{CMS Collaboration}, ``{Searches for Higgs bosons in pp
  collisions at $\sqrt{s}=7$ and 8 TeV in the context of four-generation and
  fermiophobic models}'',} \textit{ Phys. Lett. B} \textbf{ 725} (2013) 36,
  \href{http://dx.doi.org/10.1016/j.physletb.2013.06.043}{\doi{10.1016/j.physletb.2013.06.043}},
\href{http://www.arXiv.org/abs/1302.1764}{\texttt{arXiv:1302.1764}}.

\bibitem{PhysRevD.88.094010}
\hrefCMSnoop {}{J.~A. Aguilar-Saavedra, R.~Benbrik, S.~Heinemeyer, and
  M.~P\'{e}rez-Victoria, ``{Handbook of vectorlike quarks: Mixing and single
  production}'',} \textit{ Phys. Rev. D} \textbf{ 88} (2013) 094010,
  \href{http://dx.doi.org/10.1103/PhysRevD.88.094010}{\doi{10.1103/PhysRevD.88.094010}},
\href{http://www.arXiv.org/abs/1306.0572}{\texttt{arXiv:1306.0572}}.

\bibitem{DeSimone2013}
\hrefCMSnoop {}{A.~De~Simone, O.~Matsedonskyi, R.~Rattazzi, and A.~Wulzer, ``A
  first top partner hunter's guide'',} \textit{ JHEP} \textbf{ 04} (2013) 1,
  \href{http://dx.doi.org/10.1007/JHEP04(2013)004}{\doi{10.1007/JHEP04(2013)004}},
  \href{http://www.arXiv.org/abs/1211.5663}{\texttt{arXiv:1211.5663}}.

\bibitem{delAguila:1989rq}
\hrefCMSnoop {}{F.~del Aguila, L.~Ametller, G.~L. Kane, and J.~Vidal,
  ``{Vector-like fermion and standard Higgs production at hadron colliders}'',}
  \textit{ Nucl. Phys. B} \textbf{ 334} (1990) 1,
\href{http://dx.doi.org/10.1016/0550-3213(90)90655-W}{\doi{10.1016/0550-3213(90)90655-W}}.

\bibitem{EXO-11-005}
\hrefCMSnoop {}{{CMS Collaboration}, ``{Search for a vector-like quark with
  charge 2/3 in t + Z events from pp collisions at $\sqrt{s}=7$ TeV}'',}
  \textit{ Phys. Rev. Lett.} \textbf{ 107} (2011) 271802,
  \href{http://dx.doi.org/10.1103/PhysRevLett.107.271802}{\doi{10.1103/PhysRevLett.107.271802}},
\href{http://www.arXiv.org/abs/1109.4985}{\texttt{arXiv:1109.4985}}.

\bibitem{EXO-11-099}
\hrefCMSnoop {}{{CMS Collaboration}, ``{Search for pair produced
  fourth-generation up-type quarks in pp collisions at $\sqrt{s}=7$ TeV with a
  lepton in the final state}'',} \textit{ Phys. Lett. B} \textbf{ 718} (2012)
  307,
  \href{http://dx.doi.org/10.1016/j.physletb.2012.10.038}{\doi{10.1016/j.physletb.2012.10.038}},
\href{http://www.arXiv.org/abs/1209.0471}{\texttt{arXiv:1209.0471}}.

\bibitem{Aad:2012bdq}
\hrefCMSnoop {}{{ATLAS Collaboration}, ``{Search for pair production of a new
  b' quark that decays to a Z boson and a bottom quark with the ATLAS
  detector}'',} \textit{ Phys. Rev. Lett.} \textbf{ 109} (2012) 071801,
  \href{http://dx.doi.org/10.1103/PhysRevLett.109.071801}{\doi{10.1103/PhysRevLett.109.071801}},
\href{http://www.arXiv.org/abs/1204.1265}{\texttt{arXiv:1204.1265}}.

\bibitem{CMScombo2014}
\hrefCMSnoop {}{{CMS Collaboration}, ``{Search for vector-like charge 2/3 T
  quarks in proton-proton collisions at $\sqrt{s} = 8$ TeV}'',} \textit{ Phys.
  Rev. D} \textbf{ 92} (2016) 012003,
  \href{http://dx.doi.org/10.1103/PhysRevD.92.012003}{\doi{10.1103/PhysRevD.92.012003}},
  \href{http://www.arXiv.org/abs/1509.04177}{\texttt{arXiv:1509.04177}}.

\bibitem{Run1anal}
\hrefCMSnoop {}{{CMS Collaboration}, ``{Inclusive search for a vector-like T
  quark with charge $\frac{2}{3}$ in pp collisions at $\sqrt{s}$ = 8 TeV}'',}
  \textit{ Phys. Lett. B} \textbf{ 729} (2014) 149,
  \href{http://dx.doi.org/10.1016/j.physletb.2014.01.006}{\doi{10.1016/j.physletb.2014.01.006}},
\href{http://www.arXiv.org/abs/1311.7667}{\texttt{arXiv:1311.7667}}.

\bibitem{PhysRevD.92.112007}
\hrefCMSnoop {}{{ATLAS Collaboration}, ``{Search for pair production of a new
  heavy quark that decays into a $W$ boson and a light quark in $pp$ collisions
  at $\sqrt{s} = 8$ TeV with the ATLAS detector}'',} \textit{ Phys. Rev. D}
  \textbf{ 92} (2015) 112007,
  \href{http://dx.doi.org/10.1103/PhysRevD.92.112007}{\doi{10.1103/PhysRevD.92.112007}},
\href{http://www.arXiv.org/abs/1509.04261}{\texttt{arXiv:1509.04261}}.

\bibitem{Aad2015}
\hrefCMSnoop {}{{ATLAS Collaboration}, ``{Search for production of vector-like
  quark pairs and of four top quarks in the lepton-plus-jets final state in
  $pp$ collisions at $\sqrt{s}=8$ TeV with the ATLAS detector}'',} \textit{
  JHEP} \textbf{ 08} (2015) 105,
  \href{http://dx.doi.org/10.1007/JHEP08(2015)105}{\doi{10.1007/JHEP08(2015)105}},
\href{http://www.arXiv.org/abs/1505.04306}{\texttt{arXiv:1505.04306}}.

\bibitem{Aaboud:2017qpr}
\hrefCMSnoop {}{{ATLAS Collaboration}, ``{Search for pair production of
  vector-like top quarks in events with one lepton, jets, and missing
  transverse momentum in $ \sqrt{s}=13 $ TeV $pp$ collisions with the ATLAS
  detector}'',} \textit{ JHEP} \textbf{ 08} (2017) 052,
  \href{http://dx.doi.org/10.1007/JHEP08(2017)052}{\doi{10.1007/JHEP08(2017)052}},
\href{http://www.arXiv.org/abs/1705.10751}{\texttt{arXiv:1705.10751}}.

\bibitem{Aaboud:2017zfn}
\hrefCMSnoop {}{{ATLAS Collaboration}, ``{Search for pair production of heavy
  vector-like quarks decaying to high-p$_{T}$ W bosons and b quarks in the
  lepton-plus-jets final state in pp collisions at $ \sqrt{s}=13 $ TeV with the
  ATLAS detector}'',} \textit{ JHEP} \textbf{ 10} (2017) 141,
  \href{http://dx.doi.org/10.1007/JHEP10(2017)141}{\doi{10.1007/JHEP10(2017)141}},
\href{http://www.arXiv.org/abs/1707.03347}{\texttt{arXiv:1707.03347}}.

\bibitem{B2G-16-024}
\hrefCMSnoop {}{{CMS Collaboration}, ``{Search for pair production of
  vector-like T and B quarks in single-lepton final states using boosted jet
  substructure in proton-proton collisions at $\sqrt{s}=13$ TeV}'',} \textit{
  JHEP} \textbf{ 11} (2017) 085,
  \href{http://dx.doi.org/10.1007/JHEP11(2017)085}{\doi{10.1007/JHEP11(2017)085}},
\href{http://www.arXiv.org/abs/1706.03408}{\texttt{arXiv:1706.03408}}.

\bibitem{B2G-17-003}
\hrefCMSnoop {}{{CMS Collaboration}, ``{Search for pair production of
  vector-like quarks in the bW$\overline{\mathrm{b}}$W channel from
  proton-proton collisions at $\sqrt{s} =$ 13 TeV}'',} \textit{ Phys. Lett. B}
  \textbf{ 779} (2018) 82,
  \href{http://dx.doi.org/10.1016/j.physletb.2018.01.077}{\doi{10.1016/j.physletb.2018.01.077}},
\href{http://www.arXiv.org/abs/1710.01539}{\texttt{arXiv:1710.01539}}.

\bibitem{Aaboud:2018xuw}
\hrefCMSnoop {}{{ATLAS Collaboration}, ``{Search for pair production of up-type
  vector-like quarks and for four-top-quark events in final states with
  multiple $b$-jets with the ATLAS detector}'',} (2018).
  \href{http://www.arXiv.org/abs/1803.09678}{\texttt{arXiv:1803.09678}}.
Submitted to {\it JHEP}.

\bibitem{Aaboud:2018uek}
\hrefCMSnoop {}{{ATLAS Collaboration}, ``{Search for pair production of heavy
  vector-like quarks decaying into high-$p_T$ $W$ bosons and top quarks in the
  lepton-plus-jets final state in $pp$ collisions at $\sqrt{s}=13$ TeV with the
  ATLAS detector}'',}
\href{http://www.arXiv.org/abs/1806.01762}{\texttt{arXiv:1806.01762}}.

\bibitem{Aaboud:2018saj}
\hrefCMSnoop {}{{ATLAS Collaboration}, ``{Search for pair- and
  single-production of vector-like quarks in final states with at least one $Z$
  boson decaying into a pair of electrons or muons in $pp$ collision data
  collected with the ATLAS detector at $\sqrt{s} = 13$ TeV}'',}
\href{http://www.arXiv.org/abs/1806.10555}{\texttt{arXiv:1806.10555}}.

\bibitem{Chatrchyan:2008zzk}
\hrefCMSnoop {}{{CMS Collaboration}, ``The {CMS} experiment at the {CERN}
  {LHC}'',} \textit{ JINST} \textbf{ 3} (2008) S08004,
\href{http://dx.doi.org/10.1088/1748-0221/3/08/S08004}{\doi{10.1088/1748-0221/3/08/S08004}}.

\bibitem{particleflow}
\hrefCMSnoop {}{{CMS Collaboration}, ``Particle-flow reconstruction and global
  event description with the {CMS} detector'',} \textit{ JINST} \textbf{ 12}
  (2017) P10003,
  \href{http://dx.doi.org/10.1088/1748-0221/12/10/P10003}{\doi{10.1088/1748-0221/12/10/P10003}},
\href{http://www.arXiv.org/abs/1706.04965}{\texttt{arXiv:1706.04965}}.

\bibitem{Cacciari:2008gp}
\hrefCMSnoop {}{M.~Cacciari, G.~P. Salam, and G.~Soyez, ``The anti-$\kt$ jet
  clustering algorithm'',} \textit{ JHEP} \textbf{ 04} (2008) 063,
  \href{http://dx.doi.org/10.1088/1126-6708/2008/04/063}{\doi{10.1088/1126-6708/2008/04/063}},
  \href{http://www.arXiv.org/abs/0802.1189}{\texttt{arXiv:0802.1189}}.

\bibitem{Cacciari:2011ma}
\hrefCMSnoop {}{M.~Cacciari, G.~P. Salam, and G.~Soyez, ``{FastJet user
  manual}'',} \textit{ Eur. Phys. J. C} \textbf{ 72} (2012) 1896,
  \href{http://dx.doi.org/10.1140/epjc/s10052-012-1896-2}{\doi{10.1140/epjc/s10052-012-1896-2}},
\href{http://www.arXiv.org/abs/1111.6097}{\texttt{arXiv:1111.6097}}.

\bibitem{Cacciari:2008gn}
\hrefCMSnoop {}{M.~Cacciari, G.~P. Salam, and G.~Soyez, ``{The catchment area
  of jets}'',} \textit{ JHEP} \textbf{ 04} (2008) 005,
  \href{http://dx.doi.org/10.1088/1126-6708/2008/04/005}{\doi{10.1088/1126-6708/2008/04/005}},
\href{http://www.arXiv.org/abs/0802.1188}{\texttt{arXiv:0802.1188}}.

\bibitem{Chatrchyan:2011ds}
\hrefCMSnoop {}{{CMS Collaboration}, ``{Determination of jet energy calibration
  and transverse momentum resolution in CMS}'',} \textit{ JINST} \textbf{ 6}
  (2011) P11002,
  \href{http://dx.doi.org/10.1088/1748-0221/6/11/P11002}{\doi{10.1088/1748-0221/6/11/P11002}},
\href{http://www.arXiv.org/abs/1107.4277}{\texttt{arXiv:1107.4277}}.

\bibitem{JME-16-003}
\href {https://cds.cern.ch/record/2256875}{{CMS Collaboration}, ``{Jet
  algorithms performance in 13 TeV data}'',} Technical Report
  CMS-PAS-JME-16-003, CERN, 2017.

\bibitem{Khachatryan:2016bia}
\hrefCMSnoop {}{{CMS Collaboration}, ``{The CMS trigger system}'',} \textit{
  JINST} \textbf{ 12} (2017) P01020,
  \href{http://dx.doi.org/10.1088/1748-0221/12/01/P01020}{\doi{10.1088/1748-0221/12/01/P01020}},
\href{http://www.arXiv.org/abs/1609.02366}{\texttt{arXiv:1609.02366}}.

\bibitem{NNPDF30}
\hrefCMSnoop {}{{NNPDF} Collaboration, ``{Parton distributions for the LHC Run
  II}'',} \textit{ JHEP} \textbf{ 04} (2015) 040,
  \href{http://dx.doi.org/10.1007/JHEP04(2015)040}{\doi{10.1007/JHEP04(2015)040}},
  \href{http://www.arXiv.org/abs/1410.8849}{\texttt{arXiv:1410.8849}}.

\bibitem{Nason:2004rx}
\hrefCMSnoop {}{P.~Nason, ``{A new method for combining NLO QCD with shower
  Monte Carlo algorithms}'',} \textit{ JHEP} \textbf{ 11} (2004) 040,
  \href{http://dx.doi.org/10.1088/1126-6708/2004/11/040}{\doi{10.1088/1126-6708/2004/11/040}},
\href{http://www.arXiv.org/abs/hep-ph/0409146}{\texttt{arXiv:hep-ph/0409146}}.

\bibitem{Frixione:2007vw}
\hrefCMSnoop {}{S.~Frixione, P.~Nason, and C.~Oleari, ``{Matching NLO QCD
  computations with Parton Shower simulations: the POWHEG method}'',} \textit{
  JHEP} \textbf{ 11} (2007) 070,
  \href{http://dx.doi.org/10.1088/1126-6708/2007/11/070}{\doi{10.1088/1126-6708/2007/11/070}},
\href{http://www.arXiv.org/abs/0709.2092}{\texttt{arXiv:0709.2092}}.

\bibitem{Alioli:2010xd}
\hrefCMSnoop {}{S.~Alioli, P.~Nason, C.~Oleari, and E.~Re, ``{A general
  framework for implementing NLO calculations in shower Monte Carlo programs:
  the POWHEG BOX}'',} \textit{ JHEP} \textbf{ 06} (2010) 043,
  \href{http://dx.doi.org/10.1007/JHEP06(2010)043}{\doi{10.1007/JHEP06(2010)043}},
\href{http://www.arXiv.org/abs/1002.2581}{\texttt{arXiv:1002.2581}}.

\bibitem{Frixione:2007nw}
\hrefCMSnoop {}{S.~Frixione, P.~Nason, and G.~Ridolfi, ``A positive-weight
  next-to-leading-order {Monte Carlo} for heavy flavour hadroproduction'',}
  \textit{ JHEP} \textbf{ 09} (2007) 126,
  \href{http://dx.doi.org/10.1088/1126-6708/2007/09/126}{\doi{10.1088/1126-6708/2007/09/126}},
\href{http://www.arXiv.org/abs/0707.3088}{\texttt{arXiv:0707.3088}}.

\bibitem{MADGRAPH}
J.~Alwall\hrefCMSnoop {}{ {et~al.}, ``{The automated computation of tree-level
  and next-to-leading order differential cross sections, and their matching to
  parton shower simulations}'',} \textit{ JHEP} \textbf{ 07} (2014) 079,
  \href{http://dx.doi.org/10.1007/JHEP07(2014)079}{\doi{10.1007/JHEP07(2014)079}},
\href{http://www.arXiv.org/abs/1405.0301}{\texttt{arXiv:1405.0301}}.

\bibitem{FXFX}
\hrefCMSnoop {}{R.~Frederix and S.~Frixione, ``{Merging meets matching in
  MC@NLO}'',} \textit{ JHEP} \textbf{ 12} (2012) 061,
  \href{http://dx.doi.org/10.1007/JHEP12(2012)061}{\doi{10.1007/JHEP12(2012)061}},
  \href{http://www.arXiv.org/abs/1209.6215}{\texttt{arXiv:1209.6215}}.

\bibitem{MLMmatching}
J.~Alwall\hrefCMSnoop {}{ {et~al.}, ``Comparative study of various algorithms
  for the merging of parton showers and matrix elements in hadronic
  collisions'',} \textit{ Eur. Phys. J. C} \textbf{ 53} (2008) 473,
  \href{http://dx.doi.org/10.1140/epjc/s10052-007-0490-5}{\doi{10.1140/epjc/s10052-007-0490-5}},
\href{http://www.arXiv.org/abs/0706.2569}{\texttt{arXiv:0706.2569}}.

\bibitem{Sjostrand:2006za}
\hrefCMSnoop {}{T.~Sj{\"o}strand, S.~Mrenna, and P.~Skands, ``{PYTHIA} 6.4
  physics and manual'',} \textit{ JHEP} \textbf{ 05} (2006) 026,
  \href{http://dx.doi.org/10.1088/1126-6708/2006/05/026}{\doi{10.1088/1126-6708/2006/05/026}},
\href{http://www.arXiv.org/abs/hep-ph/0603175}{\texttt{arXiv:hep-ph/0603175}}.

\bibitem{Sjostrand:2014zea}
T.~Sj{\"o}strand\hrefCMSnoop {}{ {et~al.}, ``An introduction to {PYTHIA}
  8.2'',} \textit{ Comput. Phys. Commun.} \textbf{ 191} (2015) 159,
  \href{http://dx.doi.org/10.1016/j.cpc.2015.01.024}{\doi{10.1016/j.cpc.2015.01.024}},
\href{http://www.arXiv.org/abs/1410.3012}{\texttt{arXiv:1410.3012}}.

\bibitem{CUETP8M2T4}
\href {http://cds.cern.ch/record/2235192}{{CMS Collaboration},
  ``{Investigations of the impact of the parton shower tuning in Pythia 8 in
  the modelling of $\mathrm{t\overline{t}}$ at $\sqrt{s}=8$ and 13 TeV}'',}
  Technical Report CMS-PAS-TOP-16-021, CERN, 2016.

\bibitem{CUETP8M1}
\hrefCMSnoop {}{{CMS Collaboration}, ``{Event generator tunes obtained from
  underlying event and multiparton scattering measurements}'',} \textit{ Eur.
  Phys. J. C} \textbf{ 76} (2016) 155,
  \href{http://dx.doi.org/10.1140/epjc/s10052-016-3988-x}{\doi{10.1140/epjc/s10052-016-3988-x}},
\href{http://www.arXiv.org/abs/1512.00815}{\texttt{arXiv:1512.00815}}.

\bibitem{GEANT4}
\hrefCMSnoop {}{{GEANT4} Collaboration, ``{GEANT4}---a simulation toolkit'',}
  \textit{ Nucl. Instrum. Meth. A} \textbf{ 506} (2003) 250,
\href{http://dx.doi.org/10.1016/S0168-9002(03)01368-8}{\doi{10.1016/S0168-9002(03)01368-8}}.

\bibitem{TPRIMEXSEC}
\hrefCMSnoop {}{M.~Czakon and A.~Mitov, ``{Top++: A Program for the calculation
  of the top-pair cross-section at hadron colliders}'',} \textit{ Comput. Phys.
  Commun.} \textbf{ 185} (2014) 2930,
  \href{http://dx.doi.org/10.1016/j.cpc.2014.06.021}{\doi{10.1016/j.cpc.2014.06.021}},
\href{http://www.arXiv.org/abs/1112.5675}{\texttt{arXiv:1112.5675}}.

\bibitem{MITOV1}
\hrefCMSnoop {}{M.~Czakon, P.~Fiedler, and A.~Mitov, ``{Total top-quark
  pair-production cross section at hadron colliders through
  $\mathcal{O}(\alpha^{4}_{S})$}'',} \textit{ Phys. Rev. Lett.} \textbf{ 110}
  (2013) 252004,
  \href{http://dx.doi.org/10.1103/PhysRevLett.110.252004}{\doi{10.1103/PhysRevLett.110.252004}},
\href{http://www.arXiv.org/abs/1303.6254}{\texttt{arXiv:1303.6254}}.

\bibitem{MITOV2}
\hrefCMSnoop {}{M.~Czakon and A.~Mitov, ``{NNLO corrections to top pair
  production at hadron colliders: the quark-gluon reaction}'',} \textit{ JHEP}
  \textbf{ 01} (2013) 080,
  \href{http://dx.doi.org/10.1007/JHEP01(2013)080}{\doi{10.1007/JHEP01(2013)080}},
\href{http://www.arXiv.org/abs/1210.6832}{\texttt{arXiv:1210.6832}}.

\bibitem{MITOV3}
\hrefCMSnoop {}{M.~Czakon and A.~Mitov, ``{NNLO corrections to top-pair
  production at hadron colliders: the all-fermionic scattering channels}'',}
  \textit{ JHEP} \textbf{ 12} (2012) 054,
  \href{http://dx.doi.org/10.1007/JHEP12(2012)054}{\doi{10.1007/JHEP12(2012)054}},
\href{http://www.arXiv.org/abs/1207.0236}{\texttt{arXiv:1207.0236}}.

\bibitem{BARNREUTHER}
\hrefCMSnoop {}{P.~B{\"a}rnreuther, M.~Czakon, and A.~Mitov,
  ``{Percent-level-precision physics at the Tevatron: first genuine
  next-to-next-to-leading order QCD corrections to $q \bar{q} \to t \bar{t} +
  X$}'',} \textit{ Phys. Rev. Lett.} \textbf{ 109} (2012) 132001,
  \href{http://dx.doi.org/10.1103/PhysRevLett.109.132001}{\doi{10.1103/PhysRevLett.109.132001}},
\href{http://www.arXiv.org/abs/1204.5201}{\texttt{arXiv:1204.5201}}.

\bibitem{NNLL}
M.~Cacciari\hrefCMSnoop {}{ {et~al.}, ``{Top-pair production at hadron
  colliders with next-to-next-to-leading logarithmic soft-gluon
  resummation}'',} \textit{ Phys. Lett. B} \textbf{ 710} (2012) 612,
  \href{http://dx.doi.org/10.1016/j.physletb.2012.03.013}{\doi{10.1016/j.physletb.2012.03.013}},
\href{http://www.arXiv.org/abs/1111.5869}{\texttt{arXiv:1111.5869}}.

\bibitem{Chatrchyan:2014fea}
\hrefCMSnoop {}{{CMS Collaboration}, ``{Description and performance of track
  and primary-vertex reconstruction with the CMS tracker}'',} \textit{ JINST}
  \textbf{ 9} (2014) P10009,
  \href{http://dx.doi.org/10.1088/1748-0221/9/10/P10009}{\doi{10.1088/1748-0221/9/10/P10009}},
\href{http://www.arXiv.org/abs/1405.6569}{\texttt{arXiv:1405.6569}}.

\bibitem{Khachatryan:2015hwa}
\hrefCMSnoop {}{{CMS Collaboration}, ``{Performance of electron reconstruction
  and selection with the CMS detector in proton-proton collisions at $\sqrt{s}
  = 8$\TeV}'',} \textit{ JINST} \textbf{ 10} (2015) P06005,
  \href{http://dx.doi.org/10.1088/1748-0221/10/06/P06005}{\doi{10.1088/1748-0221/10/06/P06005}},
\href{http://www.arXiv.org/abs/1502.02701}{\texttt{arXiv:1502.02701}}.

\bibitem{Chatrchyan:2012xi}
\hrefCMSnoop {}{{CMS Collaboration}, ``{Performance of CMS muon reconstruction
  in $pp$ collision events at $\sqrt{s} = 7$\TeV}'',} \textit{ JINST} \textbf{
  7} (2012) P10002,
  \href{http://dx.doi.org/10.1088/1748-0221/7/10/P10002}{\doi{10.1088/1748-0221/7/10/P10002}},
\href{http://www.arXiv.org/abs/1206.4071}{\texttt{arXiv:1206.4071}}.

\bibitem{BTV-16-002}
\hrefCMSnoop {}{{CMS Collaboration}, ``{Identification of heavy-flavour jets
  with the CMS detector in pp collisions at 13 TeV}'',} \textit{ JINST}
  \textbf{ 13} (2018) P05011,
  \href{http://dx.doi.org/10.1088/1748-0221/13/05/P05011}{\doi{10.1088/1748-0221/13/05/P05011}},
\href{http://www.arXiv.org/abs/1712.07158}{\texttt{arXiv:1712.07158}}.

\bibitem{Thaler:2010tr}
\hrefCMSnoop {}{J.~Thaler and K.~Van~Tilburg, ``{Identifying boosted objects
  with N-subjettiness}'',} \textit{ JHEP} \textbf{ 03} (2011) 015,
  \href{http://dx.doi.org/10.1007/JHEP03(2011)015}{\doi{10.1007/JHEP03(2011)015}},
\href{http://www.arXiv.org/abs/1011.2268}{\texttt{arXiv:1011.2268}}.

\bibitem{PRUNING}
\hrefCMSnoop {}{S.~D. Ellis, C.~K. Vermilion, and J.~R. Walsh, ``{Techniques
  for improved heavy particle searches with jet substructure}'',} \textit{
  Phys. Rev. D} \textbf{ 80} (2009) 051501,
  \href{http://dx.doi.org/10.1103/PhysRevD.80.051501}{\doi{10.1103/PhysRevD.80.051501}},
\href{http://www.arXiv.org/abs/0903.5081}{\texttt{arXiv:0903.5081}}.

\bibitem{SOFTDROP}
\hrefCMSnoop {}{A.~J. Larkoski, S.~Marzani, G.~Soyez, and J.~Thaler, ``Soft
  drop'',} \textit{ JHEP} \textbf{ 05} (2014) 146,
  \href{http://dx.doi.org/10.1007/JHEP05(2014)146}{\doi{10.1007/JHEP05(2014)146}},
\href{http://www.arXiv.org/abs/1402.2657}{\texttt{arXiv:1402.2657}}.

\bibitem{Dasgupta:2013ihk}
\hrefCMSnoop {}{M.~Dasgupta, A.~Fregoso, S.~Marzani, and G.~P. Salam,
  ``{Towards an understanding of jet substructure}'',} \textit{ JHEP} \textbf{
  09} (2013) 029,
  \href{http://dx.doi.org/10.1007/JHEP09(2013)029}{\doi{10.1007/JHEP09(2013)029}},
\href{http://www.arXiv.org/abs/1307.0007}{\texttt{arXiv:1307.0007}}.

\bibitem{TOP-16-008}
\hrefCMSnoop {}{{CMS Collaboration}, ``{Measurement of differential cross
  sections for top quark pair production using the lepton+jets final state in
  proton-proton collisions at 13 TeV}'',} \textit{ Phys. Rev. D} \textbf{ 95}
  (2017) 092001,
  \href{http://dx.doi.org/10.1103/PhysRevD.95.092001}{\doi{10.1103/PhysRevD.95.092001}},
\href{http://www.arXiv.org/abs/1610.04191}{\texttt{arXiv:1610.04191}}.

\bibitem{B2G16005}
\hrefCMSnoop {}{{CMS Collaboration}, ``{Search for electroweak production of a
  vector-like quark decaying to a top quark and a Higgs boson using boosted
  topologies in fully hadronic final states}'',} \textit{ JHEP} \textbf{ 04}
  (2017) 136,
  \href{http://dx.doi.org/10.1007/JHEP04(2017)136}{\doi{10.1007/JHEP04(2017)136}},
\href{http://www.arXiv.org/abs/1612.05336}{\texttt{arXiv:1612.05336}}.

\bibitem{B2G16006}
\hrefCMSnoop {}{{CMS Collaboration}, ``{Search for single production of
  vector-like quarks decaying into a b quark and a W boson in proton-proton
  collisions at $\sqrt s =$ 13 TeV}'',} \textit{ Phys. Lett. B} \textbf{ 772}
  (2017) 634,
  \href{http://dx.doi.org/10.1016/j.physletb.2017.07.022}{\doi{10.1016/j.physletb.2017.07.022}},
\href{http://www.arXiv.org/abs/1701.08328}{\texttt{arXiv:1701.08328}}.

\bibitem{TTDIFFXSEC}
\hrefCMSnoop {}{{CMS Collaboration}, ``{Measurements of differential cross
  sections of top quark pair production as a function of kinematic event
  variables in proton-proton collisions at $\sqrt{s}=13$ TeV}'',} \textit{
  JHEP} \textbf{ 06} (2018) 002,
  \href{http://dx.doi.org/10.1007/JHEP06(2018)002}{\doi{10.1007/JHEP06(2018)002}},
\href{http://www.arXiv.org/abs/1803.03991}{\texttt{arXiv:1803.03991}}.

\bibitem{susy2011}
\hrefCMSnoop {}{{CMS Collaboration}, ``{Search for new physics with same-sign
  isolated dilepton events with jets and missing transverse energy at the
  LHC}'',} \textit{ JHEP} \textbf{ 06} (2011) 077,
  \href{http://dx.doi.org/10.1007/JHEP06(2011)077}{\doi{10.1007/JHEP06(2011)077}},
\href{http://www.arXiv.org/abs/1104.3168}{\texttt{arXiv:1104.3168}}.

\bibitem{LUM-17-001}
\href {http://cds.cern.ch/record/2257069}{{CMS Collaboration}, ``{CMS}
  luminosity measurement for the 2016 data taking period'',} CMS Physics
  Analysis Summary CMS-PAS-LUM-17-001, CERN, 2017.

\bibitem{SMP-14-016}
\hrefCMSnoop {}{{CMS Collaboration}, ``Measurement of the
  {${\mathrm{W}}^{+}\mathrm{W}^{-}$} cross section in pp collisions at
  $\sqrt{s}=8$ {TeV} and limits on anomalous gauge couplings'',} \textit{ Eur.
  Phys. J. C} \textbf{ 76} (2016) 401,
  \href{http://dx.doi.org/10.1140/epjc/s10052-016-4219-1}{\doi{10.1140/epjc/s10052-016-4219-1}},
\href{http://www.arXiv.org/abs/1507.03268}{\texttt{arXiv:1507.03268}}.

\bibitem{SMP-16-002}
\hrefCMSnoop {}{{CMS Collaboration}, ``{Measurement of the WZ production cross
  section in pp collisions at $\sqrt{s} =$ 13 TeV}'',} \textit{ Phys. Lett. B}
  \textbf{ 766} (2017) 268,
  \href{http://dx.doi.org/10.1016/j.physletb.2017.01.011}{\doi{10.1016/j.physletb.2017.01.011}},
\href{http://www.arXiv.org/abs/1607.06943}{\texttt{arXiv:1607.06943}}.

\bibitem{SMP-16-001}
\hrefCMSnoop {}{{CMS Collaboration}, ``{Measurement of the ZZ production cross
  section and $\mathrm{Z}\to\ell^+\ell^-\ell'^+\ell'^-$ branching fraction in
  pp collisions at $\sqrt{s}=13$ TeV}'',} \textit{ Phys. Lett. B} \textbf{ 763}
  (2016) 280,
  \href{http://dx.doi.org/10.1016/j.physletb.2017.09.030}{\doi{10.1016/j.physletb.2017.09.030}},
  \href{http://www.arXiv.org/abs/1607.08834}{\texttt{arXiv:1607.08834}}.
[Erratum: {\it Phys.~Lett.~B\/}~{\bf 772}~(2017)~884].

\bibitem{FSQ-15-005}
\hrefCMSnoop {}{{CMS Collaboration}, ``{Measurement of the inelastic
  proton-proton cross section at $\sqrt{s}=$ 13 TeV}'',} (2018).
  \href{http://www.arXiv.org/abs/1802.02613}{\texttt{arXiv:1802.02613}}.
Submitted to {\it JHEP}.

\bibitem{THETA}
\href {http://www-ekp.physik.uni-karlsruhe.de/\~ott/theta/theta-auto}{J.~Ott,
  ``\textsc{Theta}---{A} framework for template-based modeling and
  inference'',} 2010.
\newblock \url {http://www-ekp.physik.uni-karlsruhe.de/\~ott/theta/theta-auto}.

\bibitem{PDGSTATS}
\hrefCMSnoop {}{{Particle Data Group}, C.~Patrignani {et~al.}, ``Review of
  particle physics'',} \textit{ Chin. Phys. C} \textbf{ 40} (2016) 100001,
  \href{http://dx.doi.org/10.1088/1674-1137/40/10/100001}{\doi{10.1088/1674-1137/40/10/100001}}.

\bibitem{BBLITE1}
\hrefCMSnoop {}{R.~J. Barlow and C.~Beeston, ``{Fitting using finite Monte
  Carlo samples}'',} \textit{ Comput. Phys. Commun.} \textbf{ 77} (1993) 219,
\href{http://dx.doi.org/10.1016/0010-4655(93)90005-W}{\doi{10.1016/0010-4655(93)90005-W}}.

\bibitem{BBLITE2}
\hrefCMSnoop {}{J.~S. Conway, ``Incorporating nuisance parameters in
  likelihoods for multisource spectra'',} in \textit{ {Proceedings, PHYSTAT
  2011 Workshop on Statistical Issues Related to Discovery Claims in Search
  Experiments and Unfolding}, address = {CERN,Geneva, Switzerland}, month =
  {January}}, p.~115.
\newblock 2011.
\newblock \href{http://www.arXiv.org/abs/1103.0354}{\texttt{arXiv:1103.0354}}.
\newblock
\href{http://dx.doi.org/10.5170/CERN-2011-006.115}{\doi{10.5170/CERN-2011-006.115}}.

\end{thebibliography}\endgroup
\cleardoublepage \appendix\section{The CMS Collaboration \label{app:collab}}\begin{sloppypar}\hyphenpenalty=5000\widowpenalty=500\clubpenalty=5000\vskip\cmsinstskip
\textbf{Yerevan Physics Institute, Yerevan, Armenia}\\*[0pt]
A.M.~Sirunyan, A.~Tumasyan
\vskip\cmsinstskip
\textbf{Institut f\"{u}r Hochenergiephysik, Wien, Austria}\\*[0pt]
W.~Adam, F.~Ambrogi, E.~Asilar, T.~Bergauer, J.~Brandstetter, E.~Brondolin, M.~Dragicevic, J.~Er\"{o}, A.~Escalante~Del~Valle, M.~Flechl, R.~Fr\"{u}hwirth\cmsAuthorMark{1}, V.M.~Ghete, J.~Hrubec, M.~Jeitler\cmsAuthorMark{1}, N.~Krammer, I.~Kr\"{a}tschmer, D.~Liko, T.~Madlener, I.~Mikulec, N.~Rad, H.~Rohringer, J.~Schieck\cmsAuthorMark{1}, R.~Sch\"{o}fbeck, M.~Spanring, D.~Spitzbart, A.~Taurok, W.~Waltenberger, J.~Wittmann, C.-E.~Wulz\cmsAuthorMark{1}, M.~Zarucki
% ...
\vskip\cmsinstskip
% ...
1:  Also at Vienna University of Technology, Vienna, Austria\\
% ...
\end{sloppypar}
\end{document}